\documentclass[lettersize,journal]{IEEEtran}
\usepackage{amsmath,amsfonts}
\usepackage{algorithmic}
\usepackage{algorithm}
\usepackage{array}
\usepackage{cancel}
\usepackage{subfig}
\usepackage{textcomp}
\usepackage{stfloats}
\usepackage{url}
\usepackage{verbatim}
\usepackage{graphicx}
\usepackage{cite}
\usepackage{multirow}
\usepackage[table,xcdraw, dvipsnames]{xcolor}
\usepackage{colortbl}
\usepackage{booktabs}
\usepackage{forest}
\usepackage{makecell}
\usepackage{bbding}
\usepackage{hyperref}
\usepackage{tcolorbox}
\newcommand{\notcheck}{\text{\ooalign{$\checkmark$\cr\hidewidth\raisebox{0.8ex}{\rule{1.2ex}{0.1ex}}\hidewidth}}}
\usetikzlibrary{arrows.meta}

\newcommand{\ziyu}[1]{{\color{black} #1}}
\newcommand{\ziyuu}[1]{{\color{black} #1}}

\newcommand{\checkbox}[1]{
    \tikz[baseline=-0.5ex]{
        \draw[thick] (0,0) rectangle (0.3,0.3);
        \ifx#1c % checked
            \node at (0.15,0.15) {\checkmark};
        \fi
        \ifx#1x % crossed
            \node at (0.15,0.15) {\tiny$\times$};
        \fi
    }
}
\usepackage{array}
\usepackage{multirow}
\usepackage{xcolor}

\newcolumntype{L}[1]{>{\raggedright\arraybackslash}m{#1}}

\begin{document}
\bstctlcite{IEEEexample:BSTcontrol}
\title{A Survey on Reconfigurable Intelligent Surfaces \ziyu{Towards Practical Deployments}: Security and Privacy Perspectives}

\author{Ziyu Chen, Yitong Shen, Jingzhe Zhang, Yao Zheng,~\IEEEmembership{Senior Member,~IEEE,} Yili Ren,~\IEEEmembership{Member,~IEEE,} Xuyu Wang,~\IEEEmembership{Member,~IEEE,} Shiwen Mao,~\IEEEmembership{Fellow,~IEEE,} Hanqing Guo,~\IEEEmembership{Member,~IEEE} 

        % <-this % stops a space
\thanks{Ziyu Chen and Yao Zheng are with the College of Engineering, Electrical \& Computer Engineering Department, University of Hawaii at Manoa. Email: yao.zheng@hawaii.edu, ziyu89@hawaii.edu; 

Yitong Shen, Jingzhe Zhang, and Yili Ren are with the Bellini College of Artificial Intelligence, Cybersecurity and Computing, University of South Florida. Email: shen202@usf.edu, jingzhe@usf.edu, yiliren@usf.edu; 

Xuyu Wang is with the Knight Foundation School of Computing and Information Sciences, Florida International University. Email: xuywang@fiu.edu;

Shiwen Mao is with the Department of Electrical and Computer Engineering, Auburn University. Email: smao@ieee.org.

Hanqing Guo is with the Department of Computer Science, Luddy School of Informatics, Computing, and Engineering
Indiana University Bloomington. Email: guohan@iu.edu
}
}



\maketitle

\begin{abstract}
Reconfigurable Intelligent Surfaces (RIS) have emerged as a transformative technology capable of reshaping wireless environments through dynamic manipulation of electromagnetic waves. \ziyu{Although RISs have been widely studied for communication, sensing, and localization, their security and privacy implications are still fragmented across application-specific studies, prototypes, and testbeds. This survey provides a deployment-oriented review of RIS applications, with a focus on the security and privacy threats, vulnerabilities, and defensive strategies relevant to practical use.} We analyze scenarios with two types of systems (with and without legitimate RIS) and two types of attackers (with and without malicious RIS), and \ziyu{clarify how RIS introduces new attack surfaces} to practical systems, including eavesdropping, jamming, and spoofing attacks. In response, we review defenses against RIS-related attacks in these systems, such as \ziyu{protocol- and algorithm-level protections, unauthorized RIS detection, and interference/disruption-based mitigation}. We also discuss scenarios in which the legitimate user applies an additional RIS to defend against attacks. To support future research, we also \ziyu{curate papers and available open resources} at: \url{https://awesome-ris-security.github.io/}. By highlighting RIS's functionality and its security/privacy challenges and opportunities, this survey aims to guide researchers and engineers toward the development of secure, resilient, and privacy-preserving RIS-enabled practical wireless systems and environments. 

\end{abstract}

\begin{IEEEkeywords}
Reconfigurable Intelligent Surface (RIS), Physical Layer Security (PLS), Internet of Things (IoT), sub-6G, mmWave, Acoustic
\end{IEEEkeywords}

\section{Introduction}

\begin{figure}
    \centering
    \subfloat[RIS improves Wi-Fi coverage at home.]{\includegraphics[width=0.9\linewidth]{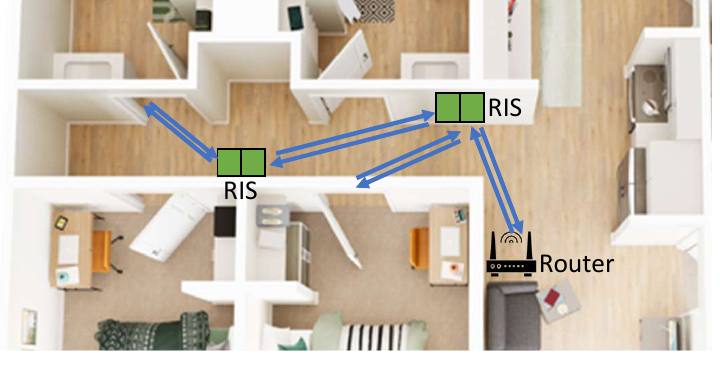}%
    \label{fig:smart_home}}
    \hfil
    \subfloat[RIS augmented in-car sensing systems.]{\includegraphics[width=0.35\linewidth]{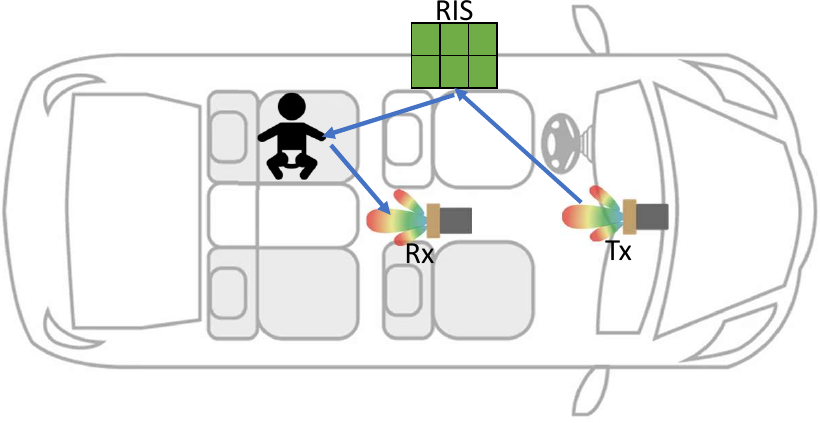}%
    \label{fig:in_car}}
    \hfil
    \subfloat[RIS-assisted industrial IoT systems.]{\includegraphics[width=0.55\linewidth]{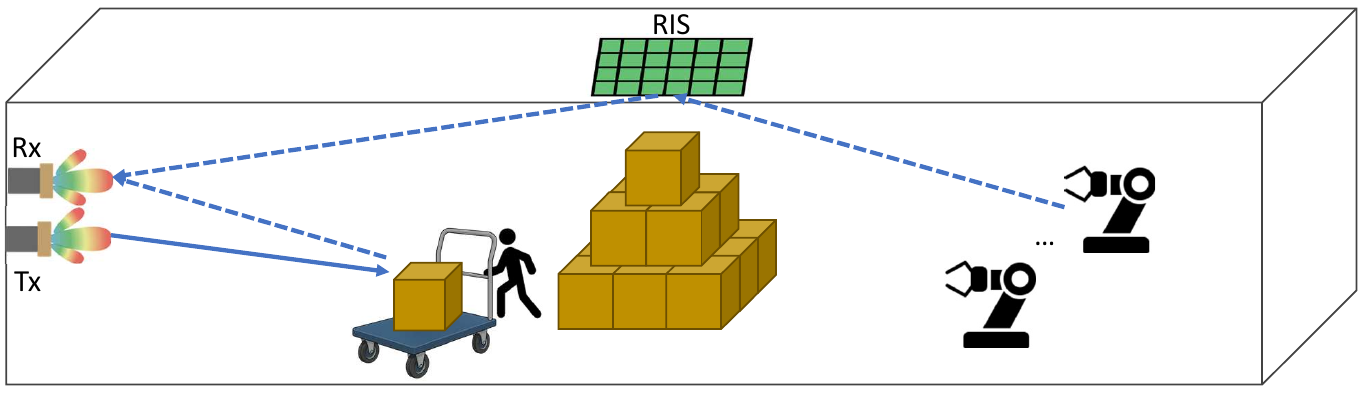}%
    \label{fig:industry}}
    \hfil
    \caption{RIS's applications in different systems.}
    \label{fig:application_in}
    \vspace{-15pt}
\end{figure}

\IEEEPARstart{R}{econfigurable} intelligent surfaces (RIS), also known as intelligent reflecting surfaces (IRS), turn traditional wireless systems into smart radio environments providing power-efficient, cost-effective services for wireless communication, sensing, and localization~\cite{di2020smart}. An RIS consists of a planar surface of numerous reflecting elements, each independently controlling the amplitude and phase of incident signals to improve coverage, signal strength, interference, and communication range~\cite{wan2025framework}. Unlike traditional relays, RIS offers fine-grained control of amplitude and phase without active RF chains~\cite{di2020reconfigurable}, allowing dynamic adjustment of the wireless propagation environment in real time and enabling unprecedented flexibility in next-generation (NextG) communication and sensing. 
\ziyu{RIS technology has been actively explored in deployment-oriented scenarios spanning home/office, vehicular, and industrial environments, as illustrated in Fig.~\ref{fig:application_in}.} Applications include indoor connectivity and coverage~\cite{huang2020holographic, pan2021reconfigurable}, in-cabin communication and autonomous driving via V2I/V2V links~\cite{naaz2024empowering, shi2024ris}, and industrial IoT deployments that must overcome metallic structures and dense machinery~\cite{yoo2025ris, elmossallamy2020reconfigurable}.

\begin{figure*}[!ht]
    \centering
    \subfloat[Attacking RIS-Assisted Systems. (Sec.~\ref{threats1})]{\includegraphics[width=0.18\linewidth]{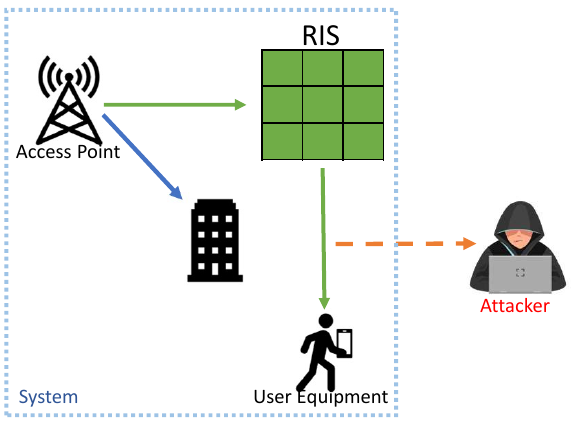}%
    \label{fig:scope_att_1}}
    \hfill
    \subfloat[Attacking Non-RIS Systems with a RIS. (Sec.~\ref{threats2})]{\includegraphics[width=0.19\linewidth]{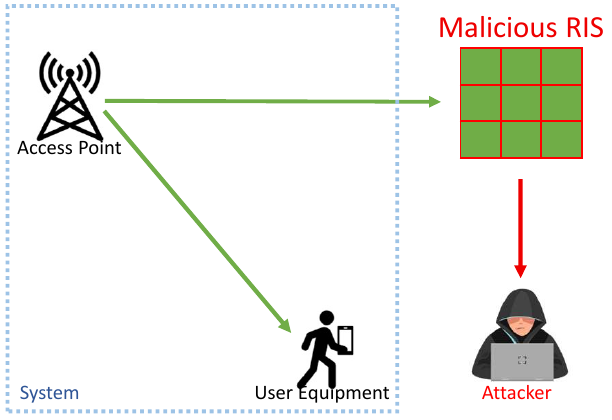}%
    \label{fig:scope_att_2}}
    \hfill
    \subfloat[Defending Attacks on RIS-Assisted Systems. (Sec.~\ref{subsec1})]{\includegraphics[width=0.18\linewidth]{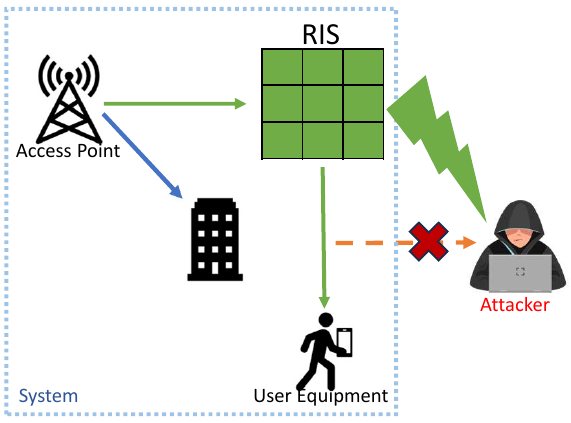}%
    \label{fig:scope_def_1}}
    \hfill
    \subfloat[Defending Non-RIS Systems against Malicious RIS. (Sec.~\ref{subsec2})]{\includegraphics[width=0.19\linewidth]{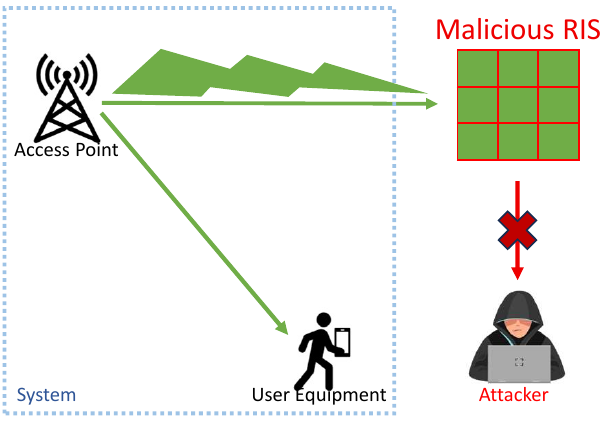}%
    \label{fig:scope_def_2}}
    \hfill
    \subfloat[Defending Non-RIS Systems with an Additional Legitimate RIS. (Sec.~\ref{subsec3})]{\includegraphics[width=0.18\linewidth]{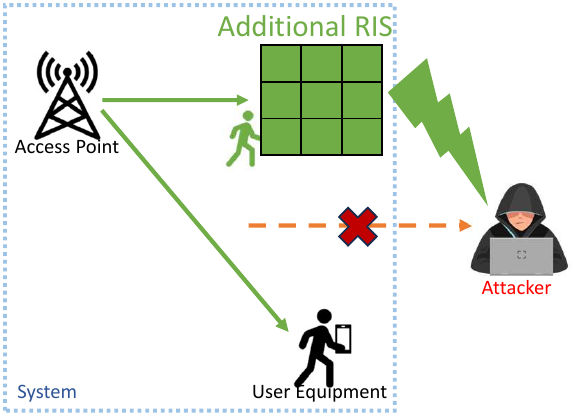}%
    \label{fig:scope_def_3}}
    \caption{Security and privacy challenges and opportunities in deployment-related RIS-related systems.}
    \label{fig:scope}
    \vspace{-10pt}
\end{figure*}

However, \ziyu{integrating RIS into deployment-oriented wireless systems introduces new security and privacy challenges as well as defensive opportunities}, as summarized in Fig.~\ref{fig:scope}.

\begin{table*}[!ht]
  \centering
  \begin{tabular}{|p{2cm}|p{1.5cm}|p{4cm}|p{1.5cm}|p{1.5cm}|p{2cm}|p{1.5cm}|}
    \hline
    \textbf{Reference} & \textbf{Year} & \textbf{RIS Usage} & \textbf{Deployment-Oriented Scenarios} & \textbf{RIS-assisted System} & \textbf{RIS for Attack and Defense} & \textbf{Security Focus} \\ \hline
    \cite{chen2016review, alexandropoulos2021reconfigurable, jian2022reconfigurable, pan2022overview, alexandropoulos2023hybrid, hassouna2023survey, elmossallamy2020reconfigurable} & 2016-2023 & Wireless communication systems & \(\times\) & \(\times\) & \(\times\) & \(\times\) \\ \hline
    \cite{aboagye2022ris} & 2022 & VLC systems & $\times$ & \checkmark & \(\times\) & \(\times\) \\ \hline
    \cite{ma2024integrated, liu2023integrated, rihan2023passive, chepuri2023integrated, tishchenko2025emergence} & 2023-2025 & ISAC systems & \notcheck & \checkmark & \(\times\) & \(\times\) \\ \hline
    \cite{li2025risbased} & 2025 & ISAC systems & \notcheck & \checkmark & \(\times\) & \checkmark \\ \hline
    \cite{naeem2023security, saeed2025comprehensive} & 2023-2025 & 6G networks & \notcheck & \checkmark & \(\times\) & \checkmark \\ \hline
    \cite{hu2020reconfigurable, zhang2022toward, zhang2023localization} & 2020-2022 & RF sensing and localization & \checkmark & \checkmark & \(\times\) & \(\times\) \\ \hline
    \cite{khoshafa2024ris} & 2024 & RF and optical communication & \notcheck & \checkmark & \(\times\) & \checkmark \\ \hline
    \cite{ahmed2025comprehensive} & 2025 & UAV-assisted networks & \checkmark & \checkmark & \(\times\) & \(\times\) \\ \hline
    \textbf{Our Survey} & \textbf{2026} & \textbf{Comprehensive Deployment-Oriented RIS Scenarios} & \textbf{\checkmark} & \textbf{\checkmark} & \textbf{\checkmark} & \textbf{\checkmark} \\ \hline
  \end{tabular}
  \caption{Comparison of existing surveys on RIS-related systems, applications, and security/privacy coverage from a deployment-oriented perspective.}
  \label{tab:comparison}
  \vspace{-10pt}
\end{table*}

\noindent\textbf{RIS-assisted systems are prone to attack.} Practical wireless systems face classical threats such as jamming, eavesdropping, and spoofing~\cite{basak2024mmwave}, and RIS reshapes the propagation environment in ways that expose additional surfaces. An RIS opens new paths that bypass obstacles and exposes receivers previously shielded by walls~\cite{jiang2024risiren, zhou2023ristealth}; its fine-grained spatial control concentrates energy into regions where nearby adversaries can align for higher SNR; and its control plane introduces a new attack surface through which a compromised controller can redirect signals, null legitimate links, or create targeted interference~\cite{cao2024security}. As shown in Fig.~\ref{fig:scope_att_1}, even an attacker without their own RIS can exploit a legitimate one, e.g., to cause near-100\% denial of service on a specific victim while leaving nearby devices unaffected~\cite{mackensen2024spatial}.

\noindent\textbf{Attackers deploy their RIS as an attacking tool.} 
Moreover, adversaries can deploy their own unauthorized RIS to attack systems without one, as shown in Fig.~\ref{fig:scope_att_2}. For example, RIStealth compromises indoor intruder detection by reflecting signals away from sensors~\cite{zhou2023ristealth}, and MetaWave attacks on-car mmWave radar using stealth metasurface tags~\cite{chen2023metawave}. Such attacks may compromise sensitive user data, disrupt critical infrastructure, or threaten safety-critical systems without the legitimate user's notice~\cite{zhou2023ristealth, jiang2024risiren, chen2023metawave}.

\noindent\textbf{Countermeasures of RIS-related attacks.} 
Defenses target both the protection of RIS-aided systems and the deterrence of malicious RIS. \ziyu{Fig.~\ref{fig:scope_def_1} corresponds to defending a system with a legitimate RIS against an adversary that does not deploy its own surface. In this case, the defender typically leverages the existing RIS together with security algorithms to improve confidentiality, reduce privacy leakage, and harden the system against interference and spoofing~\cite{magbool2025hiding}.} 
\ziyu{Fig.~\ref{fig:scope_def_2} corresponds to defending against unauthorized RIS-enabled attacks, in which defenses either mitigate the attack effect at the physical layer~\cite{gao2024benign} or detect and suppress the malicious RIS itself~\cite{henley2023detection, stamatelis2025detection}.}

\noindent\textbf{RIS for defense purposes:}
\ziyu{Meanwhile, a defender may deploy an additional RIS as a proactive defensive overlay in a non-RIS system (Fig.~\ref{fig:scope_def_3})}, programming it for secure reflection-pattern management and privacy-aware configuration against eavesdropping, sensing-based inference, and physical-layer attacks~\cite{staat2022irshield, li2022protego}. This perspective highlights an important dual role of RIS: RIS can act as both a new attack surface and an environment-level defense, which motivates a unified security and privacy perspective.

\ziyuu{
\noindent\textbf{Why RIS security differs from conventional wireless security.}
RIS does not change the underlying security objectives: confidentiality, integrity, and availability, but it changes the \emph{threat model} by making the propagation environment programmable. Conventional wireless systems generally treat the channel as an external condition rather than a directly controllable asset. An unauthorized or compromised RIS, however, can reshape the cascaded channel to redirect signal energy, create spatial nulls, or fabricate communication and sensing signatures~\cite{huang2020intelligent, wei2023adversarial}. A trusted RIS can use the same capability to strengthen legitimate links, suppress leakage, randomize channel observations, or spatially confine signal energy.

This programmability also changes attack \emph{observability and attribution}. Conventional jammers expose an RF emitter, while passive eavesdroppers must observe a sufficiently strong legitimate signal. In contrast, a passive malicious RIS can produce jamming, interception, or spoofing effects without generating its own carrier~\cite{mackensen2024spatial, chen2023metawave, lyu2020irs}. The resulting attack may therefore appear only as an altered propagation path, channel state, or sensing observation. Detection must distinguish such manipulation from ordinary fading, blockage, mobility, or hardware faults rather than merely locate an unauthorized transmitter.

Finally, RIS introduces new \emph{infrastructure and deployment dependencies}. The surface, controller, control link, and their physical placement can serve as either attack surfaces or defensive assets. Because a low-rate configuration command governs a large electromagnetic aperture, limited control-plane access can produce a spatially concentrated physical effect. Placement, cascaded CSI, calibration, synchronization, mobility, and surrounding geometry therefore become security parameters rather than only communication-design parameters. These differences in channel programmability, propagation-only observability, attribution ambiguity, and infrastructure dependence motivate the taxonomy adopted in this survey.
% These differences are summarized in Table~\ref{tab:ris_vs_conventional}.

% \begin{table}[t]
% \centering
% \caption{Comparison Between Conventional Wireless Security and RIS-Enabled Physical-Layer Security.}
% \label{tab:ris_sec_vs_conv}
% \resizebox{\linewidth}{!}{%
% \begin{tabular}{|l|p{3.8cm}|p{4.2cm}|}
% \hline
% \textbf{Dimension} & \textbf{Conventional Wireless Security} & \textbf{RIS-Enabled Security} \\ \hline
% \textbf{Channel Role} & Passive,  propagation condition not configurable. & Actively programmable medium for attackers and defenders. \\ \hline
% \textbf{Observability} & Emitter-centric (detectable active RF power/spectrum). & Propagation-only (wavefront perturbation, no RF emission). \\ \hline
% \textbf{Attribution} & Direct (locating unauthorized transmitter/source). & Ambiguous (distinguishing attack from fading/mobility). \\ \hline
% \textbf{Security Basis} & Transmit power, coding, signal keys, beamforming. & Aperture placement, cascaded CSI, geometry, control-plane. \\ \hline
% \end{tabular}%
% }
% \end{table}

}

\ziyu{
\noindent\textbf{Scope of ``practical'' deployment:} While RIS has attracted substantial interest, large-scale commercial deployment and publicly documented in-the-wild security incidents remain limited~\cite{aastrom2024ris, 3gpp_TR_21.918}. In this survey, \emph{practical} refers to deployment-oriented scenarios and prototype/testbed systems that capture realistic system constraints, such as channel estimation overhead, RIS control signaling, hardware nonidealities, mobility, and environmental dynamics, as reflected in ETSI ISG RIS reports~\cite{etsi2023ris1, etsi2023ris2, etsi2023ris3, etsi2023ris4, etsi2023ris5, etsi2023ris6}. Our goal is not to claim mature widespread deployment, but to synthesize security and privacy risks, defenses, and deployment-readiness challenges under realistic system assumptions.
}

\noindent\textbf{Motivation of this survey.} Existing surveys cover RIS from several angles: theoretical foundations and wireless communication applications~\cite{chen2016review, alexandropoulos2021reconfigurable, jian2022reconfigurable, pan2022overview, alexandropoulos2023hybrid, hassouna2023survey, elmossallamy2020reconfigurable}, VLC~\cite{aboagye2022ris}, ISAC~\cite{ma2024integrated, liu2023integrated, rihan2023passive, chepuri2023integrated, tishchenko2025emergence}, RF sensing and localization~\cite{hu2020reconfigurable, zhang2022toward, zhang2023localization}, smart cities~\cite{kisseleff2020reconfigurable}, and ML-driven RIS control~\cite{faisal2022machine, puspitasari2023survey}. From the security perspective, surveys discuss RIS-aided physical-layer security (PLS)~\cite{kaur2024survey, khoshafa2024ris, xu2023reconfiguring}, security for 6G~\cite{naeem2023security, saeed2025comprehensive}, PLS in RIS-ISAC~\cite{li2025risbased}, and UAV networks~\cite{ahmed2025comprehensive}. 
\ziyu{However, these surveys are typically organized by application domain or algorithm class, and rarely map attacks and defenses to the RIS system pipeline or discuss effectiveness under realistic constraints.} This survey fills this gap by providing a unified deployment-oriented security and privacy perspective, with explicit attack--defense correspondence. Table~\ref{tab:comparison} compares the scope of prior surveys with ours. Our main contributions are:
\ziyu{
\begin{itemize}
    \item We provide a deployment-oriented survey of RIS security and privacy across communication, sensing, localization, ISAC, industrial, and vehicular scenarios, and clarify the role of prototype/testbed studies in assessing practical relevance.
    \item We review attacks on systems with a legitimate RIS and attacks enabled by malicious/unauthorized RISs, and we organize corresponding defenses with explicit threat-model correspondence, including defender-deployed RIS as a defensive overlay.
    \item We synthesize common attack/defense mechanisms, RIS-specific attack surfaces, and deployment-related limitations (CSI uncertainty, control-plane exposure, hardware nonidealities, mobility, scalability) that govern real-world effectiveness.
    \item We maintain a public resource hub (\url{https://awesome-ris-security.github.io/}) curating papers and currently available open resources, and we identify reproducibility and benchmarking gaps for future RIS security research.
\end{itemize}
}

The remainder of the paper is organized as follows. 
Section~\ref{RIS} introduces RIS fundamentals, including far-field steering and near-field focusing. 
Section~\ref{application} surveys practical RIS applications across mmWave, sub-6G, THz, and acoustic modalities.
Section~\ref{threats} reviews attacks on RIS systems and attacks that weaponize an unauthorized RIS. 
Section~\ref{defenses} mirrors this taxonomy to review defenses for RIS-assisted systems, countermeasures against malicious RIS, and RIS used as a defensive tool. 
Section~\ref{limitations} analyzes cross-cutting limitations affecting both attacks and defenses under realistic deployment constraints. 
Section~\ref{futurework} discusses future research directions, 
and Section~\ref{discussion} \ziyu{curates public resources and identifies reproducibility gaps}.
Table~\ref{tab:acronyms} summarizes the list of acronyms used throughout this survey.

\begin{figure}
    \centering
    \includegraphics[width=0.85\linewidth]{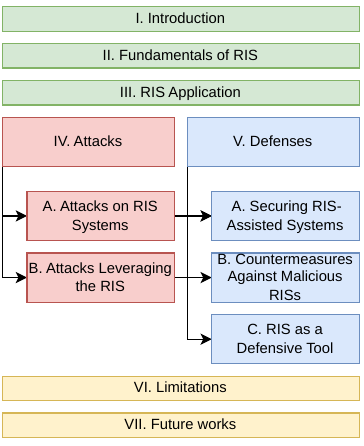}
    \caption{Structure of this survey.}
    \label{fig:structure}
    \vspace{-15pt}
\end{figure}

\begin{table}[h]
  \centering
  \begin{tabular}{|c|c|}
    \hline
    \textbf{Acronyms} & \textbf{Definitions} \\ \hline
    % 5G & Fifth Generation \\ \hline
    % 6G & Sixth Generation \\ \hline
    % AI & Artificial Intelligence \\ \hline
    AN & Artificial Noise \\ \hline
    CSI & Channel State Information \\ \hline
    DRL & Deep Reinforcement Learning \\ \hline
    DoS & Denial-of-Service \\ \hline
    EM & Electromagnetic \\ \hline
    IoMT & Internet of Medical Things \\ \hline
    IoT & Internet of Things \\ \hline
    IRS & Intelligent Reflecting Surface \\ \hline
    ISAC & Integrated Sensing and Communication \\ \hline
    LoS & Line-of-Sight \\ \hline
    LU & Legitimate User \\ \hline
    MEC & Mobile Edge Computing \\ \hline
    MIMO & Multiple-Input Multiple-Output \\ \hline
    NLoS & Non-Line-of-Sight \\ \hline
    OFDM & Orthogonal Frequency-Division Multiplexing \\ \hline
    PLS & Physical Layer Security \\ \hline
    RIS & Reconfigurable Intelligent Surface \\ \hline
    RF & Radio Frequency \\ \hline
    SISO & Single-Input Single-Output \\ \hline
    SINR & Signal-to-Interference-plus-Noise Ratio \\ \hline
    SNR & Signal-to-Noise Ratio \\ \hline
    UAV & Unmanned Aerial Vehicles \\ \hline
    V2I & Vehicle-to-Infrastructure \\ \hline
    V2X & Vehicle-to-Everything \\ \hline
    V2V & Vehicle-to-Vehicle \\ \hline
    VLC & Visible Light Communication \\ \hline
    \end{tabular}
  \caption{List of acronyms.}
  \label{tab:acronyms}
  \vspace{-15pt}
\end{table}
\section{Fundamentals of RIS} \label{RIS}

\subsection{What is RIS?}
An RIS is a planar surface composed of numerous reflecting elements, typically dense arrays of unit cells whose material, size, and number depend on the signal modality (mmWave, sub-6G, THz, or acoustic)~\cite{pan2022overview}. The RIS reflects the incident signal through a phase shift introduced by a controller; at the receiver, the reflected and direct signals combine coherently to attenuate or boost the overall signal strength. By electronically or mechanically controlling per-element phase and amplitude, an RIS dynamically shapes the propagation environment, enabling enhanced coverage, spectral efficiency, and energy efficiency.

\begin{figure*}
    \centering
    \includegraphics[width=0.8\linewidth]{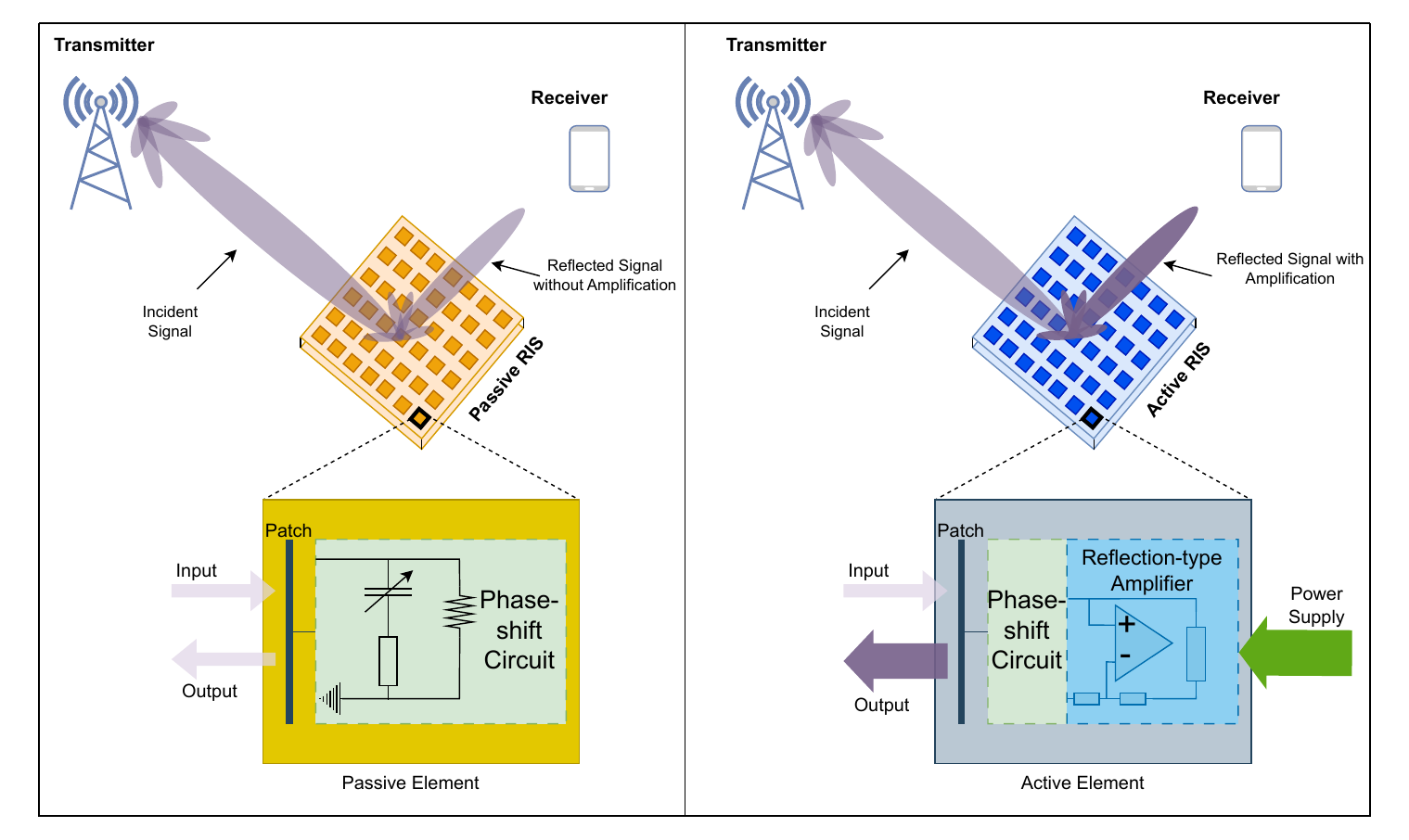}
    \caption{Hardware architecture of passive and active RISs.}
    \label{fig:RIS}
    \vspace{-10pt}
\end{figure*}

\ziyu{From the \emph{operating} perspective}, RIS can be categorized into passive, active, and hybrid RIS~\cite{zhang2022active}, as shown in Fig.~\ref{fig:RIS}. A passive RIS reflects the incident signal without amplification, \ziyu{drawing external power only for its control/biasing circuits.} An active RIS additionally amplifies the reflected signal via per-element reflection-type amplifiers, overcoming the ``multiplicative fading'' bottleneck of passive RISs~\cite{zhang2022active} at the cost of extra power and hardware complexity. 
\ziyu{Hybrid RIS (HR-RIS) combines a small number of active amplifying elements with many passive elements~\cite{taha2021enabling}.} 
Recently, simultaneously transmitting and reflecting RIS (STAR-RIS)~\cite{ahmed2023survey} is promoted to further extend coverage to both sides of the surface and can be made optically transparent for window-mounted deployments. 
Sometimes, static reflective metasurfaces are also referred to as RIS. They are fabricated with fixed reflection coefficients, offering fine-grained reflection at a lower cost but cannot adapt to the channel in real time.

Compared with relays and MIMO antennas, RISs provide \ziyu{low-cost, low-RF-complexity, low-power} wavefront control without RF chains, avoiding noise amplification and the self-interference typical of full-duplex radios. \ziyu{The compromised SNR~\cite{pan2020multicell} can be compensated by increasing the number of elements, trading deployment cost and complexity for performance.} RISs are lightweight, can be mounted on walls or ceilings, and are suitable for practical deployment. \ziyu{The remainder of this section formalizes how an RIS redirects incident signals and models the cascaded Tx--RIS--Rx channel.}

\vspace{-10pt}

\subsection{RIS as a Programmable Phase Profile}
An RIS consists of $N$ sub-wavelength elements, each imposing position-dependent phase (and possibly amplitude) shifts on an incident wavefront. The reflection coefficient at the $k$-th element is
\begin{equation}\label{eq:gamma_k}
    \Gamma_k=\rho_k e^{j\phi_k}, \quad 0\le \rho_k\le 1,\;\; \phi_k\in(-\pi,\pi],
\end{equation}
where $\rho_k$ and $\phi_k$ are the amplitude and programmable phase shift at position $x_k$, producing the element-wise mapping $s_{\mathrm{out}}(x_k)=\Gamma_k\,s_{\mathrm{in}}(x_k)$~\cite{bjornson2022reconfigurable, wu2021intelligent, di2020smart}.

\begin{figure}[!ht]
    \centering
    \includegraphics[width=0.8\linewidth]{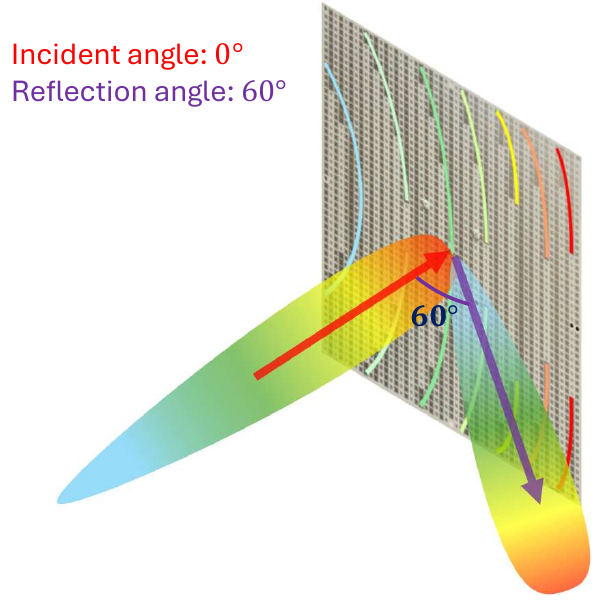}
    \caption{RIS redirecting incident wave's direction.}
    \label{fig:phase_grad}
    \vspace{-10pt}
\end{figure}

By programming a spatial phase profile $\phi(x)$ across the aperture, the incident wavefront can be redirected: a linear profile enables far-field beam steering, while a curved profile enables near-field beam focusing, as illustrated in Fig.~\ref{fig:phase_grad}. A global offset $\phi_0$ sets the phase reference and does not affect the steering angle. Unless stated otherwise, we assume phase-only control ($\rho_k\!\approx\!1$).
\vspace{-10pt}

\subsection{Near-Field and Far-Field Regimes}
\begin{figure}[!ht]
    \centering
    \includegraphics[width=0.8\linewidth]{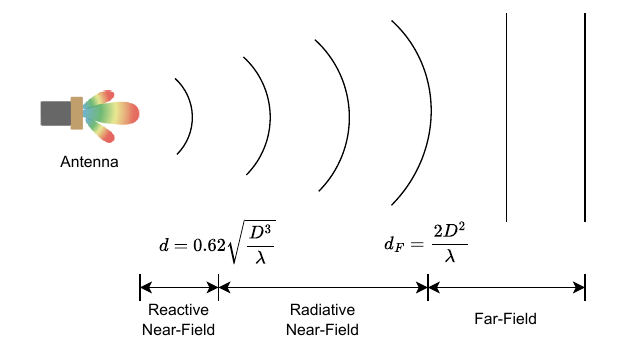}
    \caption{Field-region boundaries for an aperture of size $D$.}
    \label{fig:nf-ff}
    \vspace{-10pt}
\end{figure}

For an aperture of size $D$, the surrounding field splits into three regions (Fig.~\ref{fig:nf-ff}): the \textit{reactive near field} ($d < 0.62\sqrt{D^{3}/\lambda}$), the \textit{radiative near field} or Fresnel region ($0.62\sqrt{D^{3}/\lambda} < d < d_F$), and the \textit{far field} or Fraunhofer region ($d \gg d_F$), where $d_F \approx 2D^{2}/\lambda$~\cite{cui2024near}. We focus on the radiative near field and the far field, since the reactive near field is \ziyu{outside typical operating assumptions for communication and sensing}. In the far field the plane-wave approximation holds and a linear phase ramp suffices; in the radiative near field, wavefront curvature and distance-dependent amplitudes across the aperture matter, and a spherical-wave (Fresnel) model is required.
\vspace{-15pt}

\subsection{Far-Field Steering and Near-Field Focusing}
\label{sec:near-field}

\noindent\textbf{Far-field beam steering.} To steer the incident beam from angle $\theta_{\mathrm{in}}$ to $\theta_{\mathrm{out}}$, a linear phase profile is applied across the aperture following the generalized Snell's law~\cite{yu2011light, ding2017gradient},
\begin{equation}\label{eq:gen_snell}
    \frac{\partial \phi}{\partial x}=k_0 (\sin{\theta_\mathrm{out}}-\sin{\theta_\mathrm{in}}),\quad k_0=\tfrac{2\pi}{\lambda},
\end{equation}
which for a uniform array with spacing $d$ gives a per-element step $\Delta\phi = k_0 d(\sin\theta_{\mathrm{out}}-\sin\theta_{\mathrm{in}})$. This model applies when link distances satisfy $d\gg d_F$.

\noindent\textbf{Near-field beam focusing.} In the Fresnel region, the wavefronts are curved and a constant phase gradient no longer aligns contributions at a target point. Instead, the per-element phase is chosen to equalize the two-hop path lengths from Tx to each element to the desired focus $\mathbf r_{\mathrm f}$:
\begin{equation}\label{eq:nf_phi_star_geometry}
  \phi_k^{\star}(\mathbf r_{\mathrm f}) = k_0 \big[r_{T,k} + r_{k,R}(\mathbf r_{\mathrm f})\big] + \phi_0 \ \bmod 2\pi,
\end{equation}
where $r_{T,k}$ is the Tx-to-element distance and $r_{k,R}(\mathbf r_{\mathrm f})$ is the element-to-focus distance. Each element adds the phase needed to match its geometric path length, so all contributions arrive in phase at $\mathbf r_{\mathrm f}$, forming a tight focused spot with lateral size and depth-of-focus set by $(\lambda, D, \text{range})$~\cite{bjornson2022reconfigurable, garcia2020reconfigurable}. When the focus is far enough that the outgoing direction is approximately constant across the aperture, \eqref{eq:nf_phi_star_geometry} reduces to the far-field rule in \eqref{eq:gen_snell}.

\begin{figure}[!ht]
    \centering
    \subfloat[Near-field beam focusing.]{\includegraphics[width=0.4\linewidth]{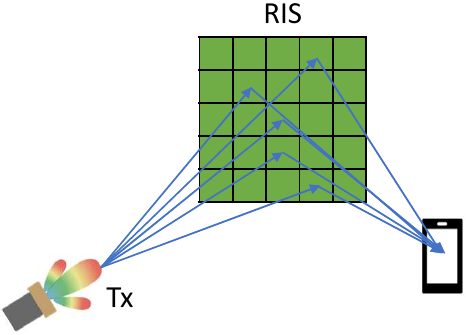}%
    \label{fig:near-field}}
    \hfil
    \subfloat[Far-field beam steering.]{\includegraphics[width=0.475\linewidth]{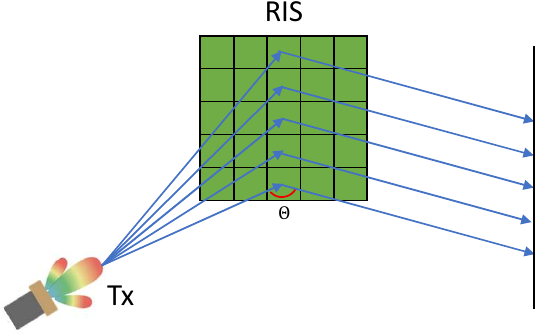}%
    \label{fig:far-field}}
    \caption{Near-field vs. far-field channel modeling with an RIS.}
    \label{fig:nf-ff-beam}
    \vspace{-10pt}
\end{figure}

\noindent\textbf{Cascaded Tx--RIS--Rx channel.} With a direct path $h_{\mathrm d}$ and one RIS, the narrowband SISO baseband model is
\begin{equation}\label{eq:hris_siso}
    y=(h_{\mathrm d}+h_{\mathrm{RIS}})\,x + n,\quad h_{\mathrm{RIS}}=\mathbf h_{\mathrm r}^{\mathsf T}\,\Phi\,\mathbf g,
\end{equation}
with $\Phi=\mathrm{diag}(\Gamma_1,\ldots,\Gamma_N)$, $\mathbf g\!\in\!\mathbb{C}^N$ the Tx$\to$RIS channel and $\mathbf h_{\mathrm r}\!\in\!\mathbb{C}^N$ the RIS$\to$Rx channel. In the far field, $\mathbf g$ and $\mathbf h_{\mathrm r}$ are plane-wave steering vectors parameterized by $(\theta_{\mathrm{in}},\varphi_{\mathrm{in}})$ and $(\theta_{\mathrm{out}},\varphi_{\mathrm{out}})$; choosing a linear phase profile makes $h_{\mathrm{RIS}}$ peak at $\theta_{\mathrm{out}}$~\cite{bjornson2022reconfigurable, wu2021intelligent, di2020smart}. In the near field, $\mathbf g$ and $\mathbf h_{\mathrm r}$ follow a spherical-wave model with element-wise amplitude $1/(r_{T,k}r_{k,R})$ and phase $k_0(r_{T,k}+r_{k,R})$, so $h_{\mathrm{RIS}}(\mathbf r)$ depends on both angle and range~\cite{bjornson2021primer, garcia2020reconfigurable, bjornson2022reconfigurable, delbari2024far, emenonye2023ris}. Fig.~\ref{fig:nf-ff-beam} contrasts the two regimes.

\begin{figure}[!ht]
    \centering
    \subfloat[Tx-Rx LoS blocked --- no link.]{\includegraphics[width=\linewidth]{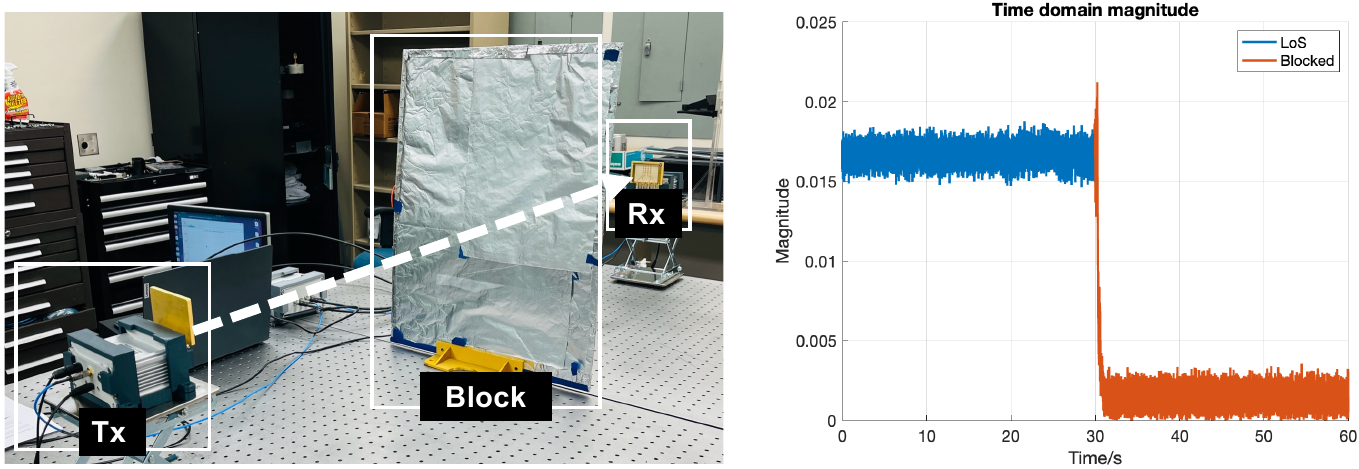}%
    \label{fig:ris_nlos_wo_ris}}
    \hfil
    \subfloat[RIS creates NLoS path --- link restored.]{\includegraphics[width=\linewidth]{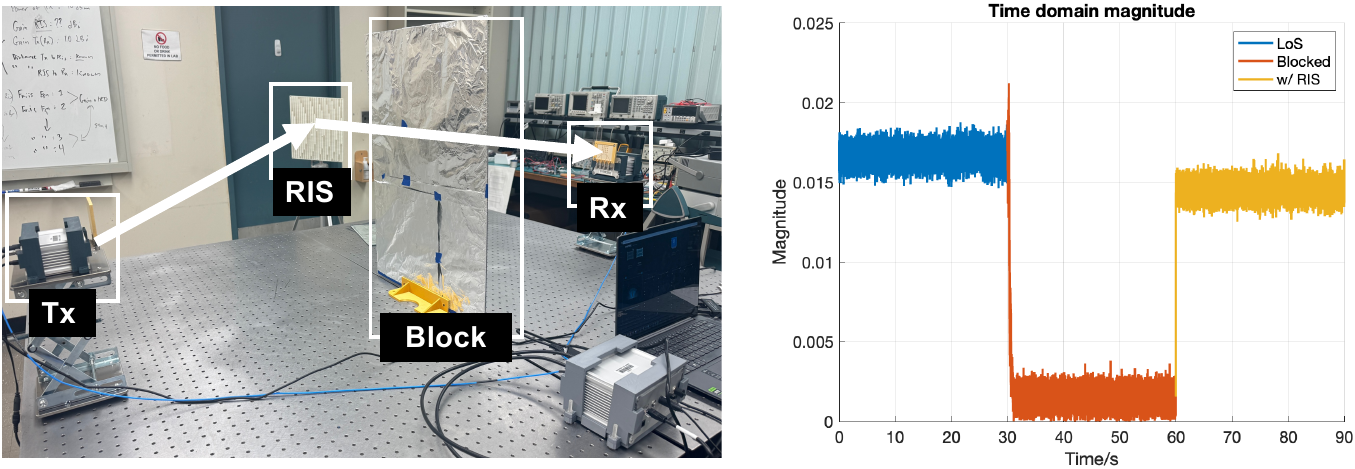}%
    \label{fig:ris_nlos_w_ris}}
    \caption{RIS recovering indoor communication under LoS blockage.}
    \label{fig:ris_nlos}
    \vspace{-15pt}
\end{figure}

\ziyu{The far-field steering abstraction underlies many RIS-enabled attacks and defenses discussed later (directional eavesdropping, selective jamming, angle-domain spoofing), while near-field focusing introduces range-selective spatial control that reshapes both the attack surface and the defense design space in localization, sensing, and short-range secure links.} Fig.~\ref{fig:ris_nlos} illustrates the baseline use case: a blockage that disrupts the direct Tx--Rx link is bypassed by a passive RIS placed 2~m from the Tx at 0\textdegree{} incidence and 60\textdegree{} reflection, restoring the link via a controlled NLoS path.

\subsection{Choice between Near-Field and Far-Field Operation}
\ziyu{
The choice between focusing and steering is governed by whether the target lies inside the Fresnel zone ($d<d_F$)~\cite{chen2024near}.
Focusing is preferred when at least one endpoint is close to the RIS (common for mmWave/THz and large apertures)~\cite{bjornson2021primer, naaz2024empowering}, when range selectivity is needed (WPT/backscatter, imaging, precise localization, multi-user separation for co-located users)~\cite{tran2022multifocus}, or when the quadratic phase variation across the aperture exceeds roughly $\pi/8$; for wideband operation it may additionally require true-time-delay compensation to avoid range squint~\cite{xu2024near}.
Far-field steering is adequate when both endpoints are sufficiently distant, when angular selectivity is sufficient~\cite{balanis2016antenna}, or when simple linear-gradient control and narrowband operation are preferred. This distinction is security-relevant: attacker observability, spoofing, and jamming leverage all depend on whether RIS control is angle-selective or range-and-angle-selective.
}

\section{RIS-Aided System Applications} \label{application}
\begin{figure*}[t]
    \centering
    \includegraphics[width=0.9\linewidth]{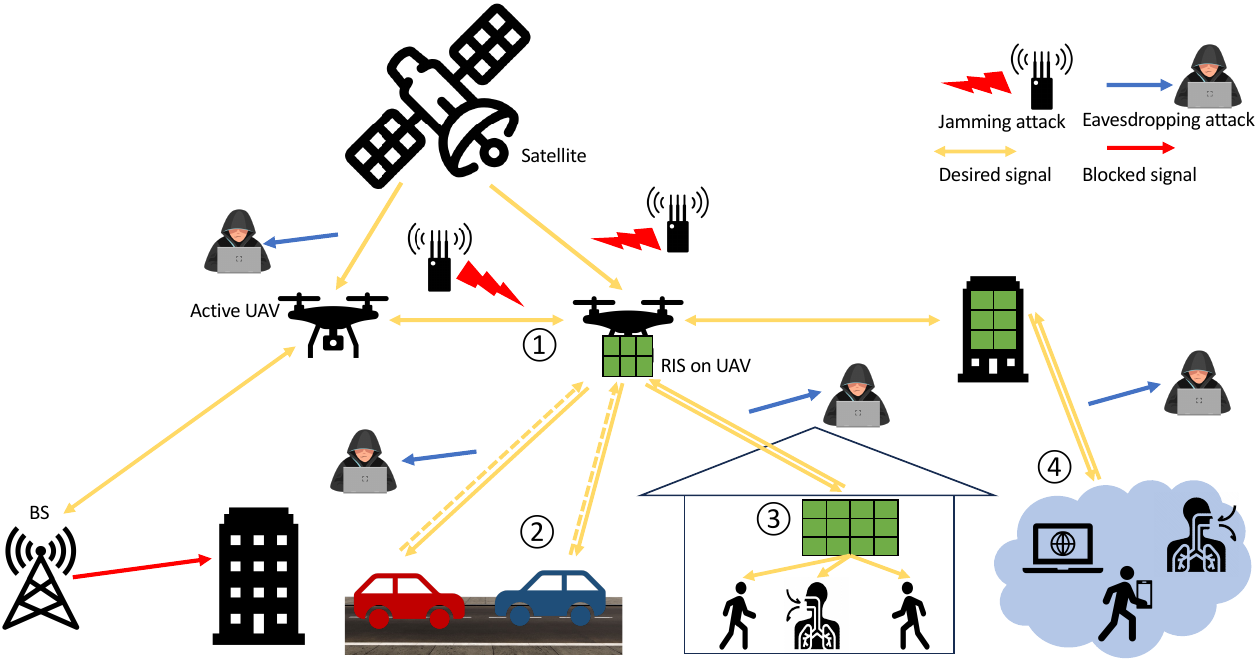}
    \caption{Overview of RIS-aided outdoor systems and their security and privacy issues. Dashed lines are echo signal and solid lines are radio signal.} 
    \label{fig:application_out}
    \vspace{-10pt}
\end{figure*}

The RIS \ziyu{can be applied} in practical systems that exist in our daily life, as shown in Fig.~\ref{fig:application_in} and Fig.~\ref{fig:application_out}. In Fig.~\ref{fig:application_out}, \textcircled{1}, \textcircled{2}, \textcircled{3}, and \textcircled{4} refer to proposed RIS applications in UAVs, vehicular networks, smart homes, and outdoor communication and sensing, respectively. 
\ziyu{Following European Telecommunications Standards Institute (ETSI) Industry Specification Group (ISG) RIS RIS reports~\cite{etsi2023ris1, etsi2023ris2, etsi2023ris3, etsi2023ris4, etsi2023ris5, etsi2023ris6}}, we elaborate on applications \ziyu{assumptions} by signal modality. Table~\ref{tab:RIS_media} summarizes the structure, control method, deployment, and application scenarios of mmWave, sub-6G, THz, and acoustic RIS.

\begin{table*}[!ht]
  \centering
  \begin{tabular}{|L{1.5cm}|L{2.3cm}|L{2.6cm}|L{2.5cm}|L{3cm}|L{3.5cm}|}
    \hline
    \textbf{Signal Modalities} & \textbf{Structure} & \textbf{Control method} & \textbf{Main Function} & \textbf{Deployment} & \textbf{Application} \\ \hline
    mmWave \cite{popov2021experimental, ma2024automs, nie2023agris, nolan2021ros, woodford2023metasight, landika2023obstruction} & 
    $\lambda/2 \text{ or } \lambda/4$ (mm-level) & 
    Electronic tuning & 
    Blockage mitigation \% NLoS link restoration &
    On walls, objects, or infrastructures & 
    Coverage expansion, ISAC, V2X, high-precision localization \\ \hline
    sub-6G \cite{heinrichs2023open, hu2020reconfigurable, li2024rfmagus, ruan2024using, li2023riscan, zhang2023passive, liu2024risar, liu2024tris, fan2026sense} & 
    $\lambda/2 \text{ or } \lambda/4$ (cm-level) & 
    Electronic tuning & 
    Coverage extension \& dead-zone mitigation &
    On walls, ceilings or UAVs & 
    Coverage expansion, RF sensing, backscatter, V2V, smart home, industrial, far-field IoT \\ \hline
    THz~\cite{yang2022terahertz, akyildiz2022terahertz} &
    Micro- or nano-scale &
    Tuning the chemical potential or materials' surface impedance &
    Ultra-high-throughput short-range link enabling&
    Within a single room, on desktop surfaces, or in micro-scale scenarios &
    Ultra-high-speed short-range communications \\ \hline
    Acoustic~\cite{tian2019programmable, zabihi2023tunable, sun2021acoustic, sun2022high, 10.1145/3643832.3661863, hu2022binary, ma2025reconfigurable, luo2025underwater} & 
    Acoustic resonators or cavities &
    Mechanical adjustments or active actuation & 
    Acoustic wavefront control &
    Underwater: on seabed or buoys \newline In-room: on walls or ceilings & 
    Underwater networks, ultrasonic imaging, non-destructive industrial testing, etc. \\ \hline
  \end{tabular}
  \caption{Comparison of mmWave, sub-6G, THz, and acoustic RIS.}
  \label{tab:RIS_media}
\end{table*}

\subsection{mmWave}
\ziyu{mmWave (FR2 in 5G~\cite{3gpp2021base}) propagation is highly directional and angularly sparse, making blockage a first-order concern}; mmWave RISs are introduced primarily to create robust NLoS paths around obstacles. Panels are extremely dense arrays of sub-wavelength elements ($\lambda/4$ to $\lambda/2$). For example, a 28 GHz RIS can integrate 1600 patch antennas on a $20 \times 20$ cm board~\cite{popov2021experimental, malik2024design}, and some designs are even smaller~\cite{chen2023metawave}. Electronic phase tuning via integrated semiconductor devices, typically controlled by FPGA/RFIC, enables fast switching ($\mu$s--ms) and agile multi-beam steering. 
\ziyu{Their typical placements include walls, building facades, indoor corridors, and street furniture, where the RIS provides a controllable NLoS path to extend coverage or bypass outside-to-inside losses~\cite{etsi2023ris1, etsi2023ris5}.}
Applications include mmWave coverage and throughput enhancement in dense urban streets and indoor hotspots~\cite{popov2021experimental, ma2024automs}, positioning and environment sensing via NLoS observability~\cite{landika2023obstruction, yan2025mmmirror, yasmeen2024around}, smart agriculture~\cite{nie2023agris}, and V2X and autonomous driving~\cite{ozcan2021reconfigurable, nolan2021ros, woodford2023metasight}. Lower-cost metasurfaces with limited functionality have also been explored~\cite{qian2022millimirror, nolan2025ricochet}.

\subsection{sub-6G}
Unlike the directional mmWave regime, \ziyu{sub-6G channels (including FR1~\cite{3gpp2021base} and Wi-Fi) are characterized by rich multipath, so sub-6G RISs both create virtual LoS links and manipulate existing multipath for spatial diversity.} Wavelengths of several centimeters require larger elements (1--3~cm) and surfaces on the order of meters in some scenarios, potentially with tiled architectures~\cite{heinrichs2023open}. Reconfiguration speed is slower than mmWave (tens of milliseconds). Sub-6G RISs are commonly deployed on walls or ceilings in smart homes, offices, or IoT settings to improve Wi-Fi coverage in dead zones~\cite{hu2020reconfigurable, li2024rfmagus, ruan2024using, zhong2023smartshell}. \ziyu{Whether the RIS operates in the near- or far-field depends on aperture size and link distance, which determines whether operation is better described as beam steering or focusing.} Sub-6G RISs are also widely used in IoT networks for localization~\cite{li2023riscan, zhang2023passive}, V2V, activity recognition~\cite{liu2024risar, liu2024tris}, and healthcare~\cite{zhu2023active}, assisting links either by creating a dominant path or by reshaping composite channels~\cite{etsi2023ris5}.

\ziyu{
\subsection{Terahertz (THz)}
THz channels rely almost exclusively on unblocked LoS propagation, making THz RISs critical for bypassing opaque obstacles~\cite{e2023survey}. Because THz wavelengths are on the order of micrometers to sub-millimeters, traditional metallic patches and PIN diodes are physically inadequate~\cite{yang2022terahertz}; THz RISs instead rely on novel nanomaterials, graphene, liquid crystals, CMOS, or plasmonic metamaterials, with reconfiguration achieved by tuning chemical potential or surface impedance. Due to severe distance attenuation, THz RISs target ultra-dense, short-range deployments such as within a single room, on desktops, or in micro-scale scenarios (intra-chip, rack-to-rack)~\cite{ahmed2024survey}. They are projected to enable Tbps-level short-range links~\cite{ahmed2024survey} for holographic communication, immersive XR, and nano-sensor networks including intra-body biomedical applications~\cite{akyildiz2022terahertz}. Deployment-oriented security and privacy for THz RIS remain largely unexplored.
}

\subsection{Acoustic}
\ziyu{Acoustic signals are mechanical waves with propagation speeds orders of magnitude slower than EM waves ($\sim$1500~m/s in water, $\sim$340~m/s in air), resulting in extreme propagation delays, severe multipath, and high Doppler sensitivity.}
Acoustic wavelengths are also much larger, so acoustic RISs use centimeter-scale unit cells (typically resonators or cavities)~\cite{tian2019programmable, luo2025underwater}. Electronic tuning is inadequate at acoustic impedances~\cite{zabihi2023tunable}, so reconfiguration relies on mechanical adjustment or active actuation (e.g., piezoelectric actuators~\cite{tian2019programmable}), making reconfiguration slower than RF RISs but adequate for acoustic channels; a recent underwater design uses a microcontroller for binary coding that more closely resembles RF RISs~\cite{luo2025underwater}. Deployment is environment-specific: underwater RISs are anchored on the seabed or buoys, while in-room metasurfaces mount on walls or ceilings. Applications span underwater communication and sensing~\cite{sun2021acoustic, sun2022high, zhang2024optimizing, rohde2018reconfigurable}, ultrasonic imaging~\cite{10.1145/3643832.3661863, hu2022binary}, and non-destructive industrial testing~\cite{ma2025reconfigurable}.
\section{Attacks in RIS-Related Systems} \label{threats}

\ziyu{
\subsection{Threat Modes Demonstration}
To systematically analyze the security and privacy vulnerabilities of RIS-aided systems, we adopt a standardized and architecture-driven approach. According to ETSI ISG on RIS, specifically outlined in the architectural and operational frameworks of ETSI GR RIS-002~\cite{etsi2023ris2}, a practical RIS deployment consists of three primary logical components: the Base Station (BS) or Access Point (AP), the \textit{RIS Controller}, and the \textit{RIS} itself. The coordination between the BS and the RIS Controller is facilitated by a dedicated \textit{Control Link}.

By anchoring our analysis in this standardized architecture, we can clearly explain the trust boundaries and attack surfaces unique to practical RIS deployments. Conceptually, the BS and the legitimate RIS operate within the network operator's trust domain. However, their physical deployment in distributed, semi-public, or outdoor environments, combined with the potentially wireless nature of the control link, heavily exposes this trust domain to adversaries. \ziyuu{More importantly, the defining property of RIS-related threats is that the adversary manipulates the propagation environment itself rather than the communicating devices. The question a threat taxonomy must therefore answer is not \emph{what} the adversary attacks (the medium) but \emph{from where} (whether the adversary operates inside or outside the network operator's trust domain as defined by ETSI GR RIS-002~\cite{etsi2023ris2}).}

Therefore, to provide higher-level insights into RIS-specific vulnerabilities, we propose a novel taxonomy that classifies security threats into two fundamental categories, based on how the adversary interacts with the ETSI reference architecture:

\begin{itemize}
    \item \textbf{Attacks \emph{on} the RIS System (Infrastructure Compromise):} In this category, the adversary targets the internal trust boundaries and components of the legitimate RIS architecture. The objective is to degrade, hijack, or destroy the RIS infrastructure itself. Attack vectors include tampering with the Control Link, launching denial-of-service (DoS) or hijacking attacks against the RIS Controller, and executing physical or hardware-level tampering on the RIS Node.
    
    \item \textbf{Attacks \emph{leveraging} the RIS (The RIS as a Weapon):} In this category, the adversary deploys an unauthorized malicious RIS or fully exploits a compromised legitimate RIS to manipulate the external wireless propagation channel. Here, the RIS is weaponized to launch advanced, programmable physical-layer attacks against the broader network, encompassing anomalous targeted eavesdropping, passive signal nulling (jamming without active RF emission), and physical-layer channel spoofing (e.g., generating ghost targets).
\end{itemize}

This standards-driven taxonomy not only aligns with ongoing industrial specifications \ziyuu{and organizes the RIS-specific vulnerabilities by a single criterion: the adversary's position relative to the operator's trust boundary}, but also explicitly isolates the unique physical-layer vulnerabilities introduced by metasurfaces, clearly distinguishing them from conventional wireless network threats.
}

\subsection{Attacks on RIS Systems} \label{threats1}
\ziyu{

\begin{figure}[!ht]
    \centering
    \includegraphics[width=0.975\linewidth]{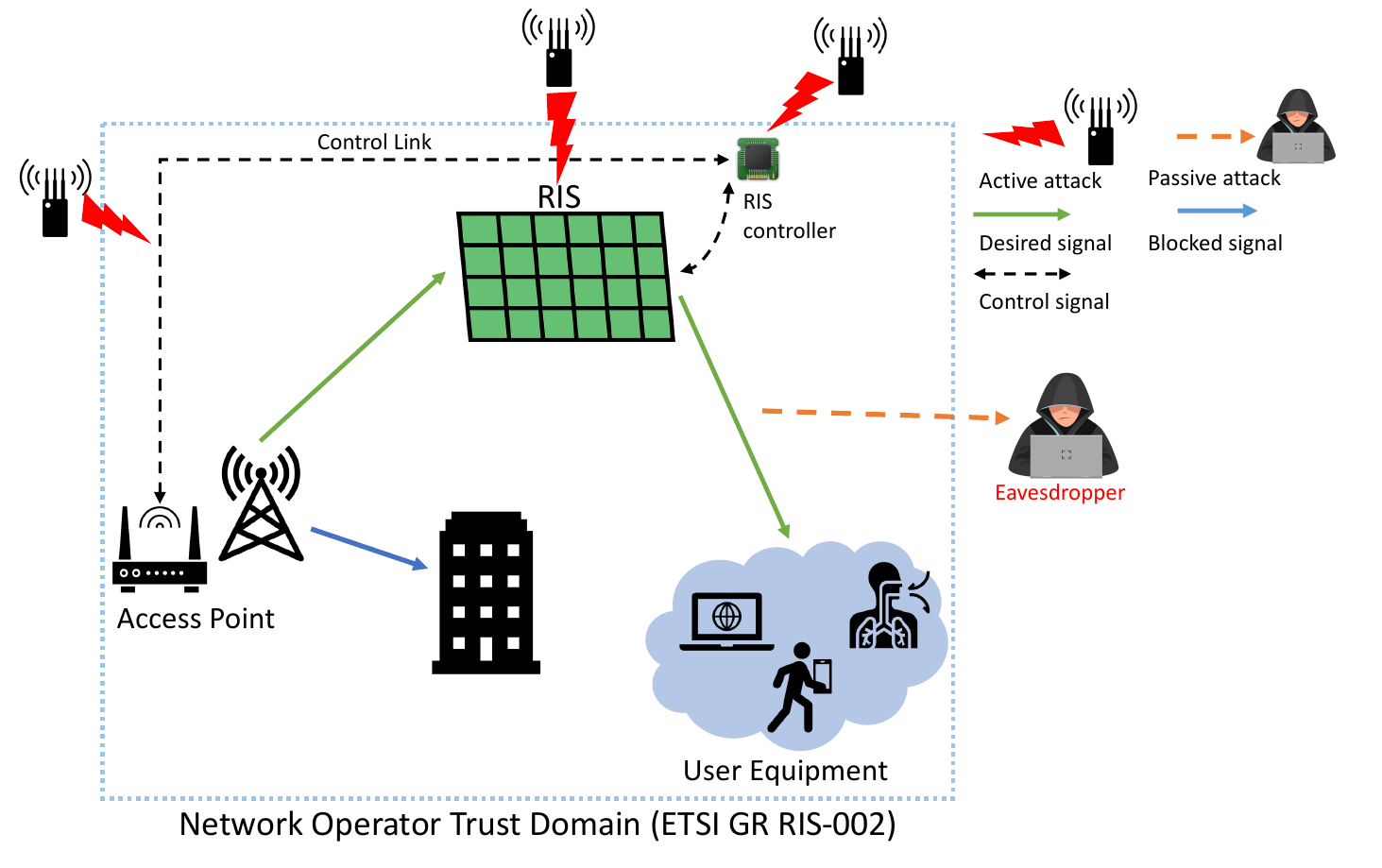}
    \caption{Attacks on RIS-assisted systems.}
    \label{fig:threat_model_etsi}
    \vspace{-10pt}
\end{figure}

\begin{table*}[!ht]
\centering
\scalebox{0.8}{
\begin{tabular}
{|L{0.04\textwidth}|L{0.08\textwidth}|L{0.09\textwidth}|L{0.06\textwidth}|L{0.07\textwidth}|L{0.06\textwidth}|L{0.08\textwidth}|L{0.145\textwidth}|L{0.06\textwidth}|L{0.06\textwidth}|L{0.09\textwidth}|L{0.12\textwidth}|}
\hline
\textbf{Ref.} &
\textbf{Target Component} &
\textbf{Threat Type} &
\textbf{Comm. Method} &
\textbf{Scenario} &
\textbf{Corres-pondence to Fig.~\ref{fig:application_out}} &
\textbf{Attack Method} &
\textbf{Required Attacker Capability} &
\ziyuu{\textbf{CSI Requirements}} &
\ziyuu{\textbf{Field Regime}} &
\textbf{Attack Performance} &
\ziyuu{\textbf{Limitations}} \\ \hline

    \cite{chen2022malicious} & 
    \multirow{3}{*}{Control Link} & 
    \multirow{2}{*}{Eavesdropping} & 
    \multirow{2}{*}{mmWave} & 
    SISO with one Eve & 
    \multirow{2}{*}{\textcircled{3}} & 
    \multirow{2}{0.08\textwidth}{Beam steering sideband to Eve.} & 
    \multirow{2}{0.145\textwidth}{Intercept link \& inject phase commands. Strict attacker-RIS alignment.} & 
    \multirow{2}{0.06\textwidth}{\ziyuu{CSI-Free}} & 
    \multirow{2}{0.06\textwidth}{\ziyuu{Far-field}} & 
    \multirow{2}{0.08\textwidth}{Secrecy capacity $\downarrow$ 81.7\%} & 
    \multirow{2}{0.13\textwidth}{\ziyuu{Blocked by authentication and upper-layer encryption.}} \\ \cline{1-1} \cline{5-5}
    
    \cite{chen2023wavefront} &
    & & & mmWave WLAN system & & & & & & & \\ \cline{1-1} \cline{3-12}
    
    \cite{staat2022mirror} &
    &
    Jamming &
    sub-6G &
    OFDM &
    \textcircled{3} &
    Reflect legitimate signals &
    Gain unauthorized access to controller software &
    \ziyuu{CSI-Free} &
    \ziyuu{Far-field} &
    Severely degrade data rates &
    \ziyuu{Depends on favorable placement or RIS size.} \\ \hline
    
    \cite{acharjee2022controller} &
    RIS Controller &
    DoS &
    sub-6G &
    SISO &
    / &
    Phase manipulation &
    Hack micro-controller/firmware &
    \ziyuu{Perfect} &
    \ziyuu{Far-field} &
    Minimized data rate &
    \ziyuu{Hardware root-of-trust required.} \\ \hline

    \cite{mughal2025malris} &
    RIS Hardware &
    Spoofing/DoS &
    sub-6G &
    Cellular &
    \textcircled{4} &
    Hardware-level manipulation &
    Supply-chain, firmware, or controller compromise enabling partial control of RIS. &
    \ziyuu{Perfect} &
    \ziyuu{---} &
    High attack success potential &
    \ziyuu{Flagged by anomaly detection.} \\ \hline
    
    \cite{lyu2020irs} &
    \multirow{2}{*}{RIS Surface} &
    Jamming &
    sub-6G, Wi-Fi &
    SISO &
    \textcircled{3} &
    Passive jamming &
    Eve controlling an RIS in NLoS &
    \ziyuu{Perfect} &
    \ziyuu{Far-field} &
    Received power$\downarrow$ 99\%, SINR$\downarrow$ 22~dB &
    \ziyuu{Relies on perfect CSI. Performance drops at long distance.} \\ \cline{1-1} \cline{3-12}
    
    \cite{mackensen2024spatial} &
    &
    Jamming &
    sub-6G &
    Wi-Fi with multiple devices &
    \textcircled{4} &
    Active selective jamming &
    An active jammer, RIS near jammer, and passive observation of target. &
    \ziyuu{Estimated} &
    \ziyuu{Far-field} &
    100\% DoS of target &
    \ziyuu{Frequency hopping mitigates threat.} \\ \hline

\end{tabular}
}
\caption{Summary of Attacks on RIS Infrastructure}
\label{tab:ris_aided_sys_security}
\vspace{-15pt}
\end{table*}

In this section, we analyze the vulnerabilities inherent to the deployment of legitimate RIS infrastructure. Based on the ETSI threat model, these attacks do not require the adversary to deploy their own metasurfaces; instead, they either passively exploit the signals reflected by the legitimate RIS or actively compromise the operational components within the network operator's trust domain, as shown in Fig.~\ref{fig:threat_model_etsi}.

\subsubsection{Passive Exploitation and Privacy Leakage}
The fundamental premise of RIS is to reshape the wireless propagation environment by creating new reflection paths or virtual LoS links. However, this architectural advantage expands the physical attack surface. In a passive exploitation scenario, adversaries do not emit any jamming signals nor do they tamper with the RIS hardware. Instead, they silently exploit the newly established, RIS-assisted multipath environment.

\paragraph{Eavesdropping on Legitimate Paths} 
Traditionally, eavesdroppers must reside within the coverage of the transmitter. With the introduction of an RIS, the signal coverage is intentionally extended to bypass blockages, which inadvertently provides eavesdroppers with high-quality reflected signals in previously secure blind spots~\cite{naeem2023security}. Furthermore, the broadcast nature of RIS-boosted signals can lead to widespread personal or operational data leakage~\cite{mao2022reconfigurable, wang2022secure}. This inadvertent signal leakage affects numerous practical applications, spanning 6G networks~\cite{khalid2025malicious}, vehicular communications~\cite{jing2024reconfigurable}, industrial IoT~\cite{liu2022security}, and healthcare systems~\cite{wang2022robust}.

\paragraph{Location and Privacy Tracking} 
Beyond data interception, the unique spatial signatures generated by RIS reflections introduce severe privacy leakage risks, particularly regarding user positioning. Because the RIS alters the channel fingerprints in a highly deterministic manner, adversaries can leverage these scattered signals for unauthorized sensing and tracking.
\begin{itemize}
    \item \textbf{High-Frequency Localization:} In mmWave and sub-6G networks, the wireless environment itself can leak location clues. For example, attackers listening to RIS-scattered signals can triangulate a user's precise position~\cite{chen2024joint, umer2023role}. In V2X scenarios, one base station with an RIS can localize platoon vehicles almost as accurately as two separate BSs, seriously undermining an individual driver's travel privacy~\cite{chen2023location}.
    \item \textbf{Indoor Occupancy and IoMT Tracking:} The rich multipath generated by sub-6G RISs allows adversaries to track human mobility inside smart homes without active radar~\cite{shenoy2022rf}, potentially leading to the severe leakage of sensitive healthcare data in IoMT networks~\cite{zhu2023active}.
    \item \textbf{Acoustic Node Exposure:} In underwater acoustic networks, location privacy is paramount. Acoustic signals propagate over extremely long distances with low attenuation. Without precautions, adversaries can passively listen to RIS-reflected acoustic signals and, via amplitude or time-difference analysis, deduce the exact position of the transmitting source nodes, compromising critical operational security~\cite{wang2024underwater}.
\end{itemize}

\subsubsection{Active Infrastructure Compromise}
Unlike passive exploitation, active infrastructure compromise involves a direct assault on the ETSI-defined logical components of the RIS. Since RIS panels are often deployed in highly accessible or outdoor environments, they face severe physical and cyber-physical threats.

\paragraph{Control Link Tampering}
The \textit{control link} bridging the BS and the RIS Controller is a critical vulnerability point~\cite{el2025study}. If this link is wireless or lacks robust encryption, adversaries can launch man-in-the-middle attacks to intercept or inject malicious phase-shift commands~\cite{mughal2025malris}. By hijacking the control link, attackers can force the RIS to intentionally scatter signals to degrade network throughput~\cite{staat2022mirror}, or elegantly manipulate the RIS's space-time freedom to generate a concealed directional sideband toward an eavesdropper while maintaining the mainband toward the legitimate client (e.g., the MeSS attack)~\cite{chen2022malicious, chen2023wavefront}.

\paragraph{Controller DoS and Hijacking}
The \textit{RIS controller} typically relies on resource-constrained microcontrollers or FPGAs. Attackers may exploit firmware vulnerabilities to launch Denial-of-Service (DoS) attacks, freezing the metasurface in a suboptimal state. In more sophisticated exploits, adversaries could flash malicious firmware via local access interfaces, overwriting the phase shifts to minimize the victim's data rate and effectively turning the legitimate RIS into a remotely controlled zombie node~\cite{acharjee2022controller, acharjee2024design}.

\paragraph{Physical, Hardware, and Supply-Chain Level Compromises}
The physical integrity of an \textit{RIS} can be compromised throughout its lifecycle.
In practical deployments, especially in harsh or hard-to-monitor conditions (e.g., underwater environments with limited physical security measures~\cite{ardizzon2024rnn}), the hardware components can be directly compromised. \ziyuu{Also, manufacturing imperfections, calibration errors, environmental stress, and component failures can cause individual RIS elements or their biasing circuits to malfunction. Such impairments introduce persistent phase/amplitude errors and unintended reflections, thereby distorting the intended beam pattern~\cite{yang2023practical, wang2025faulty}. Also, inaccurate RIS phase responses may degrade sensing and localization functions~\cite{wu2024exploit}, or cause information leakage to external eavesdroppers and degrade secrecy performance~\cite{salem2023impact}.

% supply chain attack
If such modifications are introduced deliberately, compromised elements could further cause hardware-level DoS or persistent deviations from the intended RIS behavior.
Hardware compromise can be introduced before deployment through the supply chain, where malicious modifications may be inserted during hardware design, manufacturing, assembly, or provisioning. This threat is particularly relevant to RIS because practical surfaces may integrate a large number of low-cost reflecting elements through third-party manufacturing and assembly processes.
Potential threats include manufacturing-time tampering, hardware Trojans, malicious firmware, and partial element compromise~\cite{mughal2025malris, khalid2025malicious}.
Moreover, hardware-security studies have shown that untrusted IC and FPGA design or manufacturing flows can enable stealthy Trojan insertion that may evade pre-deployment verification and activate only after deployment~\cite{akter2023survey, ahmed2024post}.}
}

\begin{figure}[!ht]
    \centering
    \includegraphics[width=0.9\linewidth]{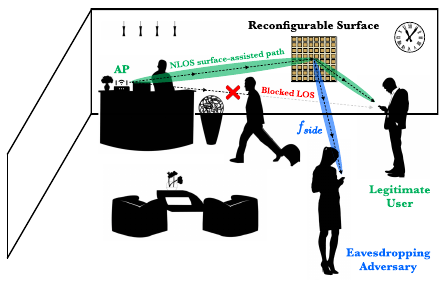}
    \caption{Attack Demo: Manipulating an existing RIS for eavesdropping while maintaining legitimate communications. In this scenario, the attacker manipulate RIS to generate a side lobe directing at the eavesdropping adversary, hence eavesdrop the communication between the AP and legitimate user. \protect\footnotemark}
    \label{fig:eavesdropping_wavefront}
    \vspace{-10pt}
\end{figure}
\footnotetext{Reproduced from~\cite{chen2023wavefront}: Chen \emph{et al.} \emph{Wavefront Manipulation Attack via Programmable mmWave Metasurfaces: From Theory to Experiments}, WiSec~'23, pp.~317--328, DOI: 10.1145/3558482.3590182. \copyright~2023 Copyright held by the owner/author(s). Publication rights licensed to ACM. Used for non-commercial purposes.}

\noindent \textbf{Pioneer Work: \textit{Wavefront Manipulation Attack via Programmable mmWave Metasurfaces: From Theory to Experiments}~\cite{chen2023wavefront}}, as shown in Fig.~\ref{fig:eavesdropping_wavefront}.

\textbf{Attack Scenario:}
This work considers a scenario where an \textbf{attacker without RIS} is \textbf{eavesdropping} an \textbf{RIS-assisted} mmWave communication system by \textbf{tampering the RIS's control link}. 
In the system, an AP sends a message to a client. When the LoS path is blocked, an RIS is set up to create an NLoS link between them.
The attacker, located in a different location from the LUs, hijacks the RIS and gains full control of it. Thus, they can manipulate the RIS to reflect incident signal to their location, and eavesdrop the signal sent by the AP. In the mean time, the victim (client) still maintains communication with the AP. 
In addition, the attacker has no prior knowledge of the LU's locations.

\textbf{Attack Goal:}
The attacker aims to eavesdrop on the communication signal sent by the AP, without affecting legitimate communication between the AP and the client.

\textbf{Attack Method:}
By periodically on/off switching the control lines of the RIS, the attacker creates a sideband channel that carries a copy of the victim’s signal.
By carefully adjusting the time-varying control signals across surface elements, the attacker steers the sideband signal towards their location while maintaining the direction of mainband toward the legitimate client.
In addition, they assume all surface elements are utilized for beamforming towards both the client and the attacker, resulting in strong directionality to both targets. Thus, the attacker can receive the signal transmitted by the AP through a concealed channel.

\textbf{Attack Result:}
(1) The attacker significantly reduces empirical secrecy capacity by 81.7\%;
(2) The attacker can steer the sideband channel toward themselves while keeping the legitimate channel pointed at the client.

\ziyu{
\textbf{Our Critical Assessment in Practical Systems:}
This article highlights that an external adversary may launch eavesdropping attack by tampering the RIS's control link in a RIS-assisted communication system. 
However, the practicality of this type of attack is often constrained by several factors. 
First, it requires the attacker to have a stable access to the legitimate RIS's control link and precise control over the RIS, which is already a strong adversarial capability assumption. 
Second, the attack performance is strongly scene-dependent, since the attack creates a decodable leakage path toward the eavesdropper while simultaneously preserving the legitimate link for the LU. This dependency is particularly significant in mmWave systems, as mmWave propagation is highly directional and sensitive to blockage, alignment errors, and environmental changes.
Third, these attacks often assume static or slowly varying geometries, so that the adversary can jointly preserve the legitimate beam and create a usable leakage path toward itself. 
Such assumptions may not hold in realistic mobile or cluttered deployments. 
Thus, while this type of attack is valuable for demonstrating a real physical-layer vulnerability, its effectiveness in practice may be limited by deployment dependence, calibration requirements, and reduced robustness under dynamic conditions.
} % from author's point of view

\begin{figure}[!ht]
\centering

\begin{tcolorbox}[
  colback=gray!12,
  colframe=black!70,
  boxrule=0.8pt,
  arc=2.5mm,
  left=6pt,
  right=6pt,
  top=4pt,
  bottom=6pt,
  title=\textbf{Lessons learned},
  coltitle=white,
  colbacktitle=black,
  fonttitle=\bfseries,
  fontupper=\normalfont
]

This type of attack reveals that the main weakness of RIS-assisted systems lies not only in the openness of the wireless medium, but also in the \emph{amplification effect} between the RIS control link and the wireless propagation plane: a small amount of malicious control over a legitimate RIS can cause large-scale wavefront manipulation and create an additional leakage path.

\begin{itemize}
    \item \textbf{Secure the control link as a high-impact actuator, not as a low-rate side interface:} 
    The root cause is that a low-complexity RIS controller governs a high-gain electromagnetic aperture. Therefore, the RIS control link should be treated as a security-critical actuator path. Future RIS designs should provide lightweight authentication, integrity protection, and access isolation for phase updates, rather than assuming the control link is trustworthy.

    \item \textbf{Add runtime channel/beam attestation rather than relying on control-link trust alone:} 
    MeSS~\cite{chen2023wavefront} works because the attacker can alter the RIS phase pattern while keeping victim-side degradation small. This means that trusted control messages alone are insufficient once the RIS is compromised. A practical defense is to let the AP/BS continuously verify whether the measured cascaded channel matches the expected RIS state, and trigger re-training, surface isolation, or fallback once a persistent mismatch is detected.
\end{itemize}

\end{tcolorbox}
\vspace{-15pt}
\end{figure} % in detail, should be about the analysis after pioneer works (why? or how to solve it.)

\subsection{Attacks Leveraging the RIS} \label{threats2}
Apart from its functionalities in practical systems, the RIS itself is an effective tool for PLS. It can be used to attack communication and sensing systems, to compromise or disrupt a target system, or as a countermeasure against attacks by LUs to protect themselves and enhance security or privacy~\cite{zhu2023active}. \ziyuu{In contrast to the attacks in the previous subsection, which compromise components inside the operator's trust domain}, in the following we will elaborbate on the systems that are prone to attacks where RIS is used and attacks that RIS is used for. The systems are summarized in Table~\ref{tab:ris_aided_sys_security}.

\begin{figure*}[!ht]
    \centering
    \subfloat[Targeted Eavesdropping.]
    {\includegraphics[width=0.35\linewidth]{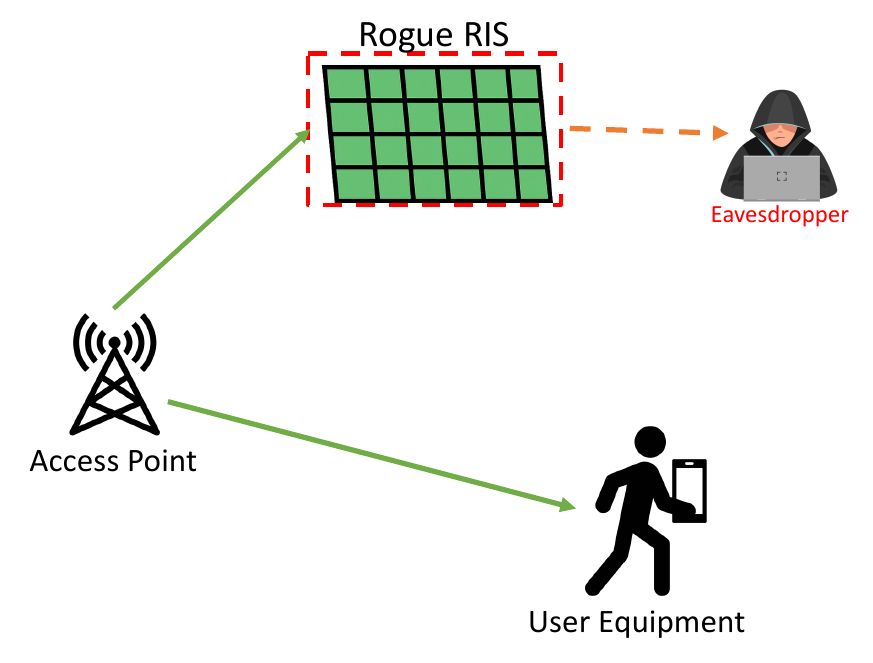}%
    \label{fig:eavesdropping}}
    \hfil
    \subfloat[Advanced Jamming.]{\includegraphics[width=0.3\linewidth]{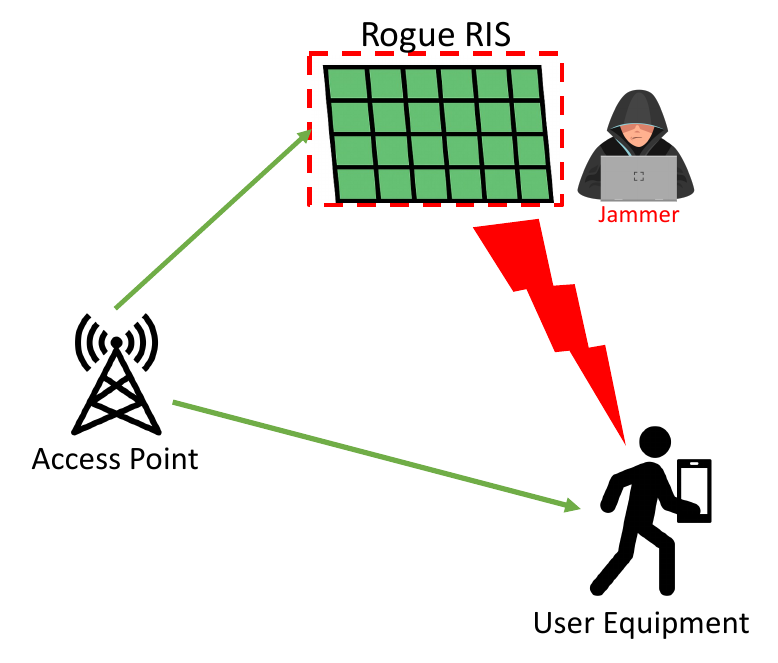}%
    \label{fig:jamming}}
    \hfil
    \subfloat[Physical-Layer Spoofing.]{\includegraphics[width=0.3\linewidth]{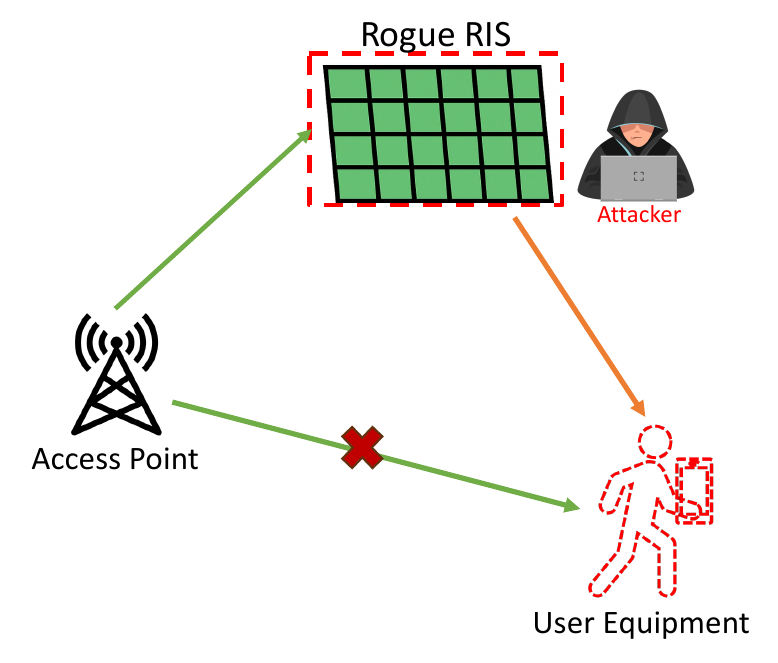}%
    \label{fig:spoofing}}
    \hfil
    \caption{Representative types of attacks leveraging a malicious RIS.}
    \label{fig:att_rogue_ris}
    \vspace{-5pt}
\end{figure*}

\subsubsection{Target systems}
Practical systems that rely heavily on physical-layer propagation are vulnerable to RIS-based attack. Three patterns recur across the system classes below. First, passivity makes detection hard: an RIS reshapes reflections without emitting new signals, so attacks often present as ordinary dead spots, sensor glitches, or connectivity issues rather than security events. Second, accessible deployment environments make it feasible to install covert surfaces close enough to the victim to matter. Third, systems that rely on the physical layer for trust, sensing-based intrusion detection, localization, ISAC, V2X beam management, are most exposed, because the RIS attacks the sensed environment rather than the protocol stack.

\textbf{(1) Smart homes:}
Smart homes constantly exchange data between cameras, smart locks, sensors, and appliances, and increasingly rely on Wi-Fi or mmWave sensing to detect intruders, falls, and motion. An RIS can passively redirect these signals for eavesdropping, or manipulate them to simulate or suppress motion and fool sensing-based security~\cite{jiang2024risiren}. Because the effects resemble ordinary Wi-Fi dead spots, sensor glitches, or random malfunctions, such attacks are difficult for residents to detect. Heavy wireless reliance, weak IoT security, and easy physical access compound the exposure.

\textbf{(2) Smart vehicular systems:}
V2X and autonomous driving systems depend on real-time, directional wireless links. Rapidly changing channels make beam control and CSI estimation difficult, and unauthorized RISs, mounted on roadside infrastructure, camouflaged in the environment, or carried on drones, can subtly inject interference, eavesdrop, or spoof vehicular sensing, especially in thin mmWave beams~\cite{chen2022reconfigurable, chen2023metawave}. A mobile RIS enables stealthy interception or error injection from a safe distance.

\textbf{(3) IoMT:}
IoMT systems consist of wireless-dependent medical devices (infusion pumps, imaging terminals, monitors, wearables) that use Wi-Fi, Bluetooth, or RFID and often run legacy OSes with weak authentication~\cite{zhu2023active, wang2022robust}. Accessible deployment environments make covert RIS placement feasible without triggering intrusion detection. A malicious RIS could block or alter wireless alerts (delaying notifications), eavesdrop on sensitive data streams, or inject false signals that cause sensors to miss critical motions. Any of these effects may have catastrophic clinical consequences.

\textbf{(4) 6G and massive MIMO networks:}
Massive MIMO serves many users simultaneously via CSI-dependent spatial multiplexing. A malicious RIS in the environment can passively induce destructive interference toward specific users by tuning its phase shifts, acting as a silent jammer that disrupts targets while leaving other users unaffected~\cite{de2024malicious}. RIS-enabled destructive beamforming degrades SNR in a targeted manner that scales with the number of RIS elements~\cite{rivetti2024malicious}.

\textbf{(5) UAV and drone control links:}
UAVs communicate with controllers over wireless links that are exposed to eavesdropping, relay manipulation, jamming, and link disruption. 
An unauthorized RIS can reflect UAV--ground signals toward external listeners, undermining the confidentiality of telemetry or media feeds. 
By concentrating a jammer's energy along the UAV direction, it can also extend jamming range and effectiveness without active transmission. Consequences include loss of control, video-feed interception, or command injection, all without the attacker emitting a recognizable signal.

\textbf{(6) ISAC systems:}
ISAC integrates sensing and communication on a shared hardware framework~\cite{ma2024integrated, liu2023integrated}. A malicious RIS simultaneously undermines sensing accuracy and communication privacy through the tightly coupled physical layer: the attacker can eavesdrop or distort communication data while corrupting sensing inputs, and these attacks may go unnoticed in systems assuming benign propagation.

\textbf{(7) IoT and low-power wireless networks:}
Off-the-shelf IoT devices use wireless channels that an RIS attacker can exploit for both eavesdropping and jamming. A covert RIS near an IoT network can redirect signals beyond their normal range, e.g., reflecting a factory sensor's path to an outside receiver~\cite{wang2022wireless}, or beamform a jammer's signal to disable one node at a time without interfering with others~\cite{mackensen2024spatial}. Because the RIS is passive, victims experience only unusual connectivity issues or battery drain while the attacker quietly manipulates the RF environment.

\subsubsection{Attacks leveraging the RIS} 
\ziyu{

In contrast to compromising legitimate infrastructure, this category of threats involves an adversary deploying an unauthorized malicious RIS or covert meta-material tags into the physical environment, as shown in Fig.~\ref{fig:att_rogue_ris}. By introducing attacker-controlled reflection paths, the adversary weaponizes the wireless channel itself. Because these attacks often do not require the adversary to emit active radio-frequency (RF) power, they are exceptionally stealthy and can easily bypass traditional spectrum monitoring systems. The attacks are summarized in Table~\ref{tab:ris_attack}

\begin{table*}[]
    \centering
    \scalebox{0.82}{
    \begin{tabular}{|L{0.7cm}|L{1.6cm}|L{1cm}|L{1.7cm}|L{1.2cm}|L{1.9cm}|L{2cm}|L{1cm}|L{1cm}|L{2.2cm}|L{2.5cm}|}
    \hline
         \textbf{Ref.} &
         \textbf{Attack type} &
         \textbf{Comm. Method} &
         \textbf{Scenario} &
         \textbf{Corres-pondence to Fig.~\ref{fig:application_out}} &
         \textbf{Attack Method} &
         \textbf{Malicious RIS Setup} &
         \ziyuu{\textbf{CSI Requirements}} &
         \ziyuu{\textbf{Regime}} &
         \textbf{Attack Performance} &
         \ziyuu{\textbf{Limitations}} \\ \hline

         % --- Eavesdropping ---
         \cite{chen2022malicious} &
         \multirow{5}{1.4cm}{Eavesdropping} &
         \multirow{2}{*}{mmWave} &
         SISO with a malicious RIS &
         \multirow{2}{*}{\textcircled{3}} &
         \multirow{2}{1.9cm}{Beam steering sideband to Eve.} &
         \multirow{2}{2cm}{Unauthorized RIS in the channel.} &
         \multirow{2}{1cm}{\ziyuu{CSI-free}} &
         \multirow{2}{1cm}{\ziyuu{Far-field}} &
         \multirow{2}{2.2cm}{Empirical secrecy capacity $\downarrow$81.7\%} &
         \multirow{2}{2.5cm}{\ziyuu{Secrecy loss assumes strict attacker-RIS alignment; vulnerable to ISAC scanning.}} \\ \cline{1-1}\cline{4-4}
         
         \cite{chen2023wavefront} &
         & & mmWave WLAN system & & & & & & & \\ \cline{1-1}\cline{3-11}
          
         \cite{shaikhanov2022metasurface} &
         &
         THz &
         Directional wireless comm. &
         \textcircled{1}\textcircled{2}\textcircled{3}\textcircled{4} &
         Redirect part of signal toward Eve &
         Unauthorized RIS in the channel. &
         \ziyuu{CSI-free} &
         \ziyuu{Far-field} &
         Reduced secrecy capacity. &
         \ziyuu{Reduced secrecy assumes strict attacker-RIS alignment near the link.} \\ \cline{1-1}\cline{3-11}
         
         \cite{wei2023metasurface} &
         &
         sub-6G &
         Network, one AP \& one+ clients &
         \textcircled{1}\textcircled{2}\textcircled{4} &
         Passive: control RIS; Active: deceive target &
         Unauthorized RIS in NLoS path. &
         \ziyuu{Estimated} &
         \ziyuu{---} &
         16 dB Eve gain; –23 Mbps LU rate. &
         \ziyuu{Geometry changes require re-optimization.} \\ \cline{1-1}\cline{3-11}
         
         \cite{ning2025stealthyvoiceeavesdroppingacoustic} &
         &
         Acoustic &
         Moving target on call, distant attacker &
         \textcircled{3}\textcircled{4} &
         Passive acoustic amplification &
         Eve at safe distance (4.5m) &
         \ziyuu{CSI-free} &
         \ziyuu{---} &
         $>$80\% accuracy; 20$\times$ magnification. &
         \ziyuu{20$\times$ gain assumes Eve within $\sim$4.5\,m.} \\ \hline

         % --- Jamming ---
         \cite{huang2023disco} &
         \multirow{2}{1.4cm}{Jamming} &
         sub-6G &
         MU-MISO systems &
         \textcircled{4} &
         Actively age the LUs' channels &
         Malicious RIS, random phase shifts. &
         \ziyuu{CSI-free} &
         \ziyuu{Far-field} &
         Large rate drop; no power/channel knowledge needed. &
         \ziyuu{Vulnerable to anomaly/Doppler detection.} \\ \cline{1-1}\cline{3-11}
         
         \cite{mackensen2024spatial} &
         &
         sub-6G &
         Wi-Fi, multiple devices &
         \textcircled{4} &
         Active selective jamming &
         Malicious RIS next to the jammer + RF jammer &
         \ziyuu{Estimated} &
         \ziyuu{Far-field} &
         100\% DoS of target; others unaffected. &
         \ziyuu{Frequency hopping mitigates threat.} \\ \hline

         % --- Spoofing ---
         \cite{chen2023metawave} &
         \multirow{7}{1.4cm}{Spoofing} &
         mmWave &
         Radar (24--77 GHz) &
         \textcircled{2} &
         Echo modulation (multipath/atten.) &
         Stealthy passive meta-material tags &
         \ziyuu{CSI-free} &
         \ziyuu{---} &
         97\% success rate. &
         \ziyuu{Needs offline RF profiling; fails with multimodal sensors.} \\ \cline{1-1}\cline{3-11}
         
         \cite{vennam2023mmspoof} &
         &
         mmWave &
         Automotive mmWave radars &
         \textcircled{2} &
         Reflect \& modulate radar signal &
         Attacker RIS on a vehicle ahead of victim &
         \ziyuu{Estimated} &
         \ziyuu{Far-field} &
         96\% success rate. &
         \ziyuu{Highly vulnerable to dynamic environments.} \\ \cline{1-1}\cline{3-11}
         
         \cite{li2025ris} &
         &
         mmWave &
         Positioning w/ legitimate RIS &
         \textcircled{3}\textcircled{4} &
         Directional beamforming &
         Unauthorized RIS with placement and codebook for strong path ambiguity. &
         \ziyuu{---} &
         \ziyuu{Far-field} &
         Poor positioning performance. &
         \ziyuu{Strong if near BS/UE. Sensitive to placement.} \\ \cline{1-1}\cline{3-11}
         
         \cite{shui2025sensing} &
         &
         mmWave &
         ISAC vehicle networks, RSU &
         \textcircled{2} &
         Dynamically adjusts phase shifts &
         At the roadside &
         \ziyuu{Perfect} &
         \ziyuu{Far-field} &
         Induces velocity \& AoD error. &
         \ziyuu{Requires phase synchronization.} \\ \cline{1-1}\cline{3-11}
         
         \cite{jiang2024risiren} &
         &
         sub-6G &
         Indoor Wi-Fi sensing &
         \textcircled{3} &
         Generate malicious multipath &
         Malicious RIS, indoor &
         \ziyuu{CSI-free} &
         \ziyuu{Far-field} &
         90\% attack success rate. &
         \ziyuu{Vulnerable to CSI anomaly detection.} \\ \cline{1-1}\cline{3-11}
         
         \cite{zhou2023ristealth} &
         &
         sub-6G &
         Indoor Wi-Fi sensing &
         \textcircled{3} &
         Motion suppression \& threshold lifting &
         Malicious RIS carried by moving person &
         \ziyuu{Estimated} &
         \ziyuu{Far-field} &
         Detection down to 16.4\%. &
         \ziyuu{Vulnerable to active environment scanning.} \\ \cline{1-1}\cline{3-11}
        
         \cite{ning2025portablestealthyinaudiblevoice} &
         &
         Acoustic &
         Voice assistants &
         \textcircled{3}\textcircled{4} &
         Inaudible attack &
         Eve 8--9\,m from system &
         \ziyuu{CSI-free} &
         \ziyuu{---} &
         76\% accuracy; 8.85\,m range. &
         \ziyuu{Vulnerable to acoustic liveness detection.} \\ \hline
         
    \end{tabular}
    }
    \caption{Summary of Attacks Leveraging the RIS.}
    \vspace{-10pt}
    \label{tab:ris_attack}
\end{table*}

\paragraph{Targeted Eavesdropping via Anomalous Reflection}
Traditional wireless security often assumes that an eavesdropper must physically reside within the transmitter's beam or coverage area. However, a malicious RIS can completely subvert this assumption by capturing side-lobe energy or intercepting the main beam, redirecting it toward an attacker located in a secure blind spot, as shown in Fig.~\ref{fig:eavesdropping}.
\begin{itemize}
    \item \textbf{RF Eavesdropping:} By deploying an unauthorized RIS in the channel, attackers can redirect a portion of the legitimate signal toward themselves, significantly reducing the secrecy capacity~\cite{shaikhanov2022metasurface}. Advanced smart wireless attacks at the physical layer can even achieve up to 16 dB eavesdropping gain by jointly manipulating radio waves and digital information~\cite{wei2023metasurface}.
    \item \textbf{Acoustic Amplification:} In acoustic channels, unauthorized metasurfaces can be used for highly targeted, long-range eavesdropping. For instance, the ``SuperEar'' attack leverages acoustic metamaterials to passively magnify speech signals by approximately 20 times, achieving over 80\% eavesdropping accuracy from a safe distance~\cite{ning2025stealthyvoiceeavesdroppingacoustic}.
\end{itemize}

\paragraph{Passive Signal Nulling and Selective Jamming}
Conventional jamming requires transmitting high-power noise, which is easily detectable. A malicious RIS, however, can act as a ``passive jammer.'' By intentionally tuning its phase shifts, the RIS can induce destructive interference (signal nulling) at the legitimate receiver's location without actively emitting any RF signals~\cite{lyu2020irs, de2024malicious, rivetti2024malicious}, as shown in Fig.~\ref{fig:jamming}.
\begin{itemize}
    \item \textbf{Spatial-Domain Selectivity:} Adversaries can utilize an RIS to launch highly precise, selective jamming attacks. For example, an attacker can completely disable the wireless communication of a specific victim device (e.g., a smart home alarm) while leaving nearby devices—even those just millimeters away—completely unaffected~\cite{mackensen2024spatial}.
    \item \textbf{Channel Aging Attacks:} Malicious RISs with random or rapidly changing phase shifts can artificially age the legitimate users' channels, severely degrading the data rate in multi-user MIMO systems without requiring additional power or prior channel knowledge~\cite{huang2023disco}.
\end{itemize}
% \begin{figure}[!ht]
%     \centering
%     \subfloat[Generating a fake target with an RIS.]{\includegraphics[width=0.75\linewidth]{Figures/new/ris_ghosting.pdf}%
%     \label{fig:ghost_attack}}
%     \hfil
%     \subfloat[RIS disappearing an object in front of the victim.]{\includegraphics[width=0.75\linewidth]{Figures/new/ris_vanishing.pdf}}%
%     \label{fig:vanish_attack}
%     \hfil
%     \caption{RIS in the environment disrupting a sensing system~\cite{chen2023metawave}.}
%     \label{fig:RIS_attack_metawave}
% \end{figure}
\begin{figure}[!ht]
    \centering
    \subfloat[A basic wireless sensing system detecting intruders.]{\includegraphics[width=0.475\linewidth]{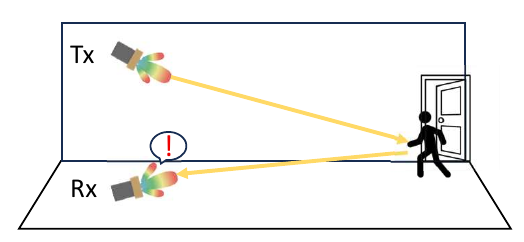}%
    \label{fig:intruder_w_out_ris}}
    \hfil
    \subfloat[RIS makes the intruder undetectable.]{\includegraphics[width=0.475\linewidth]{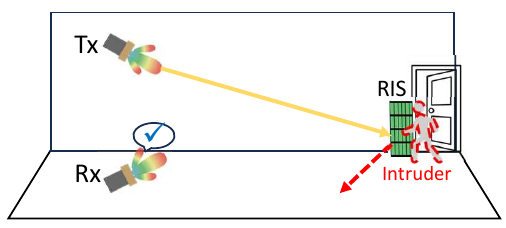}%
    \label{fig:intruder_w_ris}}
    \hfil
    \caption{RIS brings covert threats to wireless sensing~\cite{zhou2023ristealth}.}
    \vspace{-15pt}
    \label{fig:RIS_attack_wifi}
\end{figure}

\paragraph{Physical-Layer Spoofing and Ghost Targets}
As modern networks evolve towards ISAC, the physical layer is increasingly relied upon for environmental perception. Malicious RISs pose a critical threat to these sensing modalities by synthesizing fake multipath signatures or modulating echoes, as shown in Fig.~\ref{fig:spoofing}.
\begin{itemize}
    \item \textbf{Radar and ISAC Spoofing:} In mmWave and automotive networks, low-cost, stealthy passive meta-material tags (e.g., the MetaWave attack) can modulate radar echoes to create ``ghost'' vehicles out of nowhere or ``vanish'' real obstacles, achieving over 90\% success rates in angle, speed, and range manipulation~\cite{chen2023metawave, vennam2023mmspoof}. Similar vulnerabilities exist in ISAC vehicular networks~\cite{shui2025sensing} and large-scale SAR imaging systems~\cite{li2024broadband}, where adversaries dynamically adjust RIS phase shifts to induce severe estimation errors.
    \item \textbf{Indoor Wi-Fi Sensing Deception:} Sub-6G Wi-Fi sensing systems used for smart home security are highly vulnerable. Attacks like \textit{RISiren}~\cite{jiang2024risiren} (shown in Fig.~\ref{fig:RIS_attack_wifi} and \textit{RIStealth}~\cite{zhou2023ristealth} demonstrate that a malicious RIS can generate fake human-like multipath signatures or suppress real motion, dropping intrusion detection rates from 95.1\% down to 16.4\%. Additionally, unauthorized RIS panels can severely degrade the performance of wireless positioning systems~\cite{li2025ris}.
    \item \textbf{Acoustic Spoofing:} Beyond RF, malicious acoustic metamaterials can inject stealthy, inaudible voice commands into voice control systems from several meters away, bypassing traditional audio security filters~\cite{ning2025portablestealthyinaudiblevoice}.
\end{itemize}
}

\begin{figure}[!ht]
    \centering
    \subfloat[RIS disappearing an object in front of the victim.]{\includegraphics[width=0.45\linewidth]{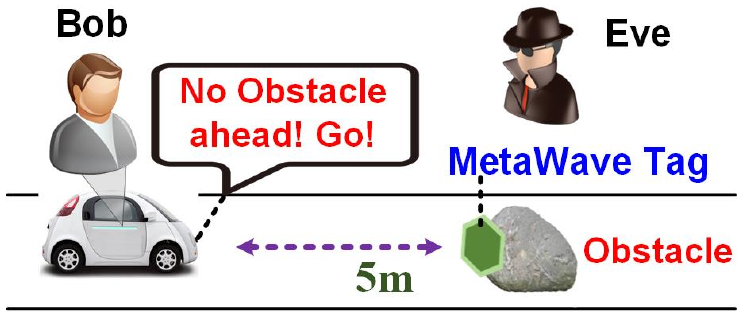}%
    \label{fig:vanish_attack_1}}
    \hfil
    \subfloat[Generating a fake target with an RIS.]{\includegraphics[width=0.45\linewidth]{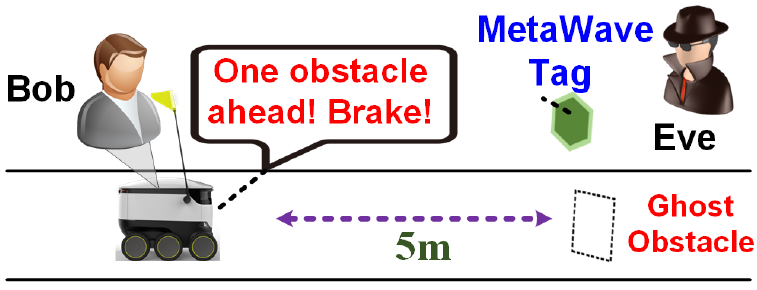}}%
    \label{fig:ghost_attack_1}
    \hfil
    \caption{Stealthy RIS tag in the environment disrupting a car radar. \protect\footnotemark}
    \label{fig:metawave}
    \vspace{-10pt}
\end{figure}

\noindent \textbf{Pioneer Work: \textit{MetaWave: Attacking mmWave Sensing with Meta-material-enhanced Tags}~\cite{chen2023metawave}}, as shown in Fig.~\ref{fig:metawave}.

\textbf{Attack Scenario:}
This work considers a scenario where an \textbf{attacker with RIS} is \textbf{spoofing} a \textbf{non-RIS} mmWave sensing system.
An attacker aims to spoof a victim that leverages mmWave sensing for measurement and detection. 
The attacker deploys low-cost and easily obtainable \textbf{passive} meta-material tags (which can be considered as RIS) at different objects in the environment.
The attacker launches 2 types of attacks: 
(1) vanish attack, which either hides an obstacle that the victim should avoid, or hide the attacker from being detected while trespassing;
(2) ghost attack, which creates fake objects out of nowhere, which triggers false alerts for obstacle detection or trespassing security alerts.
The attacker has no knowledge of the sensing system, \ziyu{but the threat model requires them to acquire detailed physical and RF parameters of the target environment for tag design}.

\textbf{Attack Goal:}
The attacker aims to cause ghost (fake objects) or vanish (hide real objects) effects to the victim. Thus, the victim will have mistaken range / angle / velocity estimates. In addition, the attack will be low-cost and stealthy.

\textbf{Attack Method:}
\ziyu{The attack relies heavily on offline optimization.} The attacker first models the target scene's parameters. They then fabricate specific passive meta-material tags customized for the sensing modality: \ziyu{using \textbf{absorption tags} to eliminate RF echoes (Vanish) and \textbf{reflection tags} to generate deceptive multi-path signals (Ghost).} Since the tags are passive, their physical parameters (size, shape, position) must be fine-tuned via offline simulation for the specific static scene prior to physical deployment.

\textbf{Attack Result:}
Across 20 environments, the attack achieves 97\% attack accuracy on range measurements, 96\% on angle measurements, and 91\% on speed measurements. Moreover, it costs 10–100$\times$ lower than active RF attacks.

\ziyu{
\textbf{Our Critical Assessment in Practical Systems:}
This type of attack reveals a real physical-layer vulnerability of mmWave sensing systems: an attacker equipped with \textbf{passive} metasurfaces can manipulate the surrounding propagation environment to create ghost targets or vanish actual targets. 
However, such attacks usually rely on geometry or deployment. 
Their effectiveness often depends on scene-aware surface/tag design, customized placement, and sufficient knowledge of the victim's radar configuration and environmental parameters, so that the induced scattering pattern can produce the desired attack results.
As a result, spoofing non-RIS sensing systems with a RIS is typically more effective in structured and relatively stable scenarios, while their performance can degrade under mobility, view-angle variation, tracking dynamics, or other scene perturbations~\cite{hunt2024madradar}. 
Therefore, this line of work should be interpreted primarily as evidence that low-cost metasurfaces can expose a practical sensing vulnerability, rather than as a robust spoofing strategy across arbitrary real-world deployments.
\vspace{-10pt}
}

\footnotetext{Reproduced from \cite{chen2023metawave}: Chen et al., MetaWave: Attacking mmWave Sensing with Meta-material-enhanced Tags, NDSS 2023, DOI: 10.14722/ndss.2023.24348. \copyright~2023 Internet Society. Used for non-commercial purposes.}

\begin{figure}[!ht]
\centering

\begin{tcolorbox}[
  colback=gray!12,
  colframe=black!70,
  boxrule=0.8pt,
  arc=2.5mm,
  left=6pt,
  right=6pt,
  top=4pt,
  bottom=6pt,
  title=\textbf{Lessons learned},
  coltitle=white,
  colbacktitle=black,
  fonttitle=\bfseries,
  fontupper=\normalfont
]

This type of attack show that the weakness of mmWave sensing lies not only in signal processing, but also in a deeper trust assumption: the system usually treats reflections as natural propagation rather than adversarially engineered scattering. As a result, an attacker with passive RIS can create ghost targets or suppress genuine reflections without emitting active RF signals.

\begin{itemize}
    \item \textbf{Do not assume the propagation environment is benign:}
    The root cause is that passive metasurface-based objects can act as low-cost wave manipulators. Therefore, future sensing systems should not treat anomalous echoes only as clutter or noise, but also consider adversarial reflectors.

    \item \textbf{Move from single-frame detection to consistency-based sensing defense:}
    Because passive spoofing can remain stealthy at the spectrum level, defenses should verify whether targets stay consistent across time, viewpoints, frequencies, or sensing modalities. Practical directions include temporal tracking, viewpoint checks, multi-frequency validation, and cross-sensor verification~\cite{hunt2024madradar}.
\end{itemize}

\end{tcolorbox}
\vspace{-10pt}
\end{figure}
\section{Countermeasures against Attacks in RIS-Related Systems} \label{defenses}
In this section, we address possible countermeasures for the potential threats to RISs and defenses provided by an extra legitimate RIS. In the following, we will elaborate on countermeasures for security and privacy threats to RIS-assisted systems, countermeasures against attacks conducted with an unauthorized RIS, as well as defenses that feature a friendly RIS. 
\vspace{-10pt}

\subsection{Securing RIS-Assisted Systems} \label{subsec1}
Securing a RIS-assisted system requires protecting both the wireless function enabled by the legitimate RIS and the RIS infrastructure itself. 
As discussed in Section~\ref{threats1}, an adversary may either passively exploit the RIS-assisted propagation environment for eavesdropping, privacy leakage, or unauthorized sensing, or actively compromise the RIS infrastructure through the control link, controller, or hardware. 
These works are summarized in Table~\ref{tab:defense_ris_aided_system}.
\ziyu{Note that in Table~\ref{tab:defense_ris_aided_system}, \ref{tab:defense_sys_design}, and~\ref{tab:ris_defense}, we qualitatively summarize defense performance using three strength levels: \emph{near-complete}, \emph{strong}, and \emph{moderate}, indicating whether the defense nearly neutralizes, substantially mitigates, or noticeably reduces the targeted threat in the evaluated setting. We additionally annotate whether a defense is \emph{selective}, i.e., it mainly suppresses the adversary while preserving legitimate functionality. We also annotate whether the defense is \emph{specialized}, i.e., it is tailored to a specific device, scenario, or attack type.}

% \begin{figure}[!ht]
%     \centering
%     \subfloat[Disrupting unauthorized sensing with artificial noise.]{\includegraphics[width=0.45\linewidth]{Figures/new/def_eave_an.pdf}%
%     \label{fig:def_eave_an}}
%     \hfil
%     \subfloat[Apply wiretap codes so that Eve is unable to decode the signal.]{\includegraphics[width=0.45\linewidth]{Figures/new/def_eave_coding.pdf}%
%     \label{fig:def_eave_coding}}
%     \hfil
%     \caption{Example of countermeasures of the attacks of RIS-assisted applications.}
%     \label{fig:def_eave}
%     \vspace{-10pt}
% \end{figure}

\begin{table*}[]
    \centering
    \scalebox{0.9}{
    \begin{tabular}{|L{0.7cm}|L{1.5cm}|L{1cm}|L{2cm}|L{1.2cm}|L{1.5cm}|L{1cm}|L{1cm}|L{2cm}|L{3.6cm}|}
    \hline
     \textbf{Ref.} &
     \textbf{Threat Type} &
     \textbf{Comm. Method} &
     \textbf{Scenario} &
     \textbf{Corres-pondence to Fig.~\ref{fig:application_out}} &
     \textbf{Defense Method} &
     \ziyuu{\textbf{CSI Require.}} &
     \ziyuu{\textbf{Field Regime}} &
     \textbf{Defense Performance} &
     \ziyuu{\textbf{Limitation}} \\ \hline

    \cite{magbool2025hiding} &
    \multirow{12}{1.5cm}{Eavesdropping} &
    sub-6G, cellular &
    RIS-aided ISAC target privacy &
    \textcircled{3} &
    Optimization &
    \ziyuu{Perfect} &
    \ziyuu{Far-field} &
    Strong, selective &
    \ziyuu{Requires adaptive RIS element assignment and structured deployment geometry; privacy tied to accurate sensing-region modeling.} \\ \cline{1-1} \cline{3-10}

    \cite{mao2022reconfigurable} &
    & sub-6G, cellular &
    RIS-assisted MEC offloading &
    \textcircled{4} &
    Optimization &
    \ziyuu{Perfect} &
    \ziyuu{Far-field} &
    Strong &
    \ziyuu{Focused on MEC utility under passive eavesdropping; optimization complexity is high.} \\ \cline{1-1} \cline{3-10}

    \cite{wang2022secure} &
    & sub-6G, cellular &
    MISO NOMA under passive eavesdropping &
    \textcircled{4} &
    Optimization &
    \ziyuu{CSI-free} &
    \ziyuu{Far-field} &
    Strong, selective &
    \ziyuu{Tailored to NOMA with an untrusted strong user; relies on artificial jamming and coordinated beamforming.} \\ \cline{1-1} \cline{3-10}

    % \cite{wang2024underwater} &
    % & Acoustic &
    % Underwater acoustic source privacy &
    % / & Optimization &
    % Estimated & --- &
    % Strong, specialized &
    % Specialized to underwater acoustic sensor networks. \\ \cline{1-1} \cline{3-10}

    \cite{jing2024reconfigurable} &
    & Optical &
    V2V VLC with roadside eavesdropping &
    \textcircled{2} &
    Optimization &
    \ziyuu{CSI-free} &
    \ziyuu{---} &
    Moderate, selective, specialized &
    \ziyuu{Effective only with favorable roadside/intersection RIS placement; specialized to VLC.} \\ \cline{1-1} \cline{3-10}

    \cite{liu2022security} &
    & sub-6G, cellular &
    UAV IIoT secure downlink &
    \textcircled{1} &
    Optimization &
    \ziyuu{Perfect} &
    \ziyuu{Far-field} &
    Moderate &
    \ziyuu{Depends on UAV deployment assumptions and secrecy-oriented channel design.} \\ \cline{1-1} \cline{3-10}

    \cite{wang2022robust} &
    & sub-6G, cellular &
    Secure IoMT with imperfect CSI &
    \textcircled{4} &
    Optimization &
    \ziyuu{Estimated} &
    \ziyuu{Far-field} &
    Moderate &
    \ziyuu{Specific to STAR-RIS-assisted IoMT and the assumed CSI uncertainty model.} \\ \cline{1-1} \cline{3-10}

    \cite{fang2024security} &
    & sub-6G, cellular &
    RIS-assisted MEC offloading &
    / &
    Optimization &
    \ziyuu{Perfect} &
    \ziyuu{Far-field} &
    Strong &
    \ziyuu{Uses task-specific secrecy-capacity/energy objectives, making comparison with standard secrecy-rate works less direct.} \\ \cline{1-1} \cline{3-10}

    \cite{chen2025secure} &
    & mmWave &
    RIS-aided ISAC against PL-sensing adversary &
    \textcircled{4} &
    Introducing randomness &
    \ziyuu{CSI-free} &
    \ziyuu{---} &
    Strong, selective &
    \ziyuu{Tailored to sensing-adversary suppression.} \\ \cline{1-1} \cline{3-10}

    \cite{xu2023deep} &
    & sub-6G, cellular &
    Secure MEC in IIoT &
    \textcircled{4} &
    Optimization &
    \ziyuu{Perfect} &
    \ziyuu{Far-field} &
    Strong &
    \ziyuu{Relies on DRL training and policy generalization; training overhead is not negligible.} \\ \cline{1-1} \cline{3-10}

    \cite{lei2025ris} &
    &
    / &
    Near-field NOMA with an untrusted near user &
    \textcircled{4} &
    Optimization &
    \ziyuu{Perfect} &
    \ziyuu{Mixed FF/NF} &
    Strong, selective &
    \ziyuu{Fixed two-user geometry; subject to a security--fairness trade-off.} \\ \cline{1-1} \cline{3-10}
    
    \cite{liang2025double} &
    &
    mmWave &
    Double-RIS near-field secure communication &
    \textcircled{4} &
    Optimization &
    \ziyuu{Perfect} &
    \ziyuu{Near-field} &
    Strong, selective, specialized &
    \ziyuu{Assumes known geometry and iterative centralized optimization.} \\ \cline{1-1} \cline{3-10}

    \cite{zhang2022deep} &
    & sub-6G &
    MEC under PLS &
    / &
    Optimization &
    \ziyuu{Perfect} &
    \ziyuu{Far-field} &
    Moderate &
    \ziyuu{Uses a mixed latency--energy objective rather than a pure secrecy metric.} \\ \hline

    \cite{cui2024xl} &
    \multirow{2}{1.5cm}{Jamming \& eavesdropping} &
    mmWave &
    XL-RIS near-field secure downlink &
    \textcircled{4} &
    Optimization &
    \ziyuu{Estimated} &
    \ziyuu{Mixed FF/NF} &
    Strong, selective &
    \ziyuu{High-dimensional optimization.} \\ \cline{1-1} \cline{3-10}
    
    \cite{zou2023ris} &
    &
    sub-6G &
    UAV anti-jamming and anti-eavesdropping link &
    \textcircled{1} & 
    Optimization &
    \ziyuu{Estimated} & 
    \ziyuu{Far-field} &
    Strong, selective &
    \ziyuu{Requires accurate CSI/UAV-state estimation and robust DRL generalization in a joint threat setting.} \\ \hline

    \cite{thanh2022anti} &
    \multirow{2}{1.5cm}{Jamming} &
    sub-6G &
    Solar-powered RIS sensor network under jamming &
    \textcircled{4} &
    Optimization &
    \ziyuu{Estimated} &
    \ziyuu{Far-field} &
    Moderate &
    \ziyuu{Depends on stable DQN training and online adaptation in an energy-constrained sensor-network setting.} \\ \cline{1-1} \cline{3-10}

    \cite{huang2023anti} &
    & sub-6G, cellular &
    MU--MISO downlink under passive jamming &
    \textcircled{4} &
    Optimization &
    \ziyuu{Perfect} &
    \ziyuu{Far-field} &
    Strong, selective &
    \ziyuu{Specifically designed for DISCO-RIS-like passive jammers; relies on statistical attacker-side knowledge.} \\ \hline

    \end{tabular}
    }
    \caption{Summary of representative defenses against attacks on RIS-assisted systems.}
    \label{tab:defense_ris_aided_system}
    \vspace{-10pt}
\end{table*}

\ziyu{
\subsubsection{Defending Against Passive Eavesdropping and Privacy Leakage}
As discussed in Section~\ref{threats1}, a legitimate RIS can inadvertently benefit an adversary by creating new reflection paths or restoring blocked links. Since the adversary does not need to jam or tamper with anything, defenses here must go beyond classical anti-eavesdropping and also address privacy leakage through unauthorized sensing, positioning, or inference over RIS-assisted channels.

\paragraph{Secrecy enhancement for communication links}
A natural starting point is to ensure that the legitimate receiver sees a strong, decodable signal while the eavesdropper does not. In practice, this is done by jointly optimizing beamforming, RIS phase shifts, power allocation, and artificial noise. 
Several works along this line optimize beamforming and RIS configurations together to improve secrecy-oriented metrics in secure offloading and multi-user systems~\cite{mao2022reconfigurable, fang2024security, wang2022secure, liu2022security, wang2022robust}. 
The same idea carries over to optical and underwater settings, where the RIS is configured to limit signal exposure towards passive listeners while keeping the link to the intended receiver intact~\cite{jing2024reconfigurable, wang2024underwater}. 
More recently, DRL-based policies have been used to replace repeated online optimization: the agent learns to adapt RIS coefficients and transmission parameters on the fly, which is particularly attractive in dynamic RIS-assisted MEC and IIoT scenarios~\cite{xu2023deep, zhang2022deep, hong2025ris}.
It is also worth noticing that multiple legitimate RISs have also been jointly optimized for mmWave secrecy and secure spectrum sharing~\cite{rafieifar2021secrecy, wang2023hybrid}.

\ziyuu{
Recent studies extend such optimization to near-field propagation, which allows LUs and eavesdroppers located in similar directions but at different distances to be distinguished. 
This capability has been exploited in RIS-assisted NOMA through range-selective focusing and far-to-near successive interference cancellation (SIC), as well as in double-RIS orbital angular momentum (OAM) systems through phase optimization, mode selection, and artificial noise~\cite{lei2025ris, liang2025double}.
However, these gains rely on accurate range-dependent channel or location information and remain largely validated through numerical evaluation.
}

\paragraph{Obfuscation against unauthorized sensing or privacy inference}
Not all threats come from decoding messages. An adversary may instead try to infer a target's presence, motion, or location from RIS-shaped reflections, and for these cases, minimizing the eavesdropper's data rate is beside the point. What matters is degrading the adversary's ability to extract useful physical-layer observations while keeping legitimate functionality intact. For example, joint BS--RIS optimization can hide a sensing target from an external observer~\cite{magbool2025hiding}, and randomized but controlled phase perturbations can reduce adversarial sensing accuracy without noticeably hurting legitimate communication or sensing~\cite{chen2025secure}. \ziyuu{Near-field focusing can also support covert communication by shaping the reflected field around an intended receiver while limiting the signal observed by a warden, even when they are located in similar directions~\cite{liu2024ris}.} Compared with secrecy-rate maximization, these methods protect privacy-sensitive quantities: motion signatures, target detectability, localization cues, rather than message content alone.

Together, these works show that defending against passive exploitation in RIS-assisted systems is more than classical anti-eavesdropping. Some methods focus on secrecy for data transmission; others suppress unauthorized observation of the physical environment. But in both cases, the underlying goal is the same: prevent the legitimate RIS from inadvertently amplifying what the adversary can see.
However, it is worth noting that most of these methods depend on at least partial channel knowledge, reasonably stable geometry, and fine-grained RIS control. Thus, their effectiveness may degrade under mobility, CSI uncertainty, or highly dynamic conditions.

\subsubsection{Countering Jamming and Link Disruption}
RIS-assisted systems are also vulnerable to attacks that directly degrade link quality rather than passively exploiting it. The defenses here prioritize keeping the communication link usable under adversarial interference, not just maximizing secrecy.

\paragraph{Joint anti-jamming and anti-eavesdropping}
In many practical settings, such as RIS-assisted UAV networks, wireless sensor deployments, and other dynamic systems, the legitimate transmitter faces both an active jammer and a passive eavesdropper at the same time. The RIS must therefore serve a dual role: boost the desired signal and suppress interference while also limiting information leakage. 
Zou et al.\ use a noisy dueling double deep Q-network (DQN) with prioritized experience replay to co-optimize RIS phases and transmit power, improving both system rate and protection level in a UAV communication scenario~\cite{zou2023ris}. 
Thanh et al.\ take a similar DQN-based approach for solar-powered RIS-aided wireless sensor networks, jointly adjusting the wireless device's transmission energy and RIS phase shifts to maintain data rate under active jamming~\cite{thanh2022anti}. 
\ziyuu{In a near-field setting, Cui et al.\ jointly optimize BS beamforming artificial noise, and RIS coefficients to defend a cooperative jammer and eavesdropper. By exploiting both angular and range selectivity, the design can maintain secure communication even when the eavesdropper is aligned with the LU and located closer to the RIS~\cite{cui2024xl}.}
Unlike pure anti-eavesdropping designs, the emphasis here falls on keeping the link alive: a high secrecy rate means little if the legitimate receiver cannot decode anything.

\paragraph{Defending against RIS-induced passive jamming}
Another threat arises when the attacker does not emit strong, easily detectable interference but manipulates the propagation environment. For instance, an attacker can create destructive combining or harmful channel fluctuations at the legitimate receiver through a DISCO-RIS with random phase shifts~\cite{huang2023disco}. Classical interference nulling struggles here because reliable attacker-side CSI is often unavailable. Huang et al.\ address this with a statistical anti-jamming precoder that leverages knowledge of how the channel varies over time rather than trying to estimate the adversarial channel on each coherence interval, reducing the impact of randomly perturbed reflections on the legitimate user~\cite{huang2023anti}. This shows that in RIS-assisted systems, anti-jamming is less about disrupting the adversarial channel and more about improving the robustness of the legitimate channel.

Overall, countering jamming and link disruption requires a robustness-first mindset. Where anti-eavesdropping defenses try to widen the gap between what the legitimate user and the adversary can observe, anti-disruption defenses try to keep the legitimate link stable despite active perturbation. Still, most existing methods assume relatively controlled conditions, such as stable geometry, decent CSI, manageable user counts, and their performance may degrade under high mobility, rapidly shifting interference, or large-scale deployments.

\subsubsection{Securing the RIS Control Plane and Trusted Infrastructure}
The defenses above assume the RIS itself is doing what it is told. But the control link, the RIS controller, and the hardware platform are all potential entry points: unauthorized access to configuration commands can reshape the propagation environment without any obvious RF emission, and compromised firmware or tampered hardware can turn the legitimate RIS into a malicious actor.

Existing work addresses different parts of this trust chain. On the control-plane side, lightweight authenticated signaling can prevent unauthorized phase-shift commands from reaching the RIS~\cite{kunz2025lightweight, chiti2023secure}, a direction that aligns with ongoing ETSI efforts to define secure RIS control channels and architecture~\cite{etsi2023ris2, etsi2023ris3}. On the hardware side, Mughal et al. propose an end-to-end framework covering trusted manufacturing, tamper-aware deployment, and runtime behavioral monitoring to guard against physical or firmware-level compromise~\cite{mughal2025malris}.
Although this line of work is still much less developed than secrecy or anti-jamming design, it addresses a distinct problem: once the RIS control plane or hardware root of trust is compromised, physical-layer optimization alone is no longer sufficient.
}

\begin{figure}[!ht]
    \centering
    \includegraphics[width=0.9\linewidth]{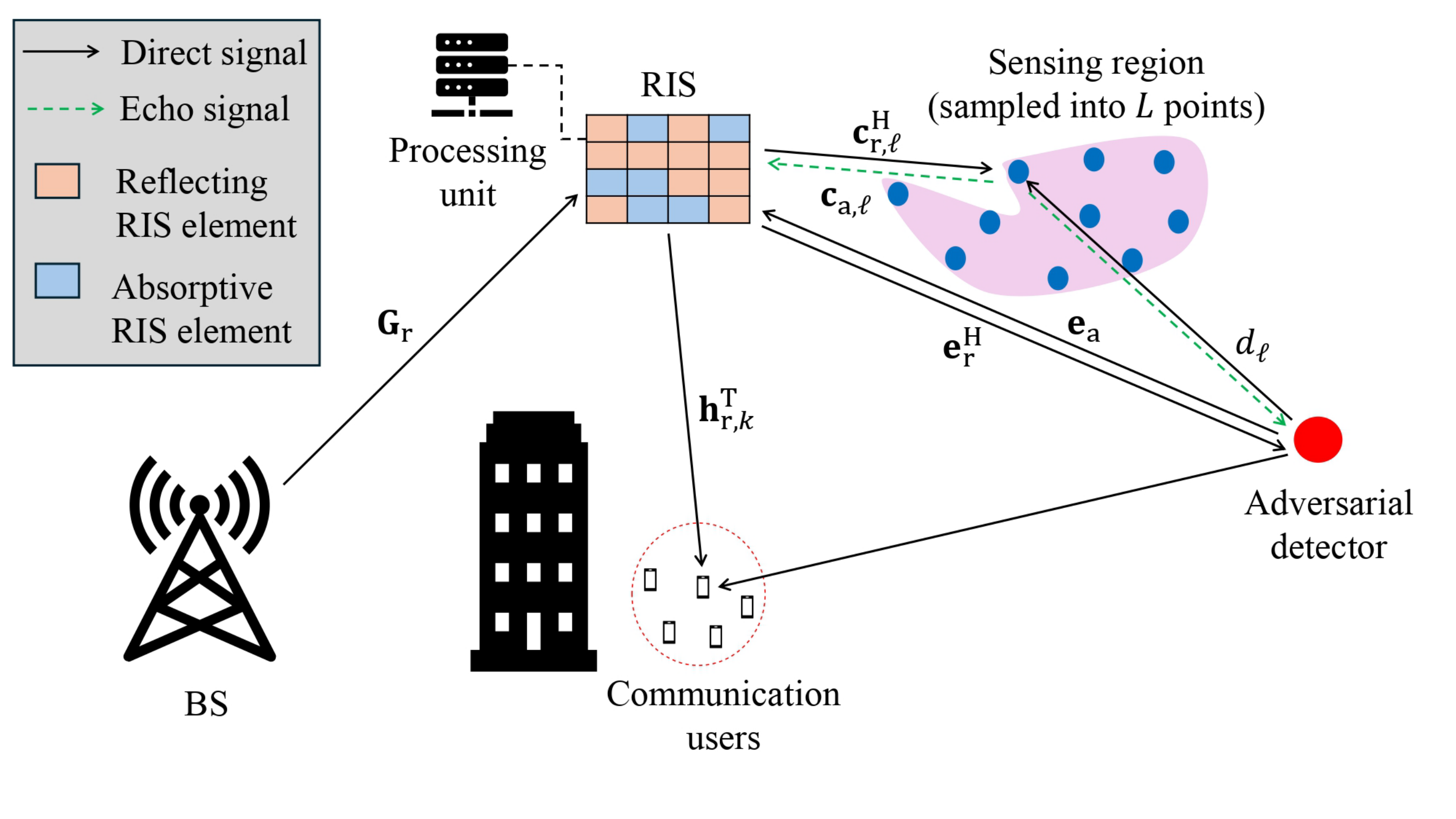}
    \caption{RIS-assisted ISAC system that (i) transmits data to multiple communication users; (ii) monitors the existence of one single target; (iii) prevents an adversarial detector from monitoring the existence of a target within the same sensing region.\protect\footnotemark}
    \label{fig:hiding}
    \vspace{-15pt}
\end{figure}
\footnotetext{Reproduced from A. Magbool \emph{et al.}, ``Hiding in Plain Sight: RIS-Aided Target Obfuscation in ISAC,'' \textit{IEEE Trans. Wireless Commun.}, vol.~25, pp.~14550--14563, 2026, doi:~10.1109/TWC.2026.3675857. \copyright~2026 The Authors, licensed under CC BY 4.0.}

\noindent\textbf{Pioneer work: \textit{Hiding in Plain Sight: RIS-Aided Target Obfuscation in ISAC}~\cite{magbool2025hiding}}, as shown in Fig.~\ref{fig:hiding}.

\textbf{Defense Scenario:}
This work considers a \textbf{RIS-assisted} downlink ISAC system in which the legitimate network simultaneously serves multiple communication users and senses a target region, while an \textbf{eavesdropper without RIS} attempts to infer the presence of the target from the same RIS-assisted propagation environment. 
Since the direct paths between the BS and the users, the sensing region, and the adversarial detector are blocked, the RIS becomes the key enabler for both legitimate sensing and potential privacy leakage.

\textbf{Defense Goal:}
The goal is not only to maintain legitimate communication and sensing performance, but also to protect \emph{target privacy}, i.e., to ensure that the target remains observable to the legitimate system while becoming difficult to detect by the adversarial observer.

\textbf{Defense Method:}
To achieve this, the defender jointly optimizes the BS transmit beamformer, the RIS phase-shift matrix, the RIS receive beamformer, and the assignment of RIS elements between reflecting and absorptive modes. 
The design minimizes the maximum sensing SINR achievable by the adversarial detector over the protected region, while enforcing minimum sensing SINR and communication SINR requirements for the legitimate system.

\textbf{Defense Result:}
Compared with the case without sensing-area protection, the proposed design reduces the maximum sensing SINR at the adversarial detector by about 25~dB. 
In addition, adaptive RIS element assignment provides about 3~dB extra protection compared with a fixed RIS element configuration~\cite{magbool2025hiding}.

\ziyu{
\textbf{Our Critical Assessment in Practical Systems:}
This work is representative because it extends RIS-aided defense beyond classical communication secrecy and explicitly addresses privacy leakage through unauthorized sensing. 
At the same time, its effectiveness depends on a structured geometry, accurate channel knowledge, and fine-grained joint control of beamforming and RIS element assignment. 
These assumptions are reasonable for a controlled ISAC prototype or deployment-oriented study, but may become harder to satisfy under mobility, fast environmental dynamics, or imperfect sensing-region knowledge. 
Therefore, this line of work should be interpreted primarily as evidence that RIS-assisted systems can protect \emph{observability} in addition to message confidentiality, rather than as a universally deployable privacy solution across arbitrary dynamic environments.
}

\begin{figure}[!ht]
\centering

\begin{tcolorbox}[
  colback=gray!12,
  colframe=black!70,
  boxrule=0.8pt,
  arc=2.5mm,
  left=6pt,
  right=6pt,
  top=4pt,
  bottom=6pt,
  title=\textbf{Lessons learned},
  coltitle=white,
  colbacktitle=black,
  fonttitle=\bfseries,
  fontupper=\normalfont
]

The main challenge of RIS-assisted privacy protection in ISAC is not merely to reduce signal exposure, but to \emph{control observability}: the wireless environment should be shaped so that legitimate users retain reliable sensing and communication capability, while unauthorized detectors observe an uninformative or misleading sensing environment.
\begin{itemize}
    \item \textbf{Defense should explicitly control adversarial observability rather than optimize only legitimate performance:}
    In RIS-assisted ISAC systems, optimizing only communication rate or legitimate sensing SINR is insufficient. 
    A more security-aware design should also consider what an unauthorized detector can infer from the same RIS-shaped environment, and treat adversarial observability as a design variable rather than a side effect.

    \item \textbf{Privacy protection must rely on selective and task-aware environment shaping:}
    A practical RIS-based defense should not simply weaken all reflections in the protected region. 
    Instead, it should preserve sufficient observability for legitimate sensing and communication while selectively suppressing or obfuscating the reflections that are useful to external detectors.
\end{itemize}
\end{tcolorbox}
\vspace{-15pt}
\end{figure}

\subsection{Countermeasures Against Malicious RISs} \label{subsec2}
\begin{figure}[!ht]
    \centering
    \subfloat[Detecting unauthorized RIS.]{\includegraphics[width=0.45\linewidth]{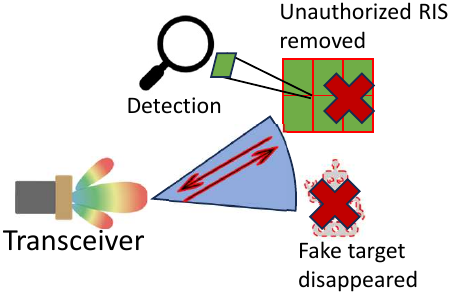}%
    \label{fig:def_metawave_detect}}
    \hfil
    \subfloat[Disrupting an unauthorized RIS with artificial noise.]{\includegraphics[width=0.5\linewidth]{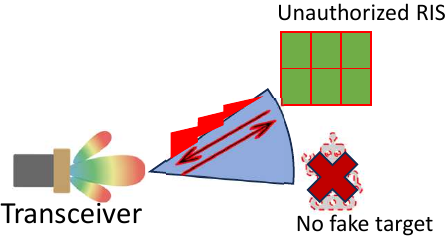}%
    \label{fig:def_metawave_jam}}
    \hfil
    \caption{Example of countermeasures of RIS for attacks.}
    \label{fig:def_metawave}
    \vspace{-10pt}
\end{figure}

Defending against an unauthorized malicious RIS remains a relatively underexplored problem compared with securing legitimate RIS-assisted systems. As discussed in Section~\ref{threats2}, the attacker deploys an external programmable reflector and uses it to manipulate wireless propagation for eavesdropping, jamming, or spoofing. 
Because such a surface could be passive, covert, and physically deployed in the environment before the attack is triggered, early detection becomes a significant part of the defense, as well as other physical layer defenses. 
Accordingly, in the following we first discuss methods for detecting and identifying malicious RISs, and then review system-level countermeasures against malicious-RIS-enabled attacks. 
An example is shown in Fig.~\ref{fig:def_metawave}.

\subsubsection{Early detection}

\begin{table*}[!ht]
    \centering
    \begin{tabular}
    {|L{1cm}|L{1.5cm}|L{2.7cm}|L{2cm}|L{2.5cm}|L{2.2cm}|L{3cm}|}
    \hline
        Reference & Target & Attacker's Goal &Scenario & Detection Method & Mechanism & Other Notes \\ \hline
        \cite{stamatelis2025detection} & 
        Non-cooperative RIS & 
        Disrupt MIMO-OFDM via stealthy reflection &
        MIMO-OFDM link & 
        Scan-B test via Deep SVDD & 
        Online change-point detection & 
        Doesn’t need layout/phase knowledge. Higher accuracy \\ \hline

        \cite{chen2025mris} &
        Malicious RIS spoofing &
        Impersonate a legitimate RIS \& inject counterfeit reflected paths &
        RIS-assisted communication system &
        Hybrid deep autoencoder with key-embedded codebook &
        Anomaly detection + key-matching validation &
        Requires secure initialization and key/codebook management. \\ \hline
        
        \cite{shui2025analysisdetectionrisbasedspoofing} & 
        Passive / stealth RIS & 
        Create deceptive path-based anomalies & 
        Multi-anchor/band system & 
        Multi-dimensional sensing (AoA/ToA/Doppler) & 
        Cross-validation of metrics & 
        Resilient to selective attacks \\ \hline
        
        \cite{zuniga2022see} & 
        Active RIS & 
        Hide the presence of active RIS while modifying the channel & 
        Interior surfaces & 
        IR thermal imaging & 
        Heat dissipation detection & 
        Ineffective for passive RIS \\ \hline
        
        \cite{henley2023detection} & 
        Reflectors / RIS surfaces & 
        Camouflage an RIS in the channel & 
        Environment surfaces & 
        Polarization imaging & 
        Polarization highlight detection & 
        Weaker if heavily camouflaged \\ \hline
    \end{tabular}
    \caption{Summary of early detection methods against unauthorized RIS.}
    \label{tab:defense_detection}
\end{table*}

Early detection is a simple yet crucial way to defend against malicious or unauthorized RIS-enabled attacks, especially because such surfaces may be passive, covert, and physically deployed in the environment before the attack is triggered~\cite{stamatelis2025detection, shui2025analysisdetectionrisbasedspoofing}. 
In this setting, the key challenge is not only to detect whether the wireless environment has changed, but also to determine whether the anomaly is caused by an attacker-controlled programmable reflector rather than by ordinary multipath, clutter, or benign environmental dynamics. A summary of these methods is shown in Table~\ref{tab:defense_detection}.

Manual inspection is the most straightforward option, but it is labor-intensive and may fail when the surface is small, well camouflaged, or integrated into surrounding objects~\cite{chen2023metawave}. Thus, several complementary detection methods can be used, such as RF detection~\cite{zhang2022active}, IR imaging~\cite{zuniga2022see}, optical detection~\cite{henley2023detection}, etc.
RF detection detects suspicious EM signals or radiation devices in the environment. Active RISs are easier to detect as they add and radiate amplifier noise, which shows up at the receiver as a direction-dependent noise term even when useful transmissions are quiet~\cite{zhang2022active}. In contrast, it is harder to detect passive RISs as they do not actively emit a signal. We have to detect their effect, such as the change of paths or CSI, to determine its presence.
IR imaging can expose a hot zone behind posters, ceiling tiles, or signage, which could signify an active RIS as they dissipate heat~\cite{zuniga2022see}, but this does not work for passive RISs.
RISs are easy to detect with polarization cameras because they exhibit strong specular highlights and distinctive polarization~\cite{henley2023detection}, while this is less effective when the RIS is heavily camouflaged. 

\ziyu{
Because of these limitations, recent work increasingly focuses on effect-based detection, i.e., detecting the anomalous channel or sensing behavior induced by a non-cooperative RIS rather than the device itself. 
For example, Stamatelis \emph{et al.} propose an online change-point detection framework based on scan-$B$ testing and Deep SVDD, which learns the normal distribution of wideband CSI and raises alarms when a non-cooperative RIS begins to reshape the channel~\cite{stamatelis2025detection}. Such methods do not require prior knowledge of the malicious RIS layout or phase configuration, and are therefore better suited to passive and silent attack surfaces. 
Similarly, Chen \emph{et al.} propose MRIS-SAD, a malicious RIS spoofing detection framework that combines dynamic key-embedded phase codebooks, hybrid deep autoencoding, and key-matching validation to distinguish legitimate and malicious RIS responses~\cite{chen2025mris}. This design explicitly targets passive malicious RISs that only manipulate reflected signals. 
In addition, physical-layer authentication can also serve as an auxiliary consistency check. 
However, unlike conventional source-authentication schemes, malicious RIS detection often requires verifying whether the reflected propagation pattern itself is legitimate, since a passive malicious RIS may not emit any distinct active signal source~\cite{huang2020intelligent}.

For spoofing attacks, detection also benefits from consistency-based sensing validation rather than single-metric thresholding. 
Recent work suggests that cross-checking angle, delay, Doppler shift, trajectory, or multi-anchor observations can expose inconsistencies introduced by malicious reflections, especially in sensing and ISAC systems~\cite{shui2025analysisdetectionrisbasedspoofing}. 
Overall, these results indicate that the most promising detection direction is to move from direct device inspection towards anomaly- and consistency-based identification of malicious propagation shaping. 
}

\subsubsection{System-level defense}
Beyond early detection, the defender can also reduce the impact of a malicious RIS at the system level. Here the attacker controls the surface, instead of the network operator, so the defender typically works with incomplete knowledge of the malicious RIS's location, configuration, and channel effect. In the following, we discuss representative system-level defenses against malicious-RIS-enabled attacks, and summarize them in Table~\ref{tab:defense_sys_design}.

\begin{table*}[!ht]
\centering
\scalebox{0.93}{
\begin{tabular}{
|L{0.8cm}
|L{1.5cm}
|L{1cm}
|L{1.8cm}
|L{1.8cm}
|L{1.7cm}
|L{1cm}
|L{1cm}
|L{1.7cm}
|L{2.6cm}|
}
\hline
\textbf{Ref.} &
\textbf{Threat Type} &
\textbf{Comm. Method} &
\textbf{Scenario} &
\textbf{Defense Method} &
\textbf{Malicious RIS Setup} &
\ziyuu{\textbf{CSI Requirements}} &
\ziyuu{\textbf{Field Regime}} &
\textbf{Defense Performance} &
\ziyuu{\textbf{Limitation}} \\ \hline

\cite{huang2020intelligent} & 
Pilot corruption & 
sub-6G, cellular & 
RIS-PCA attack & 
Detection \& optimization & 
Passive RIS & 
\ziyuu{Estimated} &
\ziyuu{Far-field} &
Strong, selective & 
\ziyuu{Requires specific pilot/training design \& communication overhead.} \\ \hline

\cite{wan2023countermeasure} &
Jamming &
sub-6G &
Key generation under reciprocity disruption with RIS &
Multipath separation \& path detection &
Malicious RIS path &
\ziyuu{Estimated} &
\ziyuu{Far-field} &
Strong, selective &
\ziyuu{Requires separable multipath structure and reliable path identification.} \\ \hline

\cite{alexandropoulos2023counteracting} & 
Eavesdropping & 
sub-6G, cellular & 
MIMO with hidden malicious RIS & 
Optimization & 
Hidden passive RIS & 
\ziyuu{Statistical} &
\ziyuu{Far-field} &
Strong, selective & 
\ziyuu{Requires a legitimate RIS; high computational complexity.} \\ \hline

\cite{de2024malicious} &
Jamming & 
sub-6G, cellular & 
Massive MIMO under RIS jamming & 
Optimization & 
Passive RIS jammer & 
\ziyuu{Estimated} &
\ziyuu{Far-field} &
Strong, selective & 
\ziyuu{Performance bounded by cascaded malicious-RIS channel estimation accuracy.} \\ \hline

\cite{gao2024benign} & 
\multirow{2}{1.5cm}{\raggedright All malicious RIS} & 
sub-6G, cellular & 
Friendly RIS deployment & 
Optimization & 
Rogue passive RIS & 
\ziyuu{Perfect} &
\ziyuu{Far-field} &
Strong, selective & 
\ziyuu{Requires favorable placement \& coordination of the legitimate RIS.} \\ 
\cline{1-1} \cline{3-10}

\cite{li2025cooperative} & 
& 
sub-6G, cellular & 
Multi-antenna system & 
Optimization & 
Interference-inducing RIS & 
\ziyuu{Estimated} &
\ziyuu{Far-field} &
Strong, selective & 
\ziyuu{Requires malicious RIS's directional knowledge and antenna-array resolution.} \\ \hline

\end{tabular}
}
\caption{Representative system-level defenses against unauthorized or malicious RIS.}
\label{tab:defense_sys_design}
\vspace{-10pt}
\end{table*}

\ziyu{
\paragraph{Anti-jamming and eavesdropping enabled with malicious RIS}
Many system-level defenses against malicious-RIS-enabled eavesdropping and jamming remain conceptually similar to those used in legitimate RIS-assisted systems, including robust beamforming, artificial-noise design, secure channel estimation, and anti-jamming precoding. 
However, the threat model is different: the attacker, not the operator, controls the RIS, and the defender may know little about where it is or how it is configured.

For malicious-RIS-enabled eavesdropping, the main objective is to reduce the attacker’s ability to exploit the manipulated propagation environment for channel learning, pilot contamination, or secrecy degradation. 
In this respect, secure training and cooperative channel-estimation schemes can help identify or suppress anomalous propagation effects introduced by the unauthorized RIS, thereby limiting the attacker’s ability to benefit from pilot corruption or favorable beam alignment~\cite{huang2020intelligent}. 
Artificial noise is also useful because it can degrade adversarial reflected paths without requiring exact attacker-side CSI~\cite{wang2022wireless}. 
More generally, secrecy-oriented joint optimization designs seek to preserve secrecy even when the malicious RIS is hidden or only partially characterized~\cite{alexandropoulos2023counteracting}.

For malicious-RIS-enabled jamming or destructive propagation shaping, the main objective is to maintain communication reliability when the attacker uses the RIS to create harmful interference or selectively degrade specific users. 
In such cases, statistical or robust precoding is often more realistic than highly tailored cancellation, since the defender may not know the exact rogue-RIS channel. 
Recent results on RIS-based jamming in massive MIMO systems further show that effective mitigation is possible even when only an imperfect estimate of the cascaded RIS channel is available~\cite{de2024malicious}. 
When partial directional knowledge of the malicious RIS is available, spatial nulling and related multi-antenna techniques can further attenuate harmful reflected components~\cite{li2025cooperative}.
If the defender also controls a friendly RIS, it can generate counteracting reflections or AN-enhanced beam patterns that overshadow the malicious RIS's effect~\cite{gao2024benign}.

Overall, these methods can still provide meaningful protection, but their selectivity and effectiveness are generally more limited than in trusted-RIS settings, especially when the malicious RIS is covert, mobile, or tightly coupled to the surrounding environment.

\paragraph{Countering spoofing enabled with malicious RIS}
Spoofing by a malicious RIS poses a distinct challenge: rather than degrading the link or stealing data, the attacker fabricates false observations that mislead sensing, positioning, or ISAC inference.
Unlike eavesdropping or jamming, spoofing does not necessarily degrade conventional signal-quality metrics such as SNR, SINR, or achievable rate. This means the received signal may look perfectly normal while the inferred target, position, or trajectory is wrong, making classical optimization objectives (maximize secrecy rate, suppress interference) largely inapplicable, and explains why existing defenses remain closely tied to detection. 
For example, Shui et al.\ show that cross-checking angle, delay, Doppler, and trajectory across multiple anchors can expose inconsistencies that a single malicious reflector cannot hide~\cite{shui2025analysisdetectionrisbasedspoofing}, and Chen et al.\ use dynamic key-embedded phase codebooks to distinguish legitimate from malicious RIS responses~\cite{chen2025mris}. Both are fundamentally detection mechanisms rather than spoofing-tolerant system designs. This is also suggested in attack-side analysis~\cite{chen2023metawave, jiang2024risiren, zhou2023ristealth}. 
Robust inference that can maintain correct sensing or localization output even under active spoofing, rather than simply flagging and discarding suspect observations, currently remain less explored.

Overall, system-level countermeasures against malicious RISs currently function more as damage control than absolute solutions. Unlike securing a trusted network, their success relies heavily on situational advantages—such as partial knowledge of the malicious surface, favorable geometry, or high spatial diversity. This fundamental limitation explains the current bottleneck in anti-spoofing research: existing methods are largely confined to anomaly detection and cross-validation. Transitioning from merely flagging fabricated observations to developing inherently spoofing-tolerant inference systems remains a critical, yet largely uncharted, frontier.
}

\noindent\textbf{Pioneer work: \textit{On the Detection of Non-Cooperative RISs: Scan B-Testing via Deep Support Vector Data Description}~\cite{stamatelis2025detection}}, as shown in Fig.~\ref{fig:def_metawave_detect}.

\textbf{Defense Scenario:}
This work considers a scenario where a defender is defending against \textbf{attackers with RIS} in a \textbf{non-RIS system} with \textbf{early detection}.
In the system, a user equipment with multiple antennas communicates in the uplink direction with a BS with multiple antennas. OFDM transmission is adopted.
There are one or more unauthorized RISs deployed in the system.
The defender does not know the characteristics of the unauthorized RISs.

\textbf{Defense Goal:}
The defender aims to detect these unauthorized RISs without assuming any model for the RIS’s phase distribution or timing.

\textbf{Defense Method:}
The defender models the problem as an online change-point detection problem.
They formulate an unsupervised distribution-free change point detection method.
In this method, they use the d-SVDD model to extract representative and low dimensional features from the BS observations, and then, feed those features to the scan B-statistic.

\textbf{Defense Result:}
This method achieves a higher detection accuracy and a lower computational complexity than existing change point detection schemes.

\ziyu{
\textbf{Our Critical Assessment in Practical Systems:}
This work is representative because it captures an important reality of malicious-RIS defense: for passive and covert attack surfaces, directly detecting the device may be less practical than detecting its impact on the propagation environment. 
At the same time, this type of defense still depends on a sufficiently stable notion of ``normal'' channel behavior, so its effectiveness may degrade in highly dynamic environments where mobility, blockage changes, or environmental variation can themselves induce frequent distribution shifts. 
Therefore, such approaches are more convincing as early warning mechanisms for anomalous propagation shaping, rather than as complete identification or localization solutions for all unauthorized RIS deployments.
}

\begin{figure}[!ht]
\centering

\begin{tcolorbox}[
  colback=gray!12,
  colframe=black!70,
  boxrule=0.8pt,
  arc=2.5mm,
  left=6pt,
  right=6pt,
  top=4pt,
  bottom=6pt,
  title=\textbf{Lessons learned},
  coltitle=white,
  colbacktitle=black,
  fonttitle=\bfseries,
  fontupper=\normalfont
]

This type of defenses shows that attacks enabled by malicious RIS are difficult to prevent only at the moment of attack. In this scenario, the malicious RIS may already be passively deployed in the environment and remain silent until it begins to reshape the propagation channel. 
As a result, the key system-design lesson is that defending against malicious RISs should not rely solely on protecting data transmission, but should also include mechanisms for monitoring whether the wireless environment itself behaves abnormally.

\begin{itemize}
    \item \textbf{Defense should move from device inspection to propagation-aware anomaly detection:}
    For passive and covert RISs, directly detecting the hardware may be difficult in practice. 
    A more realistic direction is to monitor whether the observed channel statistics, path structure, or CSI distribution deviate from the normal operating regime of the legitimate system.

    \item \textbf{Early warning is valuable even without full attacker-side knowledge:}
    This line of work suggests that useful detection can be achieved without explicitly knowing the malicious RIS layout, phase configuration, or timing model. 
    In system design, this means the defender should prioritize lightweight monitoring mechanisms that can flag suspicious propagation changes early, even if they cannot immediately localize or identify the malicious RIS in full detail.
\end{itemize}

\end{tcolorbox}
\vspace{-15pt}
\end{figure}

\ziyu{
\subsection{RIS as a Defensive Tool} \label{subsec3}
Unlike Sections~\ref{subsec1} and~\ref{subsec2}, which focus on securing RIS-assisted systems and countering malicious RISs, this part discusses a broader defensive scheme: deploying a legitimate RIS as a defensive tool for wireless systems originally without RIS. 
}
As stated previously, the RIS is also a powerful, low cost, energy-saving tool for LUs to defend against attacks and protect their security and privacy. The introduction of RIS leads to more effective and economic countermeasures, especially against stealthy and undetectable attacks. LUs gain authorized and authenticated control of the RIS in order to defend against eavesdroppers. The defenses are often white-box, which means the user jointly optimizes its beamforming, power allocation, and node trajectories to maximize security performance. The RIS are often fixed in the system, such as mounted on the wall or ceilings. \ziyu{Note that in systems already with an RIS, LUs generally do not introduce an extra RIS for defense, considering its cost and increased hardware complexity.} In the following, we discuss different defenses in which an RIS is used as a defensive tool by the types of attacks they defend against, as summarized in Table~\ref{tab:ris_defense}. 

\begin{table*}[]
    \centering
    \scalebox{0.8}{
    \begin{tabular}{|L{0.8cm}|L{1.6cm}|L{0.9cm}|L{2cm}|L{1.2cm}|L{1.6cm}|L{2.5cm}|L{1cm}|L{1cm}|L{2.2cm}|L{3.1cm}|}
    \hline
     \textbf{Ref.} &
     \textbf{Threat Type} &
     \textbf{Comm. Method} &
     \textbf{Scenario} &
     \textbf{Corres-pondence to Fig.~\ref{fig:application_out}} &
     \textbf{Defense Method} &
     \textbf{Additional RIS Setup} &
     \ziyuu{\textbf{CSI Requirements}} &
     \ziyuu{\textbf{Field Regime}} &
     \textbf{Defense Performance} &
     \ziyuu{\textbf{Limitation}} \\ \hline

    % --- Eavesdropping ---
    \cite{kompostiotis2025optimizing} &
    \multirow{10}{1.6cm}{Eavesdropping} &
    sub-6G, Cellular &
    Indoor 6G communication system &
    \textcircled{3} &
    Optimization &
    Passive, assisting indoor BS$\rightarrow$LU link &
    \ziyuu{Measured} &
    \ziyuu{Far-field} &
    Strong, selective &
    \ziyuu{Indoor measurement-specific codebooks may generalize poorly across layouts.} \\ \cline{1-1} \cline{3-11}
     
    \cite{alexandropoulos2023counteracting} &
    &
    sub-6G, Cellular &
    5G/B5G MIMO channel with a malicious RIS &
    \textcircled{4} &
    Optimization &
    Assisting LU &
    \ziyuu{Statistical} &
    \ziyuu{Far-field} &
    Strong, selective &
    \ziyuu{Dual-RIS threat model.} \\ \cline{1-1} \cline{3-11}
     
    \cite{jing2024reconfigurable} &
    &
    Optical &
    V2V communications using VLC systems &
    \textcircled{2} &
    Optimization &
    Deployed at road intersection for V2V VLC &
    \ziyuu{Perfect} &
    \ziyuu{---} &
    Moderate, selective, specialized &
    \ziyuu{Effective only with favorable roadside/intersection RIS placement.} \\ \cline{1-1} \cline{3-11}

    \cite{staat2022irshield} &
    &
    sub-6G, Wi-Fi &
    In door Wi-Fi system with a number of devices &
    \textcircled{3} &
    Introducing randomness &
    Co-located with indoor Wi-Fi devices &
    \ziyuu{CSI-free} &
    \ziyuu{Far-field} &
    Near-complete &
    \ziyuu{Needs trusted indoor infrastructure and frequent RIS reconfiguration.} \\ \cline{1-1} \cline{3-11}
     
    \cite{li2022protego} &
    &
    sub-6G, Wi-Fi &
    A low-power IoT device and a receiver &
    \textcircled{3} &
    Introducing randomness &
    One programmable metasurface near Tx for IoT PLS &
    \ziyuu{CSI-free} &
    \ziyuu{Far-field} &
    Strong, selective &
    \ziyuu{Requires attacker-direction knowledge and metasurface field-of-view coverage.} \\ \cline{1-1} \cline{3-11}
     
    \cite{xu2025chaotic} &
    &
    sub-6G &
    Broadcast wireless communication systems &
    \textcircled{3}\textcircled{4} &
    Introducing randomness &
    One information RIS for chaotic modulation (SISO) &
    \ziyuu{Measured} &
    \ziyuu{Near-field} &
    Strong, selective &
    % Needs synchronized trusted RIS modulation and compatible Rx.
    \ziyuu{Leakage possible between Tx and metasurface.} \\ \cline{1-1} \cline{3-11}
     
    \cite{shaikhanov2024audio, shaikhanov2025spoofing} &
    &
    mmWave &
    mmWave radar eavesdropping on earpiece vibrations &
    \textcircled{3}\textcircled{4} &
    Spoofing the attacker &
    Sub-THz metasurface mounted on phone &
    \ziyuu{---} &
    \ziyuu{---} &
    Near-complete, selective, specialized &
    \ziyuu{Specific to phone/earpiece scenario and specialized hardware.} \\ \cline{1-1} \cline{3-11}
     
    \cite{wang2024intelligent} &
    &
    / & 
    Electronic countermeasure against enemy radars &
    / &
    Spoofing the attacker &
    Mounted on sensing target (SISO) &
    \ziyuu{Estimated} &
    \ziyuu{Far-field} &
    Strong, specialized &
    \ziyuu{Target-mounted RIS assumption.} \\ \cline{1-1} \cline{3-11}
     
    \cite{sun2025anti} &
    &
    10.3 GHz &
    Counter multi-static radar with an RIS &
    \textcircled{4} &
    Spoofing the attacker &
    Attached to sensing target &
    \ziyuu{---} &
    \ziyuu{Far-field} &
    Strong, specialized &
    \ziyuu{Only focused on radar deception.} \\ \cline{1-1} \cline{3-11}
    
    \cite{he2026invisible} &
    &
    sub-6G, Wi-Fi &
    Privacy-preserving ISAC system &
    \textcircled{3} &
    Introducing randomness &
    RIS for masking unauthorized sensing &
    \ziyuu{Estimated} &
    \ziyuu{Far-field} &
    Strong, selective &
    \ziyuu{Requires trusted RIS coordination, authorized-Rx demasking, \& synchronization with RIS configurations.} \\ \hline

    % --- Jamming ---
    \cite{cao2024self} &
    \multirow{2}{1.6cm}{Jamming} &
    sub-6G, Cellular &
    5G wireless network &
    / &
    Optimization &
    One active RIS with self-sustainable power &
    \ziyuu{Estimated} &
    \ziyuu{Far-field} &
    Strong &
    \ziyuu{Active-RIS power supply and hardware complexity.} \\ \cline{1-1} \cline{3-11}
     
    \cite{wang2023communication} &
    &
    sub-6G, Wi-Fi &
    sub-6G wireless network &
    / &
    Optimization &
    One passive RIS with angular response &
    \ziyuu{Perfect} &
    \ziyuu{Far-field} &
    Strong &
    \ziyuu{Require perfect CSI.} \\ \hline

    % --- Jamming & eavesdropping ---
    \cite{zou2023ris} &
    Jamming \& eavesdropping &
    / &
    UAV systems &
    \textcircled{1} &
    Optimization &
    On UAV platform, assisting BS$\rightarrow$UE downlink &
    \ziyuu{Estimated} &
    \ziyuu{Far-field} &
    Strong, selective &
    \ziyuu{Requires accurate estimation and robust DRL generalization.} \\ \hline

    % --- Spoofing ---
    \cite{10100683, lloyd20233d} &
    \multirow{6}{1.6cm}{Spoofing} & 
    Acoustic &
    Voice assistants under ultrasound attack &
    \textcircled{3}\textcircled{4} &
    Attenuation &
    Passive acoustic metamaterial filter on device &
    \ziyuu{---} &
    \ziyuu{---} &
    Strong, specialized & 
    \ziyuu{Voice assistant specific passive filter} \\ \cline{1-1} \cline{3-11}
     
    \cite{ning2025metaguardian} & 
    & 
    Acoustic & 
    Voice assistants under attack & 
    \textcircled{3}\textcircled{4} &
    Attenuation &
    Acoustic metamaterial guard on smart device &
    \ziyuu{---} &
    \ziyuu{---} &
    Strong, selective, specialized &
    \ziyuu{Controlled evaluation results.} \\ \cline{1-1} \cline{3-11}

    \cite{gao2024ris} &
    &
    / &
    RIS-assisted specific-emitter identification &
    \textcircled{3} &
    PLA &
    Trusted RIS; ON for LU and OFF otherwise &
    \ziyuu{CSI-free} &
    \ziyuu{Far-field} &
    Strong, selective &
    \ziyuu{Requires trusted control \& static prototype; environment-dependent.} \\ \cline{1-1} \cline{3-11}

    \cite{meng2025efficient} &
    &
    sub-6G, Cellular &
    6G IoT uplink under spoofing &
    \textcircled{3} &
    PLA &
    One BS-controlled RIS; no direct Tx--BS link &
    \ziyuu{Estimated} &
    \ziyuu{Far-field} &
    Strong, selective &
    \ziyuu{Assumes no direct link and does not optimize phases.} \\ \cline{1-1} \cline{3-11}

    \cite{tomasin2024analysis} &
    &
    sub-6G, Cellular &
    UE challenge--response authentication &
    \textcircled{4} &
    PLA &
    LU-controlled RIS with secret random configurations &
    \ziyuu{Estimated} &
    \ziyuu{Far-field} &
    Strong, selective &
    \ziyuu{Requires trusted and secret RIS control; randomization reduces data rate.} \\ \cline{1-1} \cline{3-11}

    \cite{kaveh2025voltage} &
    &
    RFID &
    Tag-to-tag authentication in a smart home &
    \textcircled{3} &
    PLA &
    Trusted indoor RIS assisting the tag-to-tag link &
    \ziyuu{Perfect} &
    \ziyuu{Far-field} &
    Strong, selective, specialized &
    \ziyuu{Assumes perfect CSI, fixed geometry, \& a trusted RIS controller.} \\ \hline

    % --- Spoofing & eavesdropping ---
    \cite{illi2025robust} &
    Spoofing \& eavesdropping &
    / &
    Wireless mesh under impersonation and eavesdropping &
    / &
    PLA \& optimization &
    Trusted RIS assisting the authenticator &
    \ziyuu{Perfect} &
    \ziyuu{Far-field} &
    Strong, selective &
    \ziyuu{Requires perfect CSI; authentication--confidentiality tradeoff.} \\ \hline

    % --- Spoofing & jamming ---
    \cite{nasser2024online} &
    Spoofing \& jamming &
    mmWave &
    SISO mmWave system with one Eve &
    \textcircled{3} &
    Optimization &
    One partitioned RIS for beam selection &
    \ziyuu{CSI-free} &
    \ziyuu{Far-field} &
    Strong, selective &
    \ziyuu{Limited by codebook resolution and partitioned-RIS beam selection.} \\ \hline
     
    \end{tabular}
    }
    \caption{Summary of defenses where RIS acts as defensive tool.}
    \label{tab:ris_defense}
    \vspace{-15pt}
\end{table*}

\ziyu{
\subsubsection{Eavesdropping}
To defend against eavesdropping, the defender can add a friendly RIS to enhance the secrecy of communication links, or obfuscate and spoof the eavesdropper.
\paragraph{Secrecy Enhancement for Communication Links}
An additional legitimate RIS can be used to defend against eavesdropping attacks by enhancing the asymmetry between the legitimate and malicious channels. In this type of defenses, RIS is typically jointly optimized with Tx beamforming, artificial noise, and/or other environmental parameters, so that the legitimate link is strengthened while the eavesdropper-relevant channel is suppressed. 
For example, Zou et al. show that a legitimate RIS deployed on a UAV can jointly improve anti-jamming robustness and anti-eavesdropping capability through DRL-based optimization~\cite{zou2023ris}.  
Kompostiotis et al. further validate this defense direction with real indoor measurements and propose a practical RIS phase-configuration codebook design methodology for multipath environments~\cite{kompostiotis2025optimizing}. 
In a more adversarial setting with both legitimate and malicious RISs, Alexandropoulos et al. jointly optimize the legitimate precoder, artificial noise covariance, receiving combiner, and legitimate RIS reflection coefficients to maintain confidential communication even against a stronger malicious RIS~\cite{alexandropoulos2023counteracting}. 
In optical communications, Jing et al. use an RIS at a road intersection to improve legitimate VLC reception while simultaneously injecting artificial noise to degrade roadside eavesdroppers~\cite{jing2024reconfigurable}. 

Overall, this defense family is attractive because it can preserve legitimate communication quality while suppressing information leakage. However, it often depends on accurate channel or geometry knowledge, favorable RIS placement, and sometimes more complex joint optimization. 

\paragraph{Obfuscation and Deception Against Sensing-Based Eavesdropping}
Another defense against eavesdropping attacks with a RIS, especially those relying on  adversarial sensing or physical side-channel inference, is to manipulate the attacker’s observations by injecting artificial randomness, phase noise, or even misinformation. 
IRShield~\cite{staat2022irshield}, as shown in Fig.~\ref{fig:def_irshield}, is a representative example: it applies randomized RIS configurations to create wireless channel obfuscation, significantly reducing adversarial human-motion detection without modifying deployed Wi-Fi devices. 
Protego~\cite{li2022protego} similarly exploits programmable metasurfaces to introduce artificial phase noise, thereby raising the symbol error rate in unintended directions while maintaining signal integrity towards the legitimate receiver. 
Wei et al.~\cite{wei2023physical} show that an information-carrying RIS excited by random radio signals can also support secure communication, where the legitimate receiver coherently retrieves the message while the eavesdropper mainly observes noise. 
Xu et al.~\cite{xu2025chaotic} further design a chaotic information metasurface whose local reflection properties are dynamically modulated by chaotic patterns, so that legitimate receivers can directly access the intended data while eavesdroppers receive chaotically encrypted signals without any secret-key exchange. 

Beyond obfuscation, some works move towards active deception. Shaikhanov et al. propose MiSINFO~\cite{shaikhanov2024audio, shaikhanov2025spoofing}, which uses an on-phone dynamic metasurface to not only prevent mmWave radar eavesdroppers from reconstructing the true audio signal, but also inject false audio information so that the attacker is misled into believing the attack has succeeded. 

Compared with secrecy-enhancement approaches, this defense family is especially useful against sensing-based privacy leakage that conventional encryption cannot stop. However, such methods are often more scenario-dependent, and some of them may also affect legitimate sensing functions if not carefully designed.
\begin{figure}[!ht]
    \centering
    \subfloat[Before defense: the eavesdropper runs unauthorized sensing.]{\includegraphics[width=0.45\linewidth]{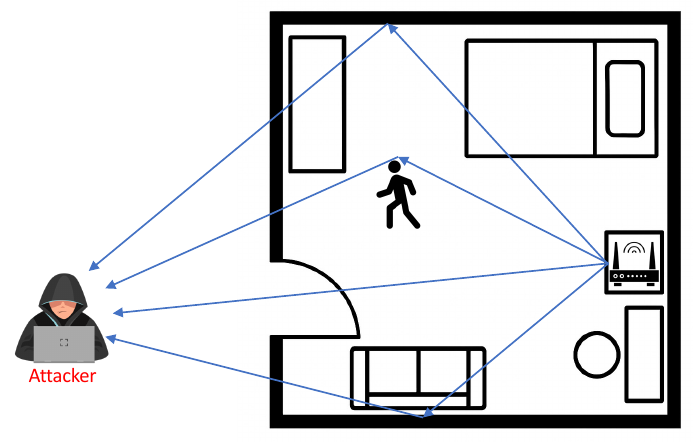}%
    \label{fig:irshield_before}}
    \hfil
    \subfloat[The dynamic RIS introduces randomness to the channel to disrupt unauthorized sensing.]{\includegraphics[width=0.45\linewidth]{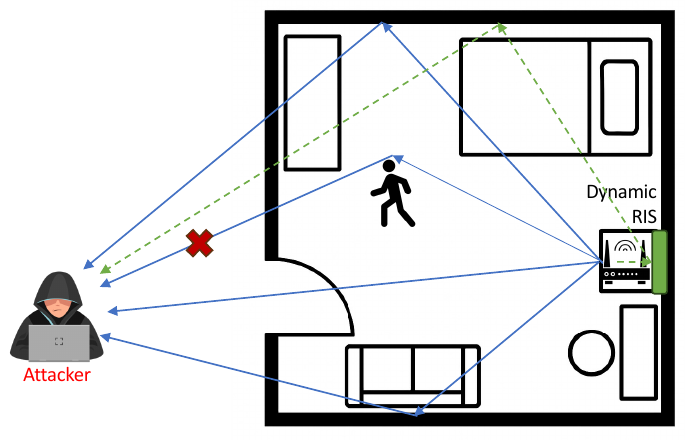}%
    \label{fig:irshield_after}}
    \hfil
    \caption{Example of RIS for defense against unauthorized sensing~\cite{staat2022irshield}.}
    \vspace{-10pt}
    \label{fig:def_irshield}
\end{figure}

\subsubsection{Jamming}
RIS is mainly used as an interference-shaping tool to defend against jamming attacks. By jointly optimizing transmit beamforming and RIS configuration, the defender can strengthen the useful link while weakening the jammer’s impact at the legitimate receiver. This idea appears in both passive and active RIS settings. 
For example, Nasser et al. propose an RL-based beam selection method for RIS-partitioning-aided mmWave physical-layer security without requiring user CSI~\cite{nasser2024online}.
Cao et al. exploit a self-sustainable \textit{active} RIS to improve anti-jamming capability in 5G networks~\cite{cao2024self}, while Wang et al. consider a \textit{passive} RIS with angular response to suppress jammer interference more effectively~\cite{wang2023communication}. 
Zou et al. show that this defense can also be extended to UAV-assisted settings where jamming and eavesdropping coexist~\cite{zou2023ris}. 
The main advantage of this type of defense is that it directly improves legitimate link's robustness and often fits naturally into RIS-assisted communication system design. However, it frequently relies on CSI or geometry estimation, making it sensitive to attacker's mobility and environmental mismatch. Besides, for active or self-powered RISs, there is additional hardware and power complexity.

\subsubsection{Spoofing}
RIS can be used to defend against spoofing attacks in different ways.
\paragraph{Attenuate malicious signals}
One of them is to use RIS-like structures or metamaterials as passive filters that attenuate malicious signals in undesired bands before they reach the victim device. This type of defense is particularly common in acoustic systems, where the goal is to suppress ultrasound, adversarial audio, or laser-induced command injection without significantly affecting normal audible signals.
For example, Joshua L. et al.~\cite{10100683, lloyd20233d} propose a compact metamaterial filter that defends voice assistants against ultrasound attacks while preserving normal audible operation. 
Ning et al. further propose MetaGuardian~\cite{ning2025metaguardian}, an acoustic metamaterial guard that protects voice assistants against inaudible, adversarial, and laser attacks without requiring software support or invasive hardware modification. Compared with deception-based methods, attenuation/filtering defenses are usually passive, compact, and easy to integrate with end devices. 
Their main limitation is that they are often highly specialized to a particular device type, signal band, or attack surface, so their application to broader spoofing scenarios may be limited.

\ziyuu{
\paragraph{Physical-layer authentication}
Another way of defense is to apply an RIS for physical-layer authentication (PLA). Depending on the design, the RIS can be controlled by the defender to enhance existing device fingerprints, create configurable channel signatures, or act as a physical challenge generator~\cite{zhang2025physical}.
To create channel fingerprints, Gao et al. install an RIS in an existing wireless system to create a distinct wireless link signature through ON/OFF control~\cite{gao2024ris}. Meng et al. introduce an RIS-assisted propagation path to improve the quality and separability of existing channel fingerprints, and use Gaussian-process classification to distinguish LUs from spoofers~\cite{meng2025efficient}.
RISs can also strengthen channel fingerprints, such as carrier-frequency offsets and circuit voltage profiles, by introducing another path, thus improving authentication accuracy~\cite{illi2025robust, kaveh2025voltage}.
Tomasin et al., however, do not introduce fingerprints, but use randomized RIS configurations as challenges and check the corresponding channel responses~\cite{tomasin2024analysis}.
Overall, these schemes extend spoofing defense from signal suppression to identity verification, but rely on trusted RIS control and sufficiently stable fingerprints.
}
}

\noindent \textbf{Pioneer work: \textit{IRShield: A Countermeasure Against Adversarial Physical-Layer Wireless Sensing}~\cite{staat2022irshield}}, as shown in Fig.~\ref{fig:def_irshield}.

\textbf{Defense Scenario:}
This work considers a scenario where a \textbf{defender with RISs} is defending against \textbf{eavesdroppers} in a \textbf{non-RIS system} with \textbf{an additional legitimate RIS as a defensive tool} by \textbf{disrupting the attacker}.
There are a number of legitimate Wi-Fi devices in the system. The devices are deployed within an ordinary indoor environment.
An attacker outside the building infers human motion by eavesdropping the signal transmitted by the Wi-Fi devices. 
The defender controls indoor infrastructure and can put the Wi-Fi devices at will. Also, they can deploy RISs within their space and apply customized configurations.

\textbf{Defense Goal:}
The goal is to obfuscate the wireless channel observed by the attacker, so that motion-induced channel variations become difficult to distinguish from RIS-induced fluctuations. 
The defense should operate continuously and unobtrusively, thereby degrading the attacker’s motion inference capability without requiring modifications to the legitimate devices.

\textbf{Defense Method:}
The defender introduces a time-varying RIS that continuously perturbs the propagation environment. 
Specifically, the RIS controller randomly changes the phase shift of a subset of elements over time, thereby injecting artificial channel variations into the eavesdropper’s observations. 
By tuning the temporal pattern of these RIS-induced perturbations, the defense masks or mimics motion-related channel changes and reduces the attacker’s ability to reliably detect human motion.

\textbf{Defense Result:}
Under the evaluated attacker strategies and experimental settings, the detection rate dropped from 90\% to 5\% or less, which means it renders motion detection largely infeasible, regardless of the adversary strategy.

\ziyu{
\textbf{Our Critical Assessment in Practical Systems:}
IRShield demonstrates that a legitimate RIS can be used as a physical-layer countermeasure to obfuscate passive wireless sensing without modifying other devices. However, its protection scope is still bounded by the assumed attacker model and deployment conditions. 
First, the evaluation mainly targets an external passive eavesdropper performing motion inference from channel observations. More capable adversaries could exploit multiple receivers or more advanced inference pipelines.
Second, the defense's effectiveness depends on deployment factors such as RIS size, anchor placement, and propagation geometry, so that the obfuscation gain is not entirely independent of the scenario. 
Third, IRShield has an undirected effect on the wireless channel and may therefore interfere with legitimate wireless sensing applications as well. 
Hence, while IRShield is a promising proof-of-concept for RIS-based sensing obfuscation, its practical deployment requires careful consideration of attacker sophistication, placement conditions, and coexistence with authorized sensing functions.
}
\vspace{-5pt}

\begin{figure}[!ht]
\centering

\begin{tcolorbox}[
  colback=gray!12,
  colframe=black!70,
  boxrule=0.8pt,
  arc=2.5mm,
  left=6pt,
  right=6pt,
  top=4pt,
  bottom=6pt,
  title=\textbf{Lessons learned},
  coltitle=white,
  colbacktitle=black,
  fonttitle=\bfseries,
  fontupper=\normalfont
]

Recent studies suggest that the main value of an additional legitimate RIS for anti-eavesdropping is not merely to inject random perturbations, but to \emph{create asymmetric observability}: the wireless environment can be reshaped so that unauthorized eavesdroppers observe an unreliable channel, while legitimate users retain a recoverable or improved link.

\begin{itemize}
    \item \textbf{Defense should evolve from undirected obfuscation to selective and task-aware RIS control:}
    A more practical direction of defenses is to exploit RIS beamforming to selectively suppress leakage in eavesdropper-relevant directions or channel dimensions, while preserving the legitimate link.

    \item \textbf{Anti-eavesdropping protection must not disrupt legitimate communication and sensing:}
    A practical RIS-based defense should not treat secrecy as the only objective. Instead, it should jointly consider privacy protection, communication quality, and, when relevant, authorized communication and sensing performance. This requires mechanisms such as receiver-aware RIS signaling, state coordination, or masking/demasking, so that legitimate users can compensate for RIS-induced perturbations while unauthorized observers cannot.

    \item \textbf{The role of RIS is shifting from passive reflector to security actuator:}
    Recent works show that a legitimate RIS can act as a physical-layer security actuator. Beyond improving secrecy rate, it can counter malicious propagation manipulation, create channel asymmetry, and support defense against more adaptive eavesdroppers.
\end{itemize}

\end{tcolorbox}
\vspace{-5pt}
\end{figure}
\section{Limitations of Existing Methods} \label{limitations}
\ziyu{
Although many RIS-related attacks and defenses show strong numerical or experimental performance, their practical effectiveness is still constrained by a number of recurring deployment bottlenecks, as shown in Table~\ref{tab:limitation}. These shared limitations reveal that both adversaries and defenders operate under the same fundamental physical and architectural constraints.

\begin{table*}[!ht]
\centering
\caption{Qualitative mapping of limitations to specific RIS-related security scenarios; $\checkmark$ means the limitation exists in the corresponding scenario, and $\times$ means it does not exist.}
\label{tab:limitation}
\begin{tabular}{| l  c  c  c  c  c  c |}
\hline
\textbf{Scenarios} & \textbf{CSI \& Est.} & \textbf{Real-Time} & \textbf{Hardware} & \textbf{Scale} & \textbf{Mobility} & \textbf{Stealth} \\
& \textbf{(\S\ref{sec:lim-csi})} & \textbf{(\S\ref{sec:lim-compute})} & \textbf{(\S\ref{sec:lim-hardware})} & \textbf{(\S\ref{sec:lim-multiuser})} & \textbf{(\S\ref{sec:lim-mobility})} & \textbf{(\S\ref{sec:lim-stealth})} \\ \hline
\multicolumn{7}{| @{\hspace{2pt}} l |}{\textit{Attacks on RIS systems (\S\ref{threats1})}} \\
Passive exploitation \& privacy leakage & $\checkmark$ & $\checkmark$ & $\checkmark$ & $\checkmark$ & $\checkmark$ & $\checkmark$ \\
Active infrastructure compromise        & $\times$     & $\checkmark$ & $\checkmark$ & $\checkmark$ & $\times$     & $\times$     \\ \hline
\multicolumn{7}{| @{\hspace{2pt}} l |}{\textit{Attacks leveraging the RIS (\S\ref{threats2})}} \\
Targeted eavesdropping                 & $\checkmark$ & $\checkmark$ & $\checkmark$ & $\checkmark$ & $\checkmark$ & $\checkmark$ \\
Passive signal nulling / jamming       & $\checkmark$ & $\checkmark$ & $\checkmark$ & $\checkmark$ & $\checkmark$ & $\checkmark$ \\
Physical-layer spoofing                & $\times$     & $\checkmark$ & $\checkmark$ & $\times$     & $\checkmark$ & $\checkmark$ \\ \hline
\multicolumn{7}{| @{\hspace{2pt}} l |}{\textit{Defending RIS-assisted systems (\S\ref{subsec1})}} \\
Defending passive eavesdropping        & $\checkmark$ & $\checkmark$ & $\checkmark$ & $\checkmark$ & $\checkmark$ & $\checkmark$ \\
Countering jamming \& disruption       & $\checkmark$ & $\checkmark$ & $\checkmark$ & $\checkmark$ & $\checkmark$ & $\times$     \\
Securing RIS control plane             & $\times$     & $\checkmark$ & $\checkmark$ & $\checkmark$ & $\times$     & $\checkmark$ \\ \hline
\multicolumn{7}{| @{\hspace{2pt}} l |}{\textit{Countering malicious RIS (\S\ref{subsec2})}} \\
Early detection                        & $\times$     & $\checkmark$ & $\checkmark$ & $\checkmark$ & $\checkmark$ & $\checkmark$ \\
System-level robust defense            & $\checkmark$ & $\checkmark$ & $\checkmark$ & $\checkmark$ & $\checkmark$ & $\times$     \\ \hline
\multicolumn{7}{| @{\hspace{2pt}} l |}{\textit{RIS as a defensive tool (\S\ref{subsec3})}} \\
Defense against eavesdropping          & $\checkmark$ & $\checkmark$ & $\checkmark$ & $\checkmark$ & $\checkmark$ & $\checkmark$ \\
Defense against jamming                & $\times$     & $\checkmark$ & $\checkmark$ & $\times$     & $\checkmark$ & $\times$     \\
Defense against spoofing               & $\times$     & $\checkmark$ & $\checkmark$ & $\times$     & $\checkmark$ & $\times$     \\ \hline
\end{tabular}
\vspace{-15pt}
\end{table*}

\subsection{CSI and Estimation Requirements}
\label{sec:lim-csi}
Most RIS attacks and defenses reduce to a phase-configuration problem that implicitly assumes accurate CSI for the cascaded Tx--RIS--Rx channel, and often for a third link to the victim or adversary. This stringent reliance on accurate CSI represents a primary vulnerability for many existing approaches. An RIS shapes the channel coherently, so small per-element phase errors add across $N$ elements and translate to large amplitude errors at the target. For phase-error variance $\sigma_\phi^2$, the expected beamforming gain degrades roughly as $N\cdot\exp(-\sigma_\phi^2)$, meaning the large apertures that motivate RIS deployment are inherently brittle under CSI noise. The effect worsens at mmWave, where beams are narrow and small angular misalignment drops gain sharply~\cite{rivetti2024malicious}. \ziyuu{Near-field security further intensifies this dependency because spherical-wave beamfocusing requires both radial and angular channel information. Range or location errors can then displace the intended secrecy focus and reduce the spatial separation between the LU and Eve~\cite{zhang20236g, zhao2025robust}.}

Attacks that rely on joint knowledge of legitimate and victim channels degrade sharply once CSI is estimated rather than assumed, such as selective jamming~\cite{mackensen2024spatial} and destructive beamforming~\cite{rivetti2024malicious}. Pre-fabricated passive tags~\cite{chen2023metawave, vennam2023mmspoof} sidestep online CSI but become tied to a specific spatial geometry. On the defense side, secrecy-rate and observability-control optimizations~\cite{magbool2025hiding, mao2022reconfigurable, jing2024reconfigurable, liu2022security} additionally require modeling the adversary's channel, which is rarely available in practice; DISCO-style anti-jamming explicitly substitutes statistical channel knowledge for exact CSI~\cite{huang2023anti}, acknowledging this limitation as a design constraint rather than attempting to solve it.

\noindent \textbf{Remarks.} While IEEE 802.11bf~\cite{ropitault2024ieee} will relax this constraint in Wi-Fi settings by standardizing sensing feedback, easier CSI access will still assist both defenders and attackers. To bypass CSI bottlenecks altogether, recent defenses explore adaptive blind beamforming~\cite{lai2024adaptive}, RL-based configurations~\cite{zou2023ris, nasser2024online}, and offline codebooks~\cite{kompostiotis2025optimizing}, though these approaches often shift the operational cost to training overhead, calibration effort, and generalization risks.

\subsection{Real-Time Constraints}
\label{sec:lim-compute}
\paragraph{Computational Latency}
Even with perfect CSI, configuring an RIS is a high-dimensional, typically non-convex optimization problem coupled with transmit beamforming and power allocation. Iterative~\cite{mackensen2024spatial}, alternating~\cite{rivetti2025destructive}, and joint secrecy-rate methods~\cite{magbool2025hiding, mao2022reconfigurable} demonstrate optimal performance in simulation but incur processing latencies that compete with the channel coherence time. This is particularly challenging at mmWave frequencies and under mobility conditions: by the time an optimal configuration is computed, the specific channel realization it was optimized for may have already expired.

Attack-side methods requiring real-time adaptation to a moving victim~\cite{mackensen2024spatial} or multiple users~\cite{rivetti2024malicious} are heavily bounded by this latency gap. In response, practical defense mechanisms often sacrifice mathematical optimality for timeliness. For instance, ML-based configurations replace iterative search with feed-forward evaluation~\cite{nguyen2021machine}; IRShield implements randomized incremental element flips~\cite{staat2022irshield}; and Protego explicitly cites the infeasibility of per-packet joint optimization by programming a 1-bit surface from precomputed patterns~\cite{li2022protego}.

\ziyuu{
\paragraph{Imperfect Synchronization}
Beyond computational latency, imperfect synchronization is also a real-time constraint for practical RIS attacks and defenses. 
Static passive reflectors do not generate an independent carrier. However, attacks based on coordinated active jamming or rapid RIS reconfiguration may require knowledge of the legitimate system timing. 
In particular, channel-aging attacks must change the RIS configuration between channel estimation and data transmission. Inaccurate pilot/data-phase identification, switching jitter, or control delay can weaken the induced mismatch~\cite{huang2023disco, khalid2025malicious}.

On the defensive side, channel-dependent secure beamforming and anti-jamming schemes must be applied before the estimated channel or interference state becomes outdated~\cite{huang2023anti}. 
Similarly, RIS-based challenge-response authentication requires each RIS configuration to be correctly associated with its corresponding probe and channel response, and the misalignment may distort the intended fingerprint and increase authentication errors~\cite{tomasin2024analysis}. 
These timing effects remain insufficiently quantified in practical security evaluations.
}

\noindent \textbf{Remarks.} An architectural bottleneck persists across the literature: the optimizer, the RIS controller, and the physical channel operate on misaligned timescales, forcing practical deployments to compromise on either responsiveness or optimality. \ziyuu{Evaluations of attacks and defenses should record their tolerance to imperfect synchronization, including timing, frequency, and phase offsets.} Reported results that assume these timescales align, such as a static channel during a long optimization, should be interpreted as upper bounds rather than deployment-ready performance.
\vspace{-10pt}

\subsection{Hardware and Physical-Layer Constraints}
\label{sec:lim-hardware}
Hardware realities impose limitations that are frequently abstracted away in theoretical models but dominate in physical deployment. Passive RIS architectures lack power amplification, subjecting the cascaded path to double path loss~\cite{zhang2022active}. Furthermore, practical reflecting elements utilize coarsely quantized phases (often 1-bit) with frequency-dependent responses, \ziyuu{limiting the accuracy of security beams and spatial nulls. Their reflection amplitude and phase are also coupled, thus changing the phase profile can alter the reflected power~\cite{abeywickrama2020intelligent}. Mutual coupling makes each element's response dependent on neighboring elements, while calibration errors create persistent mismatches between optimized and implemented reflection profiles~\cite{qian2021mutual, yang2023practical}. These non-ideal factors can distort intended beams and nulls, increase unintended sidelobes, and thereby reduce secrecy performance~\cite{salem2023impact}. 
Besides, these non-idealities could broaden or displace the focal region in near-field operation, which may cause unintended leakage outside the desired range-angle location~\cite{zhang2025near}.}

The double-path-loss penalty inherently restricts the usable range of passive RISs, a limitation that is exacerbated at higher frequencies. At mmWave, the surface must typically be positioned within a few meters of either the transmitter or receiver to maintain a viable link budget. While active RIS architectures mitigate this attenuation, their internal amplifier noise, static power draw, and hardware complexity scale with $N$, eventually negating the energy-efficiency advantages over traditional relays~\cite{zhang2022active}. 
\ziyuu{Controller latency is also non-negligible: channel estimation, optimization, control signaling, and element switching can all cause the end-to-end reconfiguration delay~\cite{etsi2023ris4}. If this delay is comparable to the channel coherence time, the implemented security beam or spatial null could already be outdated.
Recent ETSI testing guidance further defines common procedures for characterizing RIS beam patterns, operating bandwidth, EVM, radar cross section, and static power consumption, providing a useful hardware baseline for security experiments~\cite{etsi2026ris8}.}
Frequency selectivity further bounds reported results for passive-RIS eavesdropping~\cite{chen2022malicious, chen2023wavefront}, selective jamming~\cite{mackensen2024spatial}, and radar spoofing tags~\cite{chen2023metawave, vennam2023mmspoof}. On the defense side, measured indoor secrecy spectral efficiency drops substantially under wideband excitation relative to narrowband continuous waves~\cite{kompostiotis2025optimizing, vordonis2025evaluating}. This constraint is particularly acute in acoustic systems, where existing metasurface designs are rigidly constrained to narrow operational bands~\cite{10.1145/3458864.3467880, 10.1145/3643832.3661882, 10100683}.

\noindent \textbf{Remarks.} Reported attack ranges and defense secrecy rates should always be read alongside the operating frequency, bandwidth, element count, \ziyuu{phase resolution, calibration accuracy, and reconfiguration latency}; narrowband single-frequency results do not transfer to wideband operation, and fixed-$N$ results do not scale linearly. 
Co-designing the phase-profile algorithm with the underlying hardware, rather than treating the RIS as an ideal phase shifter, is the direction most likely to close this gap. 
\vspace{-5pt}

\subsection{Scale and Multi-Entity Coupling}
\label{sec:lim-multiuser}
A significant portion of the literature evaluates RIS security assuming a simplified topology of one legitimate user and one adversary. However, scaling these models to multiple users and multiple adversaries fundamentally alters the optimization structure and the achievable secrecy capacity.

A single RIS possesses finite aperture and phase-control resolution. Serving multiple spatial targets necessitates either partitioning the surface elements or time-sharing configurations, both of which degrade per-target efficacy. When jamming energy is distributed across multiple victims, per-target denial weakens~\cite{mackensen2024spatial}; similarly, steering destructive beamforming toward multiple locations reduces the suppression depth and intensifies inter-target coupling~\cite{rivetti2024malicious}. Furthermore, joint optimization over $K$ users and $M$ adversaries scales the CSI acquisition burden to $\mathcal{O}(KM)$ cascaded channels, where the estimation errors discussed in Section~\ref{sec:lim-csi} compound rapidly, and joint designs can unintentionally increase leakage toward specific eavesdroppers. Fairness remains an under-addressed metric, as sum-secrecy-rate objectives frequently leave spatially correlated users in chronic outage~\cite{li2022protego, cui2019secure, wafai2025opportunistic}. Recent works attempt multi-user, multi-adversary joint optimization~\cite{zou2023ris, tang2025joint}, but the resulting designs are high-dimensional and assume CSI quality that is unrealistic at scale~\cite{monemi2025practical}.

\textbf{Remarks.} Single-user, single-adversary results should not be extrapolated to dense deployments without re-evaluation. The research gap here is not only algorithmic but also evaluative: benchmarks that stress multiple users, multiple adversaries, fairness, and outage, rather than reporting average sum-secrecy rate, would give a much clearer picture of what RIS-based security actually delivers at scale.

\subsection{Mobility and Environmental Dynamics}
\label{sec:lim-mobility}
The optimization frameworks underlying most RIS attacks and defenses predominantly assume static or quasi-static geometries, which rarely reflect operational environments.

Mobility introduces rapid variations in Tx--RIS--target path lengths, dynamic multipath structures, and Doppler shifts that broaden the frequency response~\cite{wang2025wi}. Irregular target movement disrupts the path-length consistency essential for precise beam- or null-steering~\cite{ren2022gopose}. 
\ziyuu{Additionally, near-field focusing is even more sensitive to radial motion: even when the user's angle changes only slightly, displacement along the propagation direction may move the receiver outside the intended focal depth~\cite{monemi2024characterization, rahal2024ris}.}
Consequently, spoofing attacks designed for structured environments~\cite{chen2023metawave, vennam2023mmspoof} and selective jamming schemes~\cite{mackensen2024spatial} exhibit significant performance degradation when subjected to target tracking dynamics or view-angle variations~\cite{hunt2024madradar}.

\ziyuu{
Note that mobility does not affect all attack and defense classes uniformly. 
For CSI-dependent coherent attacks and defenses, Doppler, feedback, and RIS reconfiguration latency may make the applied phase profile unfit for the updated channel profile. The resulting performance depends on the CSI update interval and mobility level~\cite{shah2024optimizing}. 
Conversely, CSI-free channel-aging attacks create a pilot-data channel mismatch, although natural mobility may reduce their controllability and selectivity~\cite{huang2023disco}.
}

Environmental dynamics compound these tracking challenges. Background clutter, physical obstacles, and dynamic multipath inject CSI variations that mirror the signatures monitored by anomaly-based detectors, leading to elevated false-positive rates in practical deployments. Additionally, common physical barriers such as walls or curtains heavily attenuate injected attack signals~\cite{ning2025portablestealthyinaudiblevoice, ren2023person}.

\noindent \textbf{Remarks.} Evaluations constrained to stationary, line-of-sight laboratory setups generally represent an upper bound on performance. For both attacks and defenses, reporting results under a range of mobility profiles and clutter conditions, rather than a single stationary setup, is needed to distinguish methods that work only in the lab from those that survive deployment.
\vspace{-5pt}

\subsection{The Stealthiness--Effectiveness Tradeoff}
\label{sec:lim-stealth}
A unique characteristic of malicious RIS deployments is that stealthiness and operational effectiveness exhibit a fundamental tradeoff. Because passive RISs emit no active RF signals, they are inherently difficult to detect via standard spectrum monitoring. However, maintaining this covert profile imposes severe constraints on the attack vector.

To evade anomaly detection, an adversary must restrict the magnitude of the injected channel perturbation to mimic benign environmental scattering, which caps the maximum attack depth and operational range~\cite{ning2025stealthyvoiceeavesdroppingacoustic}. Minimizing observable side effects on un-targeted devices necessitates highly precise spatial focusing, which compounds the CSI acquisition burden (Section~\ref{sec:lim-csi}) and computational overhead (Section~\ref{sec:lim-compute}). Consequently, stealthy configurations are highly brittle to environmental variations.

Defenders encounter a corresponding challenge. Effect-based detection algorithms~\cite{stamatelis2025detection, shui2025analysisdetectionrisbasedspoofing} operate on the premise that any effective RIS attack must alter the propagation environment. While a highly sensitive detector can identify these subtle channel variations, increasing the detection sensitivity inevitably amplifies the false-positive rate triggered by the benign environmental mobility and clutter discussed in Section~\ref{sec:lim-mobility}.

\noindent \textbf{Remarks.} Stealthiness should be treated as a design parameter with quantifiable cost, not a binary property. Reporting attack effectiveness at matched detection-evasion rates, and reporting defense detection rates at matched false-positive rates, would make results across this literature directly comparable. While currently this is difficult because each work fixes its own operating point.
}
\section{Future Works} \label{futurework}
In recent years, RIS has emerged to be a key enabling technology to next generation systems such as 6G, ISAC, etc. In the following, we will introduce how RIS can be integrated into next generation applications such as AI, 6G, movable antenna, near-field configurations and real-world scenarios. Also, we will discuss potential security problems in these scenarios, as well as under multiple RIS configurations. 

\subsection{AI-Aided RIS Systems}
\ziyuu{AI/ML methods are increasingly used to map channel, user, and network states to RIS configurations, reducing the need for repeated online optimization. Existing studies have mainly considered performance-oriented applications, including beamforming, resource allocation, and LLM-assisted RIS control in dynamic environments~\cite{ahmed2024comprehensive, liu2024llm}. From a security perspective, this introduces two related questions: whether the learning pipeline controlling an RIS can itself be manipulated, and whether learned policies can be used to configure an RIS for attack or defense~\cite{wang2025navigating}.

\textbf{Adversarial learning against RIS-assisted AI systems.}
At inference time, crafted perturbations to CSI, pilot signals, or sensing inputs may cause an AI-enabled RIS controller to produce an ineffective or malicious configuration. The vulnerability of AI-powered RIS models to adversarial examples has already been demonstrated~\cite{catak2022security}. Related studies in wireless sensing and localization also show that learning models can be sensitive to adversarially modified wireless observations~\cite{yin2025evasion, wang2022adversarial}, although these studies do not directly model closed-loop RIS control. Future evaluations should therefore consider physically realizable perturbations constrained by wireless propagation, waveform structure, and RIS hardware, rather than only modifying model inputs numerically.

\textbf{Data poisoning.}
RIS controllers trained or updated using online observations and reinforcement-learning feedback may also be compromised during training. An adversary could manipulate CSI samples, training labels, replay buffers, reward signals, or distributed model updates, causing the controller to learn persistently biased configurations. Reward poisoning has recently been studied for DRL-based RIS control, together with reward clipping and anomaly filtering as countermeasures~\cite{tashman2026trustworthy}. Other poisoning settings, particularly those involving online channel observations and distributed RIS coordination, remain insufficiently studied.

\textbf{Model inversion.}
AI-assisted RIS controllers may encode information about user locations, channel maps, trajectories, or environmental geometry. General model-inversion attacks have shown that sensitive training information can sometimes be inferred from model outputs or parameters~\cite{fredrikson2015model}; however, this threat has not yet been systematically demonstrated for RIS controllers. Future work should clarify the adversary's required level of model access, determine which channel or location attributes can be recovered, and quantify the tradeoff between privacy protection and RIS control performance.

\textbf{AI-driven RIS attacks and defenses.}
Existing malicious-RIS studies have demonstrated destructive beamforming, eavesdropping, falsification, and spatially selective jamming~\cite{wei2023metasurface, rivetti2024malicious, mackensen2024spatial}, while malicious manipulation of channel reciprocity can also undermine physical-layer key generation~\cite{wei2023adversarial}. Most of these attacks use model-based optimization rather than AI. More recently, DRL has been applied to optimize RIS-assisted selective multi-user jamming~\cite{sabud2025mujam}, indicating that learning may make malicious configurations more adaptive to changing targets and channel conditions. Conversely, defensive learning could identify abnormal CSI or RIS behavior and select configurations that suppress interference or signal leakage. Non-reciprocal D-RIS links provide a model-based defensive baseline~\cite{chen2024defensive}, while learning-based RF fingerprinting may provide useful identification features~\cite{zhao2024cross}. However, recent backdoor attacks against RF-fingerprinting models show that the defensive model may itself become an attack surface~\cite{zhao2025protocol, zhao2024explanation}. These methods should therefore be evaluated under mobility, CSI errors, unseen environments, and adaptive attackers, with explicit reporting of false alarms and generalization failures.}

\vspace{-10pt}

\ziyuu{
\subsection{Near-Field RIS Security}
Previous analysis shows that existing RIS security studies generally assume far-field plane wave propagation and beam steering. However, this assumption is not suitable in near-field scenarios. The spherical wavefront and beam focusing nature in the near field allows LUs and adversaries located in similar directions but at different distances to be distinguished, as suggested in Section~\ref{RIS}. Because of this, there are some distinctive security problems and opportunities in the near field. 
Existing studies have already exploited range-selective defenses. When the LU and the attackers are angularly aligned, covert transmission can be enabled, and jamming and eavesdropping can be suppressed through beamfocusing~\cite{lei2025ris, liang2025double, liu2024ris, cui2024xl}. These results show the potential of focal-depth isolation and near-field spatial confinement with an RIS, although most of them are based on optimized and relatively static geometries. 
Besides, recent studies show that an RIS at the near-field can also reshape the rank and correlation structure of the cascaded channel~\cite{wu2026engineering}. This suggests that RIS placement may also affect the spatial separability in security designs. Recently ETSI has also privided a guidance on dedicated procedures for RIS near-field channel modeling, simulation, and measurement~\cite{etsi2026ris7}.

However, range-selective attacks with malicious RIS remain largely unexplored. Theoretically, a malicious RIS can exploit near-field beamfocusing to redirect legitimate signal to an eavesdropper at the same angle as the LU, concentrate reflected jamming energy at a victim's range, or induce destructive combining at a selected location while limiting its effects at nearby locations. Future attack models should quantify the attacker's required range or location knowledge, CSI, deployment, achievable focal resolution, and sensitivity to hardware and control errors. 
In addition, existing evaluations rarely report focal width and depth, the minimum safe LU-eavesdropper distance and angle, or signal leakage outside the intended range-angle region~\cite{monemi2024characterization}. Future work should evaluate these security boundaries under location uncertainty, radial mobility, mixed near-/far-field links, and hardware-induced grating lobes~\cite{zhao2025robust, zhang2025near, rahal2024ris}. 
}

\subsection{Real-World Deployment}
RIS attack and defense research should transition from idealized simulations to real-world deployments. There is an urgent need in realistic channel and environmental modeling, including mobility, dynamic multipath, partial line-of-sight, near-field effects, and hardware impairments, as identified in recent surveys~\cite{kaur2024survey, zhang2025systematic, li2025risbased}. Because practical systems exchange CSI and control information over time, temporal attack strategies and cross-packet defense mechanisms (e.g., anomaly detection across CSI time-series) will be essential. 

Next, the impact of hardware constraints, such as phase quantization, switching latency, mutual coupling, and imperfect reflectivity, on both attack feasibility and defense effectiveness deserves systematic study. On the authentication front, designing joint RIS-user authentication protocols (including RIS fingerprinting or physical-layer cryptographic methods) is increasingly important, particularly for vehicular and IoT networks that leverage RIS to assist authentication~\cite{shawky2025reconfigurable}. With the rise of AI/ML-driven RIS control frameworks, it becomes imperative to architect robust learning frameworks capable of resisting adversarial perturbations (e.g., from compromised RIS units) and detecting anomalous reflection patterns through spatio-temporal or graph-based models. 

Finally, there is an urgent need to establish experimental testbeds and open-source benchmark datasets for RIS security research, covering both legitimate deployments and malicious RIS-controlled threat models, thereby bridging the gap between theoretical analysis and field deployment.

\ziyuu{
\subsection{Security under Multi-RIS Scenarios}
% Distributed/multiple RIS, Multi-hop RIS, Element allocation
Recent RIS deployment studies characterize multi-RIS architectures along two distinct but related dimensions: the spatial distribution of the surfaces and the reflection order. In distributed single-reflection deployments, several RISs assist different users, links, or coverage regions, but each RIS-assisted path involves only one surface. These architectures have been studied in point-to-point, point-to-multipoint, and point-to-area scenarios. In multi-reflection deployments, the same signal is successively reflected by two or more RISs. The cascaded path's design then involves placement, inter-RIS connectivity, element allocation, and route selection~\cite{mei2022intelligent, wu2025intelligent, yuan2025reconfigurable}. Centralized and distributed architectures trade aggregate beamforming gain against spatial diversity and multiplexing~\cite{chen2024intelligent}. Orthogonal multi-RIS placement, element allocation, and joint multi-BS/multi-RIS optimization further extend these decisions to spatial-multiplexing and network-level deployment~\cite{chen2025spatial, yu2025green}.

Security research on these systems has also begun, including the joint optimization of multiple legitimate RISs for mmWave secrecy rate~\cite{rafieifar2021secrecy}, secure spectrum sharing~\cite{wang2023hybrid}, VLC authentication, secrecy, and integrity~\cite{soderi2024multi}, and double-RIS near-field secure links~\cite{liang2025double}. These studies, however, generally treat every participating surface, controller, and control link as trusted, and assume the number and locations of the RISs to be fixed.

This trusted-infrastructure assumption leaves an important gap. Each additional surface introduces another physically exposed node, and may require an additional controller, control link, and maintenance interface. An adversary controlling one distributed RIS could degrade a particular user cluster, redirect energy toward an eavesdropper, or shape interference while the rest of the network operates normally. Such localized effects may resemble blockage, fading, or hardware faults. In a multi-reflection route, compromising an intermediate surface may be more consequential: because the cascaded channel is multiplicative, the attacker could attenuate the entire route, redirect its signal, or disrupt the joint configuration of the participating RISs. Moreover, the receiver observes only their combined response, so a loss of rate or secrecy does not directly reveal which surface caused it.

Multi-RIS operation may also complicate the detection strategies discussed in Section~\ref{defenses}. When many legitimate surfaces continuously reshape the channel, the normal propagation state becomes less stationary, and the effects of a rogue or compromised RIS may be masked by legitimate reconfiguration-induced variations. Cross-domain deployments sharpen this problem further because RISs operated by different network operators, infrastructure providers, or property owners may not share the same authentication and configuration policies, blurring the trust boundary underlying the taxonomy in Section~\ref{threats}.

Future threat models should therefore admit partial compromise, including one or more compromised RISs, impersonated controllers, forged coordination messages, coherently colluding surfaces, and RISs spanning different trust domains. They should also distinguish between compromising a local distributed RIS and an intermediate RIS shared by several reflective routes. Deployment design should then treat trust and resilience as variables alongside placement, element allocation, association, and routing. Potential mechanisms include authenticated inter-RIS coordination, configuration attestation and provenance, compromised-surface localization, revocation, and rerouting through trusted alternatives when route diversity permits. These mechanisms should additionally tolerate imperfect inter-RIS CSI, asynchronous configuration updates, and controller latency.
Meanwhile, evaluation should accordingly report resilience to partial compromise, attack-localization accuracy, rerouting overhead, and recovery time rather than secrecy rate alone. Measurement-based multi-RIS evaluations under mobility, asynchronous control, imperfect CSI, and hardware nonidealities are needed to determine whether the predicted security and route-diversity gains persist in practical deployments.
}

\ziyu{
\subsection{6G Usage Scenarios}

\begin{table*}[!ht]
\centering
\caption{Role of RIS and corresponding security concerns in 6G usage scenarios.}
\label{tab:future_work_6g}
\begin{tabular}{|p{3cm}|p{6cm}|p{7cm}|}
\hline
\textbf{Scenario} & \textbf{RIS Role} & \textbf{Security/Privacy Concern} \\ \hline
Immersive communication~\cite{shen2023toward, liu2021learning} & Bypass blockages for THz/XR links; sustain high-rate holographic and multi-sensory traffic. & Biometric/behavioral leakage from telepresence and holography; deepfake impersonation, eavesdropping, and DoS on latency-sensitive media~\cite{naeem2023security}. \\ \hline
Massive communication~\cite{shi2024risaided, zhao2019survey, sanguinetti2019toward} & Large reconfigurable arrays integrated with massive MIMO; meet high-density, heterogeneous-rate, low-power requirements. & Expanded attack surface from resource-constrained IoT; large-scale surveillance via continuous data collection; RIS redirection for covert channels. \\ \hline
Hyper-Reliable and Low-Latency Communications (HRLLC)~\cite{LANG2025110916, 10255487, 11052282, el2024multi, 10333792, hassan2021key} & Alternative reflection paths on walls, tunnels, and industrial structures; maintain reliability under blockage. & Controller manipulation causing sudden drops in latency-critical links; precision jamming via focused RIS reflections; exposure of control signals in previously shielded regions. \\ \hline
Ubiquitous connectivity~\cite{xiao2024space, saeed2021point, toka2024ris, geraci2022will, jiang2022green, song2024high, 11010845, bariah2022ris} & Redirect weak satellite, UAV, or HAPS signals into underserved regions in space--air--ground integrated networks. & Malicious RIS redirecting non-terrestrial signals to unintended users or forging coverage; difficulty of verifying legitimate RIS at wide-area scale. \\ \hline
AI and communication~\cite{6g_white_paper, mao2021ai, raihan2023overview, padmapriya2022ai} & AI-driven RIS control for adaptive beamforming, mobility tracking, and network self-adaptation. & AI models vulnerable to adversarial inputs~\cite{ambalkar2021adversarial}; RIS can be weaponized for physical-layer attacks against AI-driven control loops. \\ \hline
ISAC~\cite{6g_white_paper, magbool2025survey} & Dual-functional waveform design; improved communication-sensing trade-offs through propagation shaping. & Shared frequency bands heighten eavesdropping and unauthorized-access risk~\cite{naeem2023security, magbool2025survey}; malicious RIS configurations inject falsified sensing data into both legitimate and adversarial observers. \\ \hline
\end{tabular}
\vspace{-10pt}
\end{table*}

According to the ITU-R M.2160 framework~\cite{6g_white_paper}, 6G envisions six primary usage scenarios, in all of which RIS is expected to play a role. Each scenario also introduces distinct security and privacy concerns that arise from, or are amplified by, RIS integration. Table~\ref{tab:future_work_6g} summarizes the RIS role and corresponding security challenge for each. 
\ziyuu{RIS-enabled satellite and UAV-integrated systems deserve particular attention~\cite{zhai2022energy}. Long propagation distances, rapidly changing geometries, and intermittent control links make cascaded CSI acquisition and RIS authentication more complicated~\cite{xiao2024space, bariah2022ris, toka2024ris}. UAV-mounted RISs additionally expose their locations, trajectories, and controllers to physical capture, navigation spoofing, and unauthorized reconfiguration. This risk motivates joint platform--RIS authentication and mobility-aware configuration verification.}

Two recurring patterns emerge. First, RIS acts as a dual-use enabler: the same propagation-shaping capability that restores blocked links, boosts throughput, or extends coverage also gives adversaries new leverage for eavesdropping, selective jamming, spoofing, and covert surveillance. Second, the security implications compound with scale: ultra-dense IoT deployments, non-terrestrial integration, and AI-driven control all enlarge the attack surface and make authentication, anomaly detection, and trust-chain management central to future 6G-RIS research.
\vspace{-10pt}
}

\subsection{Incorporating Movable Antennas}
Movable antennas (MA)~\cite{zhu2025tutorial}, also known as fluid antenna systems (FAS)~\cite{new2024tutorial}, exploit spatial diversity by reconfiguring element positions within a predefined region, introducing additional degrees of freedom for beamforming, polarization control, and frequency tuning at lower hardware cost than massive MIMO. 

\ziyu{
Element-wise, MA can be realized in three ways. \textbf{Mechanical movable antennas}~\cite{zhu2023movable} use stepper motors to reposition antennas in a coordinate system, but the movement area and speed are limited and the system is prone to wear. \textbf{Liquid or fluid elements}~\cite{paracha2019liquid, huang2021liquid, martinez2022toward} offer flexible reconfigurability in frequency, radiation pattern, and polarization, and can greatly improve outage probability and multiplexing gain in mmWave systems~\cite{shen2024design}; however, position-flexible liquid-based MA is still emerging and sensitive to fluctuations~\cite{new2024tutorial}. \textbf{Electronically reconfigurable elements} use switches to reconfigure within microseconds, similar to typical RIS, but are costly in circuit complexity and limited in phase-center offset and coverage range~\cite{ning2025movable}.

MA concepts can be combined with RIS in two ways: element-level, where RIS elements themselves become position-reconfigurable, and surface-level, where the RIS is made sliding~\cite{zhang2025movable}, rotatable~\cite{shi2024capacity}, or foldable~\cite{chahat2017deep}. Surface-level movable RIS is more mature and reduces hardware cost relative to element-level MA.
From a security perspective, MA/FAS integration introduces a new attack surface: adversaries can exploit position control to reshape the channel geometry without modifying phase shifts, potentially bypassing defenses that monitor only the RIS phase configuration. 
Conversely, defenders can use position reconfiguration as an additional randomization degree of freedom against eavesdropping and sensing-based inference.
\ziyuu{Recent studies jointly optimize antenna positions, RIS coefficients, beamforming, and artificial noise to limit information leakage~\cite{ma2025movable} and improve secrecy in RIS-assisted ISAC and directional-modulation systems~\cite{li2025secure}. However, this research remains at an early stage, with most results relying on numerical evaluation, channel or location knowledge, accurate antenna positioning, and timely joint optimization. The effects of mechanical latency, calibration errors, mobility, and position-control attacks require further investigation.}
\vspace{-10pt}
}

\subsection{Low-Cost RIS}
% MobiCom articles
% - alumin paper
% - 3D printing
% Low-cost + RIS + ATTACK & defense
% TODO(Thomas): need revise
While recent investigations have illustrated the potent threat and defense potential of RIS in wireless security, several critical directions remain underexplored. 
First, attack modeling for systems with coarse (e.g., 1-bit) phase resolution or minimal element count requires rigorous quantification: prior work has demonstrated spatially selective jamming under lab conditions~\cite{mackensen2024spatial}, yet the boundary conditions for inexpensive hardware (element count, quantization error, placement constraints) remain unknown. 
Second, defensive countermeasures such as spatio-temporal consistency checks, RIS-aware physical-layer authentication, and side-channel detection of RIS switching lack systematic evaluation under realistic constraints. 
Third, most studies assume ideal or high-cost RIS elements, whereas practical setups built from inexpensive FR-4 boards, PIN diodes, slow biasing networks, and commodity controllers would enable reproducible benchmarks. 
Fourth, metrics and datasets for RIS-enabled attack and defense need standardization: element count, switching rate, geometry, secrecy-rate gain, and publicly shared CSI/RF traces would facilitate cross-study comparison. 
Finally, the dual-use nature of low-cost RIS raises the need for integrated attack-defense co-design frameworks that account for quantization, placement, and hardware errors. These gaps align with recent survey calls for robust, scalable, hardware-aware RIS-PLS frameworks~\cite{li2025risbased, iqbal2025comprehensive}.
\section{Public Resources and Reproducibility Gaps}\label{discussion}
To facilitate follow-up research, we maintain a public website, \emph{Awesome RIS Attacks and Defenses}, which curates RIS-security-related papers and currently available public resources.\footnote{\url{https://awesome-ris-security.github.io/}} Table~\ref{tab:open-resources} summarizes the security-relevant demos, datasets, and code that are publicly available at the time of writing.

\begin{table}[!ht]
    \centering
    \begin{tabular}{|c|c|p{4cm}|c|}
        \hline
        \textbf{Reference} & \textbf{Year} & \textbf{Category} & \textbf{Resources} \\ \hline
        \cite{shaikhanov2022metasurface} & 2022 & Attacks Leveraging the RIS & \href{https://drive.google.com/file/d/1hu5ivAArYmeul0-GmbCC0SM9leYhDIeJ/view}{Demo} \\ \hline
        \cite{vennam2023mmspoof} & 2023 & Attacks Leveraging the RIS & \href{https://wcsng.ucsd.edu/mmspoof/}{Demo} \\ \hline
        \cite{huang2023disco} & 2023 & Attacks Leveraging the RIS & \href{https://github.com/huanhuan1799/Disco-Intelligent-Reflecting-Surfaces-Active-Channel-Aging-for-Fully-Passive-Jamming-Attacks}{Code} \\ \hline
        \cite{jiang2024risiren} & 2024 & Attacks Leveraging the RIS & \href{https://www.youtube.com/watch?v=yY80IhLvz3Q}{Demo} \\ \hline
        \cite{ning2025stealthyvoiceeavesdroppingacoustic} & 2026 & Attacks Leveraging the RIS & \href{https://github.com/helo0507/Super-Ear}{Code} \\ \hline
        \cite{chen2025secure} & 2025 & Securing RIS-Assisted Systems & \href{https://www.youtube.com/watch?v=oii3aj86tP0}{Demo} \\ \hline
        \cite{shaikhanov2025spoofing} & 2025 & RIS as a Defensive Tool & \href{https://sites.google.com/view/zhambyl-shaikhanov/spoofing-eve}{Demo} \\ \hline
        \cite{sun2025anti} & 2025 & RIS as a Defensive Tool & \href{https://static-content.springer.com/esm/art\%3A10.1038\%2Fs41467-025-62633-w/MediaObjects/41467_2025_62633_MOESM3_ESM.mp4}{Demo} \\ \hline
        \cite{staat2022irshield} & 2022 & RIS as a Defensive Tool & \href{https://zenodo.org/records/6367411}{Dataset} \\ \hline
        \cite{ning2025metaguardian} & 2025 & RIS as a Defensive Tool & \href{https://github.com/Meta-Guardian/MetaGuardian}{Code} \\ \hline
    \end{tabular}
    \caption{Open resources on RIS security.}
    \label{tab:open-resources}
    \vspace{-5pt}
\end{table}
\vspace{-5pt}

\ziyu{
Despite this effort, public resources for RIS security and privacy research remain limited. In particular, the community still lacks standardized attack implementations, defense benchmarks, shared CSI/RF traces, and reproducible evaluation scenarios dedicated to RIS-enabled security and privacy problems. General-purpose RIS simulation, channel-modeling, deployment, and implementation tools may help, as shown in Table~\ref{tab:tools}, but they are not security-specific benchmarks. Closing these reproducibility gaps is important for enabling more rigorous cross-study comparison and hardware-aware evaluation.
}

\begin{table}[!ht]
    \centering
    \begin{tabular}{|c|p{2cm}|p{5cm}|}
        \hline
        \textbf{Year} & \textbf{Provider / Publication} & \textbf{Resources} \\ \hline
        2019 & ASU Wireless Intelligence Lab & \href{https://deepmimo.net/}{DeepMIMO} \\ \hline
        2020 & Ko\c{c} University CoreLab & \href{https://github.com/ifyildirim/SimRIS-Channel-Simulator}{SimRIS Channel Simulator v2.0} \\ \hline
        2022 & NVIDIA & \href{https://github.com/NVlabs/sionna/releases/tag/v0.19.2}{Sionna RT v0.19.2 (RIS-supported release)} \\ \hline
        2023 & ISAP 2023 & \href{https://github.com/mheinri/OpenSourceRIS}{OpenSourceRIS} \\ \hline
        2024 & MathWorks & \href{https://www.mathworks.com/help/phased/ug/introduction-to-reconfigurable-intelligent-surfaces.html}{Introduction to RIS} \\ \hline
        2024 & MathWorks & \href{https://www.mathworks.com/help/phased/ug/reconfigurable-intelligent-surfaces-ris-aided-sensing.html}{RIS-Aided Radar Sensing} \\ \hline
        2024 & MathWorks & \href{https://www.mathworks.com/help/5g/ug/model-reconfigurable-intelligent-surfaces-with-cdl-channels.html}{RIS Modeling with CDL Channels} \\ \hline
        2024 & InterDigital & \href{https://github.com/InterDigitalInc/NeoRadium}{NeoRadium} \\ \hline
        2024 & University of Trento & \href{https://github.com/michele-segata/cooperis}{CoopeRIS} \\ \hline
        2025 & Telef\'onica Scientific Research & \href{https://github.com/Telefonica-Scientific-Research/DDRD}{Data-Driven RIS Deployment (DDRD)} \\ \hline
        Ongoing & GitHub community & \href{https://github.com/ken0225/RIS-Codes-Collection}{RIS-Codes-Collection} \\ \hline
        \end{tabular}
    \caption{Open RIS simulation, channel-modeling, deployment, and implementation resources.}
    \label{tab:tools}
    \vspace{-10pt}
\end{table}

\section{Conclusion} \label{conclusion}

In this paper, we have presented a comprehensive deployment-oriented survey of RIS in the perspective of security and privacy. We began with a detailed technical overview of RIS, followed by an examination of its practical applications, with particular emphasis on security and privacy challenges that arise in RIS-based environments. Subsequently, we conducted an in-depth analysis of the roles of RIS in attack, defense, and countermeasure strategies, and distilled prescriptive design lessons from representative pioneer works. Based on these insights, we discuss the limitations of existing solutions and identify several promising directions for future research, with reproducibility and benchmarking identified as a central need.
Our analysis underscores the critical role of RIS in enhancing the security and privacy of practical systems while also revealing open research challenges in this domain. We envision that this survey will serve as both a practical reference for engineers aiming to secure emerging 6G and IoT infrastructures and a research roadmap for scholars addressing the unresolved issues outlined here.

% In this paper, we present a survey of RIS in real-world systems. We first introduce a comprehensive technical overview of the RIS. Then, starting from real-world applications, we highlight the security and privacy issues in the real-world RIS-based applications. Next, we focus on and conduct an in-depth analysis of functionalities of RIS in attack, defense, and countermeasure strategies. Lastly, we analyze the limitations of current solutions and present some future directions. These analyses show that RIS plays a significant role in practical systems in security and privacy domain and point out open problems in this topic. We hope this survey will be a practical handbook for engineers hardening forthcoming 6G and IoT systems, and a road‑map for researchers tackling the open problems listed above.

\bibliographystyle{IEEEtran}
\bibliography{ref}

\end{document}